\pdfoutput=1
\documentclass[aps,prd,reprint,twocolumn,showpacs,superscriptaddress,nofootinbib]{revtex4-1}  
\usepackage{tabularx}
\makeatletter
\def\hlinewd#1{%
\noalign{\ifnum0=`}\fi\hrule \@height #1 %
\futurelet\reserved@a\@xhline}
\makeatother
\usepackage{scrextend}
\usepackage{amsmath}
\usepackage{amssymb}
\usepackage{epsfig}
\usepackage{graphicx}
\usepackage{hyperref}
\usepackage{dcolumn}   
\usepackage{slashed}
\usepackage{color}
\usepackage{rotating}
\usepackage[margin=0.9in,a4paper]{geometry}
\usepackage[table,xcdraw,dvipsnames]{xcolor}
\usepackage[utf8]{inputenc}
\usepackage{colortbl}
\usepackage[normalem]{ulem}
\usepackage{mathrsfs} 

\definecolor{nicered}{rgb}{0.7,0.1,0.1}
\definecolor{nicegreen}{rgb}{0.1,0.5,0.1}
\definecolor{red}{rgb}{1.0, 0, 0}

\renewcommand{\[}{\left[}
\renewcommand{\]}{\right]}
\renewcommand{\(}{\left(}
\renewcommand{\)}{\right)}

\newcommand{\bdm}{\begin{displaymath}}
\newcommand{\edm}{\end{displaymath}}
\newcommand{\bea}{\begin{eqnarray}}
\newcommand{\eea}{\end{eqnarray}}

\newcommand{\nn}{\nonumber}

\newcommand{\dmu}{\partial_\mu}

\newcommand{\pim}{\pi^-}
\newcommand{\pip}{\pi^+}
\newcommand{\piz}{\pi^0}

\definecolor{nicered}{rgb}{0.7,0.1,0.1}
\definecolor{nicegreen}{rgb}{0.1,0.5,0.1}
\definecolor{red}{rgb}{1.0, 0, 0}
\definecolor{niceblue}{rgb}{0,0,0.8}
\definecolor{red}{rgb}{1.0, 0, 0}
\hypersetup{colorlinks,citecolor= nicegreen,linkcolor= nicered,urlcolor=nicered}

\def\eq#1{{Eq.~(\ref{#1})}}

\def\fig#1{{Fig.~\ref{#1}}}

\def\sect#1{{Sect.~\ref{#1}}}

\def\app#1{{Appendix~\ref{#1}}}

\def\vev#1{\left\langle #1\right\rangle}

\def\Im{\mbox{Im}\,}
\def\Re{\mbox{Re}\,}

\def\gsim{\raise0.3ex\hbox{$\;>$\kern-0.75em\raise-1.1ex\hbox{$\sim\;$}}}
\def\lsim{\raise0.3ex\hbox{$\;<$\kern-0.75em\raise-1.1ex\hbox{$\sim\;$}}}

\def\mb[#1]{\mathbf{#1}}

\renewcommand{\bar}{\overline}

\definecolor{LightCyan}{rgb}{0.88,1,1}
\definecolor{piggypink}{rgb}{0.99, 0.87, 0.9}
\definecolor{applegreen}{rgb}{0.55, 0.71, 0.0}
\definecolor{darkpastelgreen}{rgb}{0.01, 0.75, 0.24}
\definecolor{green-yellow}{rgb}{0.68, 1.0, 0.18}

\newcommand{\beq}{\begin{equation}}
\newcommand{\eeq}{\end{equation}}
\newcommand{\beqa}{\begin{eqnarray}}
\newcommand{\eeqa}{\end{eqnarray}}

\begin{document}



\title{Axion-pion 
thermalization rate
in unitarized NLO chiral perturbation theory}

\author{Luca Di Luzio}
\email{luca.diluzio@pd.infn.it}
\affiliation{\small \it 
Dipartimento di Fisica e Astronomia `G.~Galilei', Universit\`a di Padova, Via F.~Marzolo 8, 35131 Padova, Italy}
\affiliation{\small \it 
INFN Sezione di Padova, Via F.~Marzolo 8, 35131 Padova, Italy}
\author{Jorge Martin Camalich}
\email{jcamalich@iac.es}
\affiliation{\small \it 
Instituto de Astrof\'iısica de Canarias, C/ V\'ia L\'actea, s/n E38205 - 
La Laguna, Tenerife, Spain}
\affiliation{\small \it 
Universidad de La Laguna, Departamento de Astrof\'isica - La Laguna, Tenerife, Spain
}
\author{Guido Martinelli}
\email{guido.martinelli@roma1.infn.it}
\affiliation{\small \it Physics Department and INFN Sezione di Roma La Sapienza, Piazzale Aldo Moro 5, 00185 Roma, Italy
}
\author{Jos\'e Antonio Oller}
\email{oller@um.es}
\affiliation{\small \it 
Departamento de F\'isica,
Universidad de Murcia,
E-30071 Murcia, Spain
}
\author{Gioacchino Piazza}
\email{gioacchino.piazza@ijclab.in2p3.fr}
\affiliation{\small \it 
IJCLab, P\^{o}le Th\'{e}orie (B\^{a}t.~210), CNRS/IN2P3 et Universit\'{e}  Paris-Saclay, 91405 Orsay, France}

\begin{abstract}
\noindent
We compute the axion-pion scattering $a \pi \to \pi \pi$, 
relevant for the axion thermalization rate in the early universe, 
within unitarized NLO chiral perturbation theory. 
The latter extends the range of validity of the chiral expansion of axion-pion scattering 
and thus provides a crucial ingredient 
for the reliable determination of the 
relic density of
thermal axions, whenever the 
axion decoupling temperature 
is below that of 
the QCD phase transition. 
Implications for cosmological 
observables are briefly discussed.

\end{abstract}

\maketitle

\section{Introduction} 
\label{sec:intro}

The QCD axion is a well-motivated new physics paradigm 
which provides at the same time a solution to the strong CP 
problem \cite{Peccei:1977hh,Peccei:1977ur,Wilczek:1977pj,Weinberg:1977ma} and a cold dark matter candidate \cite{Preskill:1982cy,Abbott:1982af,Dine:1982ah,Davis:1986xc}. 
Additionally, a thermal population of relativistic 
axions \cite{Turner:1986tb}, behaving as 
dark radiation or 
hot dark matter, 
might further contribute to the energy density of the universe.  
Thermally produced axions can be probed 
by cosmic microwave background (CMB) experiments, 
such as the Planck satellite \cite{Akrami:2018vks,Aghanim:2018eyx},   
as well as planned CMB Stage 4 (CMB-S4) 
surveys \cite{Abazajian:2016yjj}, 
which provide an observational 
window on 
the axion couplings to the 
Standard Model (SM) fields. 

Depending on the axion decay constant $f_a$ 
(or equivalently the axion mass $m_a \simeq 5.7 \times10^6 \text{GeV}/f_a$ eV) whose inverse sets the strength of axion couplings, 
there are several processes 
stemming from the model-independent axion coupling to gluons, 
$\frac{\alpha_s}{8\pi} \frac{a}{f_a} G \tilde G$, 
which can keep 
the axion in thermal equilibrium with the SM thermal bath. 
For $m_a \lesssim 10$ meV, 
thermal axion production 
dominantly 
proceeds 
via its scatterings with gluons 
\cite{Masso:2002np,Graf:2010tv}, 
corresponding to a 
decoupling temperatures, $T_D$, above the GeV scale. 
On the other hand, for heavier axions 
one has $T_D \lesssim 1$ GeV and hence
also processes involving pions and nucleons  
must be considered \cite{Berezhiani:1992rk,Chang:1993gm,Hannestad:2005df}. 
Although this transition region cannot be precisely determined 
due to the complications of the quark-hadron 
phase transition,\footnote{It was 
recently proposed in Refs.~\cite{DEramo:2021psx,DEramo:2021lgb} 
to interpolate the axion thermalization rate 
by matching the known high- and low-temperature asymptotic regions.} 
for axions approaching the eV scale  
the main thermalization channel is 
provided by the scattering 
$a\pi \to \pi\pi$ 
\cite{Chang:1993gm,Hannestad:2005df}, 
with $T_D \lesssim T_c$, 
where $T_c \simeq 155$ MeV~\cite{Aoki:2006br,Borsanyi:2010bp,Bazavov:2011nk} 
is the QCD deconfinement temperature. 
The highest attainable axion mass 
from cosmological constraints 
on thermally produced axions 
is known as the axion hot dark matter bound 
(for recent analyses, see Refs.~\cite{Caloni:2022uya,DEramo:2022nvb}),  
and it is mainly set by the axion-pion thermalization rate. 

The scattering $a \pi \to \pi \pi$ can be computed 
at low energies
within 
chiral perturbation theory (ChPT). 
The LO calculation was performed in Refs.~\cite{Chang:1993gm,Hannestad:2005df}, 
while Ref.~\cite{DiLuzio:2021vjd} considered 
the impact of NLO corrections in order to 
assess the convergence of the chiral expansion. 
In this paper, we correct a mistake 
of Ref.~\cite{DiLuzio:2021vjd} 
regarding the evaluation of the loop function in the NLO contribution. 
As discussed in the following, 
with the corrected result 
it can still be argued that the 
temperature where the 
chiral expansion of the 
axion-pion thermalization rate breaks down is  
$T_\chi \sim 70$ MeV, 
and hence it remains a 
crucial question to extend the validity of ChPT between 
$T_\chi$ and $T_c \simeq 155$ MeV. 
This is actually the main goal of the present work, 
that is to extend the chiral description of axion-pion scattering 
above the validity region of standard ChPT, 
by employing a unitarization technique 
known as the Inverse Amplitude Method (IAM) \cite{Lehmann:1972kv,Truong:1988zp,Oller:2020guq}. This method 
restores exact elastic unitarity attached to the so-called unitarity or right-handed cut of the amplitude, while preserving crossing symmetry perturbatively. 

The paper is structured as follows: 
In \sect{sec:ChPT} 
we recall the basic ingredients 
of the axion-pion chiral Lagrangian 
and update the NLO 
correction to axion-pion scattering in ChPT.    
Along \sect{sec:unitarization} 
we present the new calculation of the axion-pion scattering 
within unitarized NLO ChPT, 
whose impact 
on the axion-pion thermalization rate is 
subsequently discussed in 
\sect{sec:pheno}. 
In \sect{sec:breakdownChPT} we discuss the convergence 
of the chiral expansion, while cosmological implications 
are considered in \sect{sec:cosmo} and 
we finally conclude in \sect{sec:concl}. 
More technical details are deferred to a set of Appendices. 

\section{Axion-pion scattering in 
ChPT}
\label{sec:ChPT}

At the LO in the chiral expansion, the axion-pion 
effective Lagrangian is described by the contact interactions 
(see e.g.~\cite{Georgi:1986df,DiLuzio:2020wdo})
\begin{align}
\label{eq:axionpionLOexp}
\mathscr{L}^{\rm LO}_{a\text{-}\pi} 
&\supset 
\frac{C_{a\pi}}{f_a f_\pi} \partial^\mu a
[2 \dmu \piz \pip \pim -\piz \dmu \pip \pim \nonumber \\
&-\piz \pip \dmu\pim]
\, ,
\end{align}
and coupling strength 
\beq 
\label{eq:defCapi}
C_{a\pi} = \frac{1}{3}\( \frac{m_d-m_u}{m_u+m_d} +c_d^0-c_u^0\) \, .
\eeq
Here, $c_{u,d}^0$ are model-dependent coefficients which 
depend on the axion UV completion. 
For instance, $c^0_{u,d} = 0$ in the KSVZ model \cite{Kim:1979if,Shifman:1979if}, 
while $c^0_{u} = \frac{1}{3} \cos^2\beta$ and $c^0_{d} = \frac{1}{3} \sin^2\beta$ 
in the DFSZ model \cite{Zhitnitsky:1980tq,Dine:1981rt}, 
with $\tan\beta$ the ratio between the 
vacuum expectation values 
of two Higgs doublets.

For temperatures below the QCD phase transition, 
the main processes relevant for the 
axion thermalization rate are 
$a (p_1) \piz (p_2)\rightarrow\pip (p_3) \pim (p_4)$, 
whose amplitude at LO reads 
\beq 
\label{eq:Mapi0pippimLO} 
\mathcal{M}^{\rm LO}_{a \piz \rightarrow \pip \pim} =
\frac{C_{a\pi}}{f_\pi f_a} \frac{3}{2}\left[m_\pi^2-s\right] \, , 
\eeq
with $s= (p_1 + p_2)^2$, 
together with the crossed channels 
$a \pim\rightarrow\piz \pim$ and $a \pip\rightarrow\pip \piz$.   
The amplitudes of the latter 
are obtained by replacing 
$s \leftrightarrow t=(p_1 - p_3)^2$ 
and $s \leftrightarrow u=(p_1 - p_4)^2$, respectively. 
Taking equal masses for the neutral and charged pions, 
one finds the squared matrix element (summed over the three channels above) \cite{Hannestad:2005df} 
\begin{equation}
\sum|\mathcal{M}|_{\rm LO}^2= \left(\frac{C_{a\pi}}{f_a f_\pi}\right)^2\frac{9}{4}\left[s^2+t^2+u^2-3m_\pi^4\right] \, . 
\end{equation}
The formulation of the axion-pion chiral Lagrangian 
including axion derivative terms 
at the NLO 
was worked out in Ref.~\cite{DiLuzio:2021vjd} 
(see also \cite{DiLuzio:2022tbb}). 
The main ingredients are the axion-dressed $\mathcal{O}(p^4)$ 
terms of the standard chiral Lagrangian \cite{Gasser:1983yg} and  
the NLO pion axial current to which the axion couples derivatively. 
A non-trivial aspect, compared to the standard 2-flavour chiral Lagrangian, 
consists in the mixing between the axion 
and the neutral pion, 
which can be dealt with either by diagonalizing the 
axion-pion propagator at the 
NLO or by explicitly retaining the mixing in the 
Lehmann-Symanzik-Zimmermann reduction formula \cite{Lehmann:1954rq} 
for the $a\pi \to \pi\pi$ scattering amplitude. 
For more details, we refer the reader directly to Ref.~\cite{DiLuzio:2021vjd}. 

However, 
Ref.~\cite{DiLuzio:2021vjd} contained a mistake in the 
loop function of the NLO scattering amplitude, related to a wrong choice 
of the branch cut of the two-point unitary loop 
function
that affects the results
for negative $u$ and $t$. 
The corrected $a \pi^0 \to \pi^+\pi^-$ NLO amplitude is 
given in \app{app:NLOampl}, together with that for 
$a \pi^0 \to \pi^0 \pi^0$ which enters the cross-section 
only at NNLO order (being this channel absent at LO), but which will be needed for the nonperturbative unitarization method of the NLO ChPT $a\pi\to\pi\pi$ amplitudes 
to be discussed in \sect{sec:unitarization}. 

For the numerical evaluation of the perturbative ChPT rates discussed in this work we use the central values of the standard 
low-energy constants (LECs): 
$\overline{\ell_1} = -0.36(59)$ \cite{Colangelo:2001df}, 
$\overline{\ell_2} = 4.31(11)$ \cite{Colangelo:2001df},
$\overline{\ell_3} = 3.53(26)$ \cite{FlavourLatticeAveragingGroupFLAG:2021npn},
$\overline{\ell_4} = 4.73(10)$ \cite{FlavourLatticeAveragingGroupFLAG:2021npn},
$\ell_7 = 7(4) \times 10^{-3}$ \cite{diCortona:2015ldu}, along with
$m_u / m_d = 0.50(2)$ \cite{FlavourLatticeAveragingGroupFLAG:2021npn}, 
$f_\pi = 92.1(8)$ MeV \cite{Zyla:2020zbs} 
and $m_\pi = 137$ MeV 
(corresponding to the average neutral/charged pion mass).

\section{Unitarized axion-pion scattering} 
\label{sec:unitarization}

Partial wave amplitudes (PWAs) are the most adequate method to impose unitarity constraints to amplitudes at low energies. As it is also conventional in studies of $\pi\pi$ scattering, we start our analysis by projecting the  amplitudes $\mathcal M$ from the charge basis to a basis with well-defined total isospin $I$, giving rise to the amplitudes $A_I$. 
For $a\pi^0\to\pi^+\pi^-$ and $a\pi^0\to\pi^0\pi^0$ scattering  
(see Appendix~\ref{sec:IAMdetails} for conventions),  
\begin{align}
A_{0}&=-\frac{1}{\sqrt{3}}\left(2\mathcal M_{+-}+\mathcal M_{00}\right) \, ,\nonumber\\
A_{2}&=\sqrt{\frac{2}{3}}\left(\mathcal M_{00}-\mathcal M_{+-}\right) \, ,\label{eq:isospin_neutral}
\end{align}
where we have simplified the notation by indicating the charges of the two final pions as subscripts of the amplitudes in the charge basis. We have also used that $\mathcal M_{+-}=\mathcal M_{-+}$ because of charge conjugation symmetry.   

For $a \pi^+\to\pi^0\pi^+$ scattering,  
\begin{align}
A_{1}&=-\frac{1}{\sqrt{2}}\left(\mathcal M_{+0}-\mathcal M_{0+}\right) \, ,\nonumber\\
A_{2}'&=-\frac{1}{\sqrt{2}}\left(\mathcal M_{+0}+\mathcal M_{0+}\right) \, .
\label{eq:isospin_charged}
\end{align}
The amplitudes with definite isospin for $a\pi^-\to\pi^0\pi^-$ differ from $A_1$ and $A_2'$ only by a global minus sign.  Note that $A_{2}$ and $A_{2}'$ are different because the coupling of the axion with pions violates isospin.

The projection of these amplitudes into a basis of states with well-defined total angular momentum $J$ is obtained 
by means of the usual formulae for the PWAs of the scattering of spin zero particles,
\begin{align}
&A_{IJ}(s)=\frac{1}{2}\int_{-1}^{+1}dx P_J(x)A_I(s,x)\, ,\nn\\
&A_{I}(s,x)=\sum_{J=0}^\infty(2J+1)P_J(x)A_{IJ}(s)\, ,\label{eq:PWAs}
\end{align}
where $x=\cos\theta$ is the scattering angle in the center of mass and $P_J(x)$ are Legendre polynomials. 

\begin{figure*}[t!]
\centering
\begin{tabular}{cc}
        \label{fig:d11}
		\includegraphics[width=0.45\textwidth]{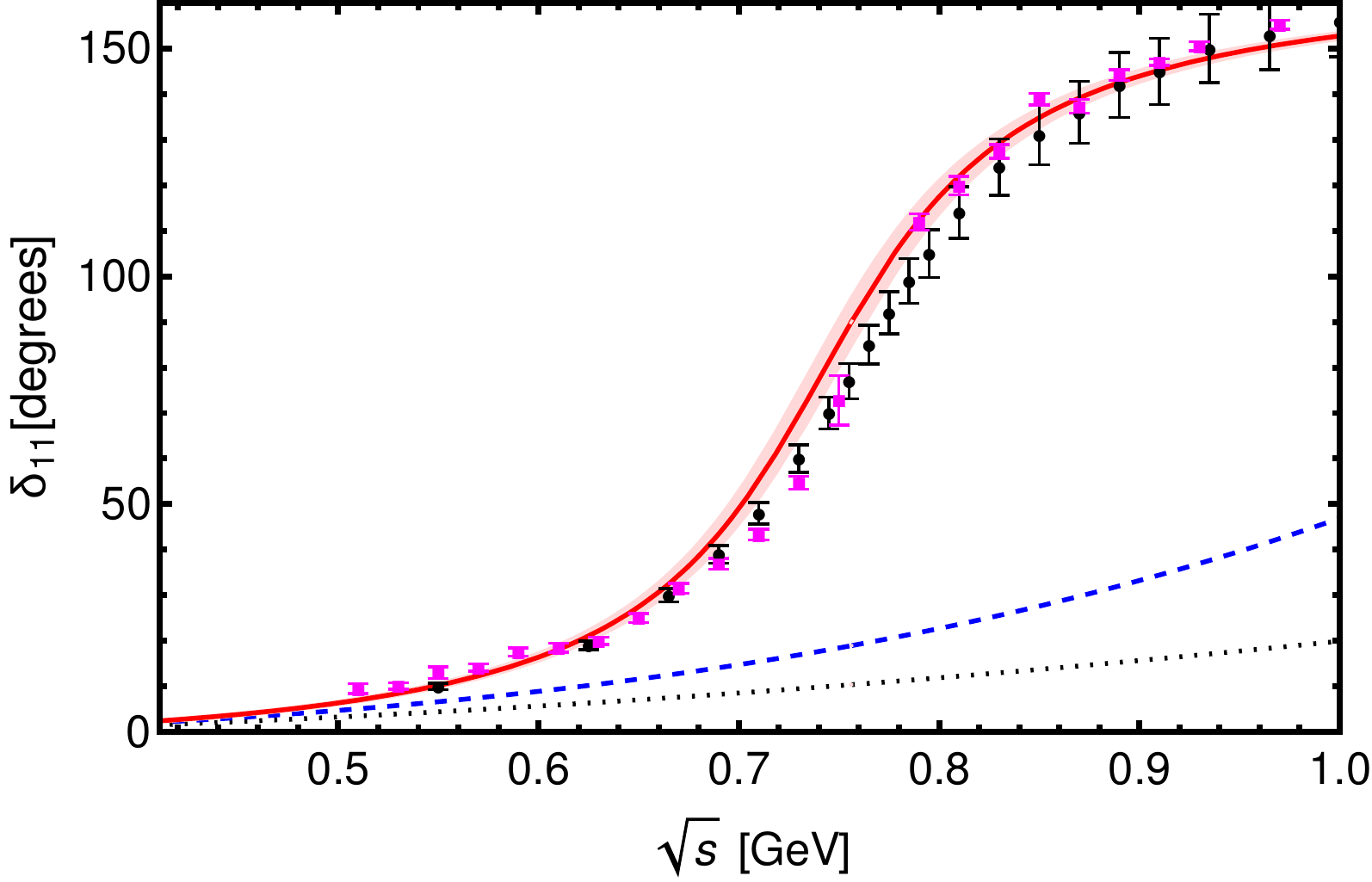}
&
        \label{fig:di0}
		\includegraphics[width=0.45\textwidth]{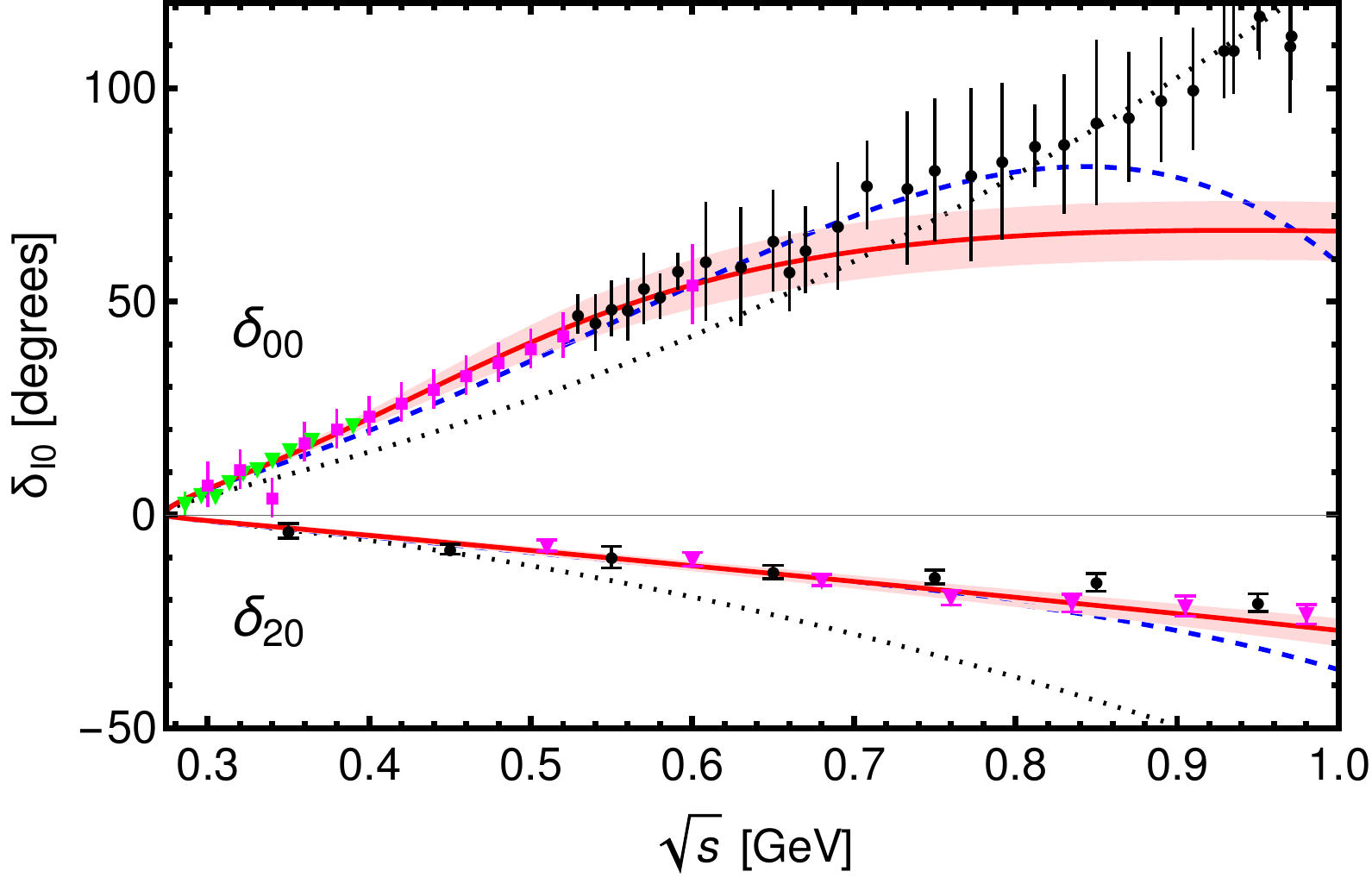}
\end{tabular}
\caption{Experimental data for the $\pi\pi\to\pi\pi$ phase shifts in the relevant channels 
compared to the theoretical $a \pi\to\pi\pi$ phase shifts in IAM (solid red), and the  $\pi\pi\to\pi\pi$ predictions at LO ChPT (dotted black) and NLO ChPT (dashed blue). The IAM predictions include the 1$\sigma$ confidence level regions that stem from the uncertainties in the LECs. The references for the data of the phase shifts for the $\pi\pi$ PWAs are given next: $\delta_{11}$,  \cite{Lindenbaum:1991tq} (pink squares) and \cite{Estabrooks:1974vu} (black circles); $\delta_{20}$, \cite{Losty:1973et} (pink triangles) and \cite{Hoogland:1977kt} (black circles); $\delta_{00}$ \cite{NA482:2007xvj} (green triangles), \cite{Froggatt:1977hu} (pink squares), and the average data from Refs.~\cite{Ochs:thesis1974,Hyams:1973zf,Protopopescu:1973sh,Estabrooks:1973,Grayer:1972,Kaminski:1996da} (black circles). The average procedure is explained in the  $\delta_{11}^{00}$ subsection of Ref.~\cite{Oller:1998zr}.
}
\label{fig:phase_shifts}
\end{figure*}

As long as inelasticities in $a\pi\to\pi\pi$ scattering can be neglected (see  discussion below), unitarity implies the following algebraic constraint for its PWAs~\cite{Oller:2020guq,Oller:2019opk},
\begin{align}
\Im A_{IJ}(s)=\frac{\sigma(s)}{32\pi}A_{IJ}(s)T_{IJ}^*(s)\theta(s-4m_\pi^2)\, ,\label{eq:unitarity}
\end{align}
where $\sigma(s)$
is the phase-space factor defined below 
\eq{eq:Mapi0pippimNLO} and 
$T_{IJ}(s)$ are the \textit{strong} PWAs of $\pi\pi$ scattering in the isospin basis. 
In Eq.~\eqref{eq:unitarity} we are using the conventions for the normalization of the states in the Appendix~\ref{sec:IAMdetails}
and have included a Bose-symmetric factor 1/2 that appears in the isospin basis. From the unitarity relation it follows that the continuous phases of $A_{IJ}(s)$ and $T_{IJ}(s)$ (i.e.~\textit{phase shifts})  are the same, which is the Watson's theorem for final state interactions~\cite{Watson:1952ji}. 

Unitarity is fulfilled only perturbatively in ChPT. Indeed, if we denote the amplitudes calculated up to $\mathcal O(p^{2n})$ in the chiral expansion by $A^{(2n)}_{IJ}$ and $T^{(2n)}_{IJ}$ then Eq.~\eqref{eq:unitarity} implies\footnote{We have explicitly checked that the imaginary parts of our NLO results fulfill perturbative unitarity in the PWAs studied in this work.}
\begin{align}
 \label{eq:unitarity_pert}
\Im A_{IJ}^{(4)}(s)=\frac{\sigma(s)}{32\pi}A_{IJ}^{(2)}(s)T_{IJ}^{(2)}(s)\theta(s-4m_\pi^2) \, .
\end{align}
Different methods have been proposed to impose exact elastic unitarity in scattering amplitudes that match to the perturbative ChPT predictions at low energies. 
These have seen multiple applications and led to very significant progress in the understanding of the hadronic phenomena (see Refs.~\cite{Pelaez:2015qba,Oller:2019rej,Oller:2020guq,Oller:2019opk} for recent reviews). In fact, $\pi\pi$ scattering, with the characterization of the  $\sigma$ or $f_0(500)$ resonance, stands as one of the first successful applications of these methods~\cite{Dobado:1996ps,Oller:1997ti,Oller:1998hw,Nieves:1998hp,Pelaez:2015qba,Oller:2020guq}. Given that the unitary corrections to the ChPT NLO calculation of $a\pi \to \pi\pi$ scattering will be given by the pion's final-state interactions, we expect the unitarization methods to provide a realistic amplitude in the energy region relevant for the axion hot dark matter bound. 

In our analysis we focus on the 
IAM technique 
which adopts the form,
\begin{align}
A_{IJ}(s)&=\frac{A_{IJ}^{(2)}(s)}{1-A_{IJ}^{(4)}(s)/A_{IJ}^{(2)}(s)}\, , \label{eq:IAM_Standard}
\end{align}
and can also be regarded as a Pad\'e approximant of the NLO ChPT amplitude~\cite{Dobado:1989qm}. The IAM formula can be formally derived using a dispersion relation~\cite{Truong:1988zp,Truong:1991gv,Dobado:1992ha,Oller:2020guq} and the different caveats and uncertainties of the method have been thoroughly studied 
in Ref.~\cite{Salas-Bernardez:2020hua}. 
One particular caveat concerns the validity of the two-body unitarity relation for $s$ above the four-pion threshold. However, as discussed and estimated quantitatively for $\pi\pi$ scattering in~\cite{Salas-Bernardez:2020hua}, these inelastic contributions to the imaginary part are suppressed and can be neglected for the energies of interest.  

\begin{figure*}[t!]
\centering
\begin{tabular}{cc}
\includegraphics[width=0.45\textwidth]{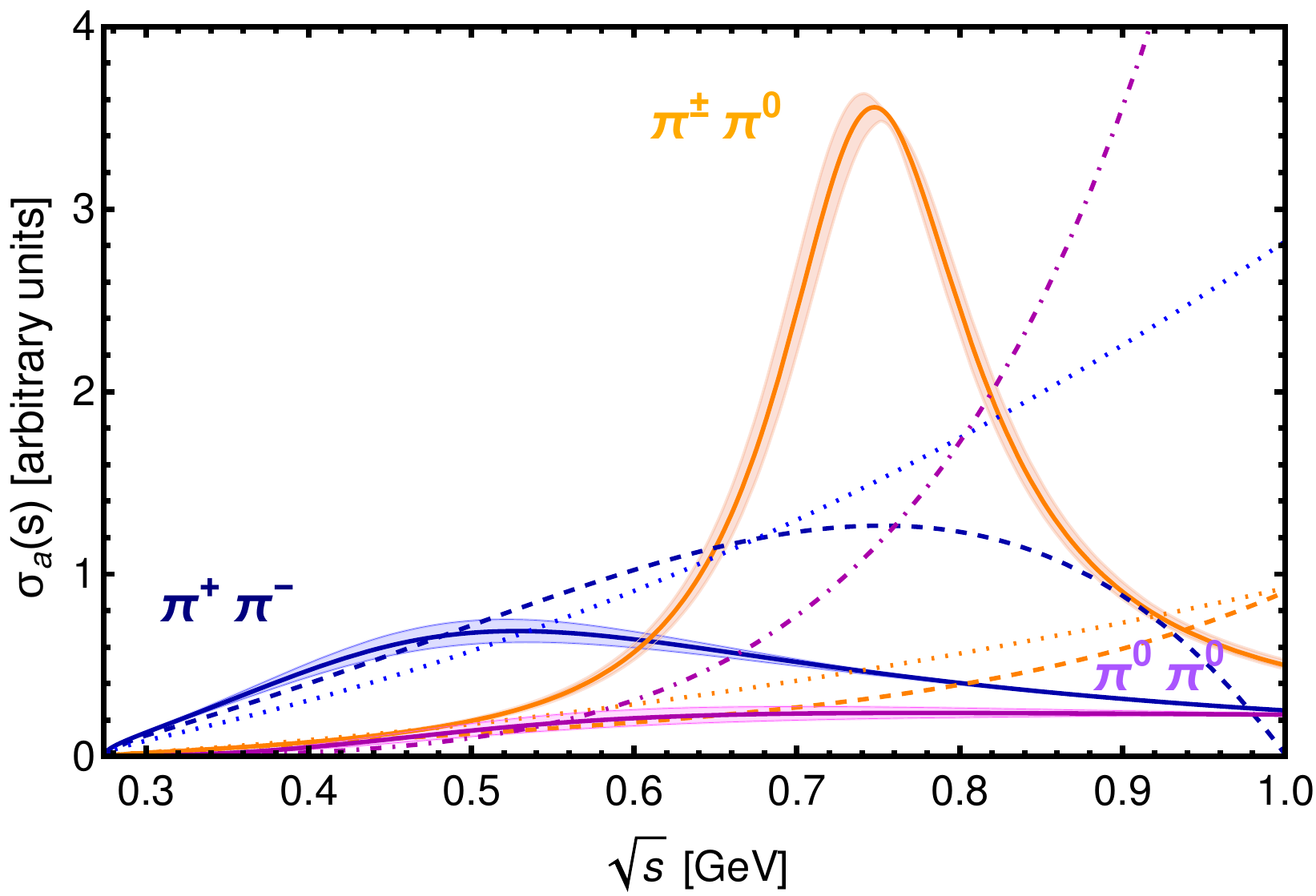}
&
\includegraphics[width=0.44\textwidth]{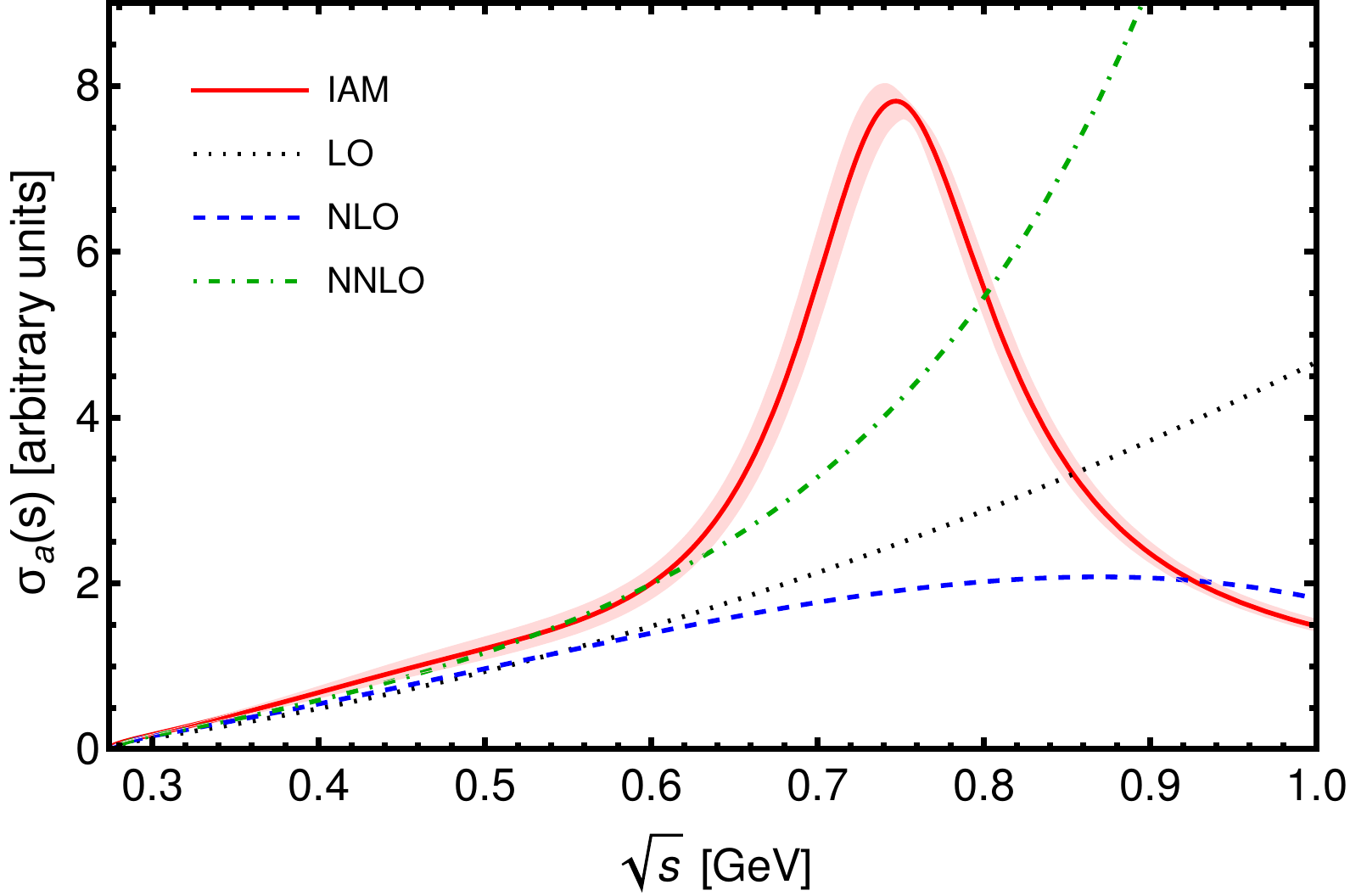}
\end{tabular}
\caption
{Cross sections $\sigma_a(s)$ 
for axion-pion scattering in units of mbarn for $f_a=f_\pi$, so they scale as $\propto f_a^{-2}$. 
{\it Left:} Plots for $a\pi^0\to\pi^+\pi^-$ (blue), $a\pi^0\to\pi^0\pi^0$ (magenta) and $a\pi^\pm\to\pi^\pm\pi^0$ (orange). Solid lines are the predictions in IAM, dashed in NLO ChPT and dotted in LO ChPT. We also include a dot-dashed magenta line describing the rate for the $a\pi^0\to\pi^0\pi^0$ channel in ChPT which is a pure NLO$²$ contribution (the amplitude is zero at LO \cite{DiLuzio:2021vjd}).
{\it Right:} Sum of all the cross-sections predicted in the IAM (solid, red) and in ChPT at LO (dotted, black), NLO (dashed, blue) and including the squared NLO pieces (NNLO) in the cross-section (dot-dashed, green). Uncertainties in the IAM predictions are 1$\sigma$ C.L. regions stemming from the errors in the LECs.} 
\label{fig:xsecs0ALL}       
\end{figure*}

An obvious benefit of expanding the inverse of the $A_{IJ}$ instead of the latter is that  $A_{IJ}^{-1}$ has a zero at a resonance pole, while $A_{IJ}$ becomes infinity. This makes 
the IAM, 
in the form Eq.~\eqref{eq:IAM_Standard},
a suitable method to address resonance dynamics below the chiral expansion scale $\Lambda_{\rm ChSB}\simeq4\pi f_\pi$ \cite{Salas-Bernardez:2020hua}.  
This is also reflected in the two-body elastic unitarity relation  for the inverse amplitude which reads
\begin{align}
\label{unit_inv}
\Im A_{IJ}^{-1}(s)=-\frac{\sigma(s)}{32\pi}\frac{T_{IJ}(s)}{A_{IJ}(s)}\,,
\end{align}
as it can be easily deduced from Eq.~\eqref{eq:unitarity}. Therefore, a resonance pole, which appears both in $T_{IJ}$ and $A_{IJ}$,  cancels in their ratio.

For our analysis we implement the IAM for the PWAs in the $S$-wave ($J=0$, $I=0$, $2$) and $P$-wave ($J=1$, $I=1$). The cases ($I=J=0,1$) are of special interest since they correspond to the quantum numbers of the prominent $f_0(500)$ (also known as $\sigma$) 
and $\rho(770)$ resonances \cite{Workman:2022ynf}, respectively, driving to large (unitarity)  corrections to $\pi\pi$ scattering in the low-energy energy region of interest below 1~GeV. The infinite tower of PWAs with $J\geq2$ can be included perturbatively in ChPT. Indeed,  we have checked that their contribution is only of a few percent relative to the $S$- and $P$-waves in the low-energy 
region.   
Therefore, we neglect them in the following. 


In Fig.~\ref{fig:phase_shifts} we show the phase shifts $\delta_{IJ}(s)$ of the different $a\pi\to\pi\pi$ PWAs compared to the experimental data from $\pi\pi$ scattering, which should be identical as per Watson's theorem. 
Besides the prediction in the IAM we show, for comparison purposes, the $\pi\pi$ scattering phase shifts obtained from perturbative ChPT at LO and NLO. The latter is derived using the results in Ref.~\cite{Gasser:1983yg} and the standard values for the LECs introduced above in Sec.~\ref{sec:ChPT}. The perturbative expressions for the phase shifts are described in Appendix~\ref{sec:pert_phase_shifts}. 
The LECs in IAM can be slightly different to those of ChPT. In particular, for the IAM calculations we use the combinations $\overline{\ell_1}-\overline{\ell_2}=-5.95(2)$, 
with $\overline{\ell_1}+\overline{\ell_2}=4.9(6)$, determined from $\pi\pi$ scattering to fit the pole position and width of the $\rho$ resonance precisely~\cite{Dobado:1996ps}.     
This is illustrated on the left panel of Fig.~\ref{fig:phase_shifts} by the good agreement of $\delta_{11}(s)$ with data across the resonance region. 
 
For the case of the phase shifts of $a\pi^0$ scattering the IAM
also agrees with the experimental data in both the $I=0$ and $I=2$ channels. In particular, the amplitudes describe the structure induced by the $\sigma$ resonance in $\delta_{00}(s)$. As expected, the phase shifts obtained for the $a\pi$ scattering amplitudes are equivalent to those calculated in~\cite{Dobado:1996ps} for the $\pi\pi$ scattering amplitudes using the IAM. 
Note that the worsening of the agreement in $\delta_{00}$ starting at $\sqrt{s}\gtrsim0.8$ GeV is an effect induced by the raise of the $f_0(980)$ resonance and the subsequent strong coupling to the $K\bar K$ channel with a prominent threshold effect~\cite{Janssen:1994wn,Oller:1997ti,Oller:1998hw}, which are omitted in our $SU(2)$ analysis. In fact, our results for $\delta_{00}(s)$ are in very good agreement with those obtained in Ref.~\cite{Albaladejo:2012te} by unitarizing $\pi\pi$ scattering calculated at NLO in $SU(2)$ ChPT. On the other hand, the energy range of applicability of the IAM framework can be in principle improved by unitarizing the coupled $\pi\pi$, $K\bar{K}$ and $\eta\eta$ interactions predicted by NLO $SU(3)$ ChPT, as first shown in Ref.~\cite{Guerrero:1998ei}. 
 

In Fig.~\ref{fig:xsecs0ALL}, left, we present our theoretical predictions for the $a\pi\to\pi\pi$ cross sections in the different channels of the charge basis, obtained in the IAM by inverting Eqs.~\eqref{eq:isospin_neutral}, \eqref{eq:isospin_charged} and \eqref{eq:PWAs}. ChPT departs from the IAM results at low energies, $\sqrt{s}\simeq0.5$ GeV.
In case of the $\pi^+\pi^-$ channel this is the typical scale at which unitarity corrections become large due to the $\sigma$ resonance in the $I=J=0$ channel. In case of the $\pi^\pm\pi^0$ channel
the disagreement is due to the prominent structure of the $\rho$ resonance emerging in the amplitude.

In the right panel of Fig.~\ref{fig:xsecs0ALL} we show the predictions in IAM and ChPT for the sum of cross sections, which is the quantity most closely related to the thermal rate to be calculated in the next \sect{sec:pheno}. 
NLO and higher order corrections of size estimated by including the NNLO pieces (from the squared NLO contributions to the rate), start to get very large around $\sqrt{s}\simeq0.6$ GeV.
In Appendix~\ref{sec:IAMdetails} we present a more detailed comparison between ChPT at different orders and the IAM for the cross sections and also the absolute values of the PWAs.

\section{Axion-pion thermalization rate}
\label{sec:pheno}

The axion-pion thermalization rate is 
defined via the phase-space integral \cite{Chang:1993gm,Hannestad:2005df}
\begin{align}
\label{Gamma1}
    \Gamma_a &= \frac{1}{n_a^{\rm eq}} \int\frac{d^3 \mathbf{p}_1}{(2\pi)^3 2 E_1}\frac{d^3  \mathbf{p}_2}{(2\pi)^3 2 E_2}\frac{d^3  \mathbf{p}_3}{(2\pi)^3 2 E_3}\frac{d^3  \mathbf{p}_4}{(2\pi)^3 2 E_4} \nonumber \\
    &\times \sum |\mathcal{M}|^2 (2\pi)^4 \delta^4\left( p_1+p_2-p_3-p_4\right) \nonumber \\
    &\times f_1 f_2 (1 + f_3)(1 + f_4) \, , 
\end{align}
where $n_a^{\rm eq} = (\zeta_3 / \pi^2) T^3$ 
and $f_{i} = 1/(e^{E_{i}/T} - 1)$.  
Here we neglect thermal corrections to the scattering matrix element, which is a good approximation for   
$T \lesssim m_\pi$ \cite{Gasser:1986vb,Gasser:1987ah,Gerber:1988tt}. 
The integration of the thermal rate has been performed following the same procedure presented in Ref.~\cite{DiLuzio:2021vjd} 
(see also \cite{Hannestad:1995rs}). 
  
\begin{figure}[t!]
\centering
\includegraphics[width=0.45 \textwidth]{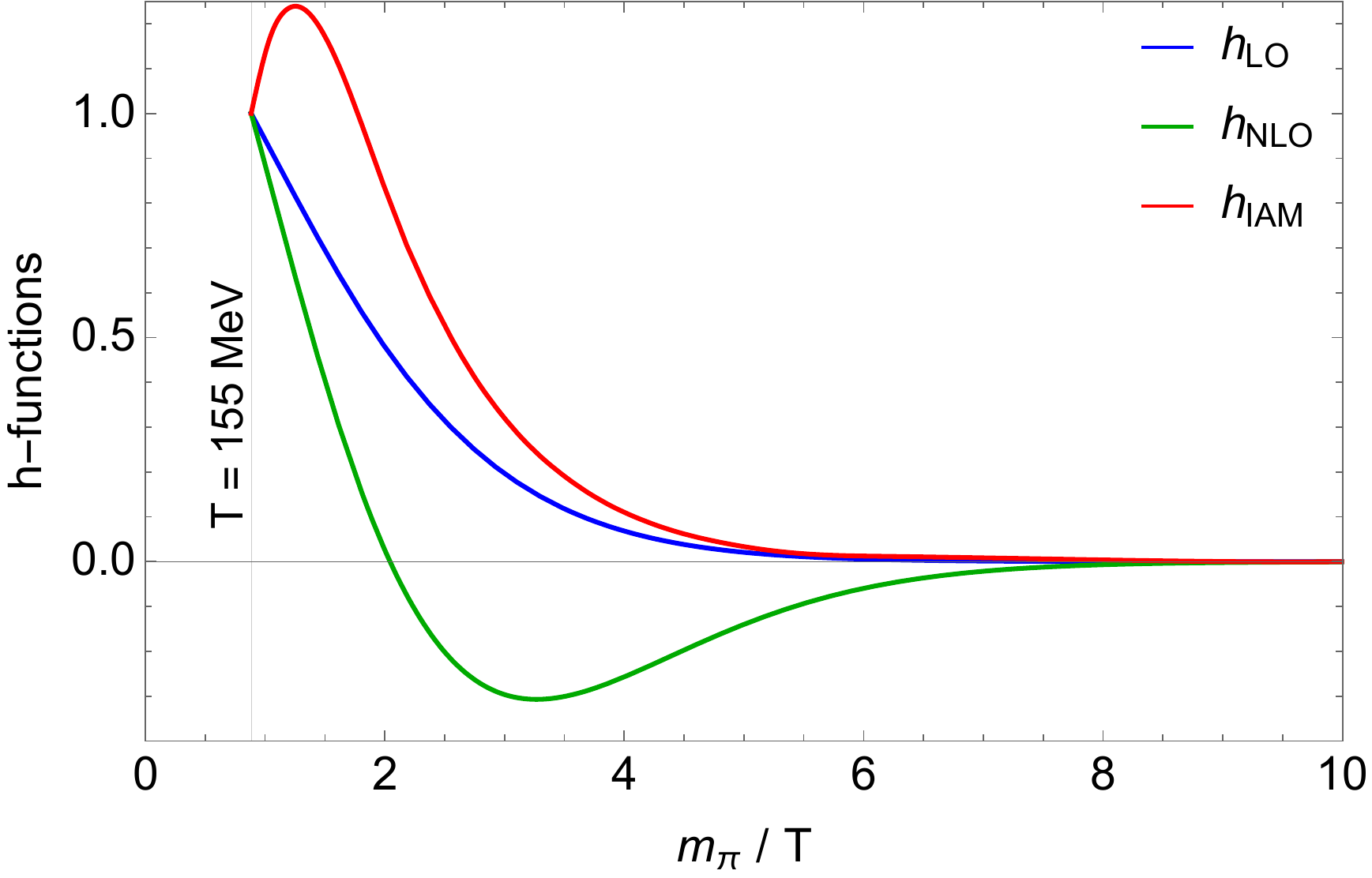}
\caption{
Profile of the $h_{\rm LO}$, $h_{\rm NLO}$ and $h_{\rm IAM}$ functions, normalized to $1$ at the value 
$m_\pi/T_c \simeq 0.88$. 
}
\label{fig:hfuncs}       
\end{figure}

The perturbative result, $\Gamma_a = \Gamma_a^{\rm LO} 
+ \Gamma_a^{\rm NLO}$, is obtained by expanding 
the amplitude squared in ChPT as 
$\sum |\mathcal{M}|^2\simeq 
\sum |\mathcal{M}|_{\rm LO}^2 +  \sum 2\Re[\mathcal{M}_{\rm LO} \mathcal{M}^*_{\rm NLO}]$ 
and it can be cast into the following way   
\begin{align}\label{gammaPert}
  \Gamma_a(T) &= \left(  \frac{C_{a\pi}}{f_a f_\pi}\right)^2 
0.163 \ T^5 \Big[ h_{\rm LO}(m_\pi/T)   \nonumber \\   & - 0.251 \frac{T^2}{f_\pi^2}\ h_{\rm NLO}(m_\pi/T)\Big] \, ,
\end{align}
where the $h$-functions are shown in \fig{fig:hfuncs}. 
Note that we normalized 
$h_{\rm LO}(m_\pi/T_c)=h_{\rm NLO}(m_\pi/T_c)=1$, 
with $m_\pi/T_c \simeq 0.88$.  
In fact,  
the $h$-functions are meaningful only for $T \lesssim 
T_c$, since for higher temperatures pions are deconfined.

On the other hand, 
the thermal rate obtained via the unitarized IAM 
amplitude, is given by 
\begin{align}\label{gammaIAM}
  \Gamma_a^{\rm IAM}(T) &= \left(  \frac{C_{a\pi}}{f_a f_\pi}\right)^2 
  0.137\ T^5 h_{\rm IAM}(m_\pi/T)  \, ,
\end{align} 
where we factored out a $T^5$ dependence, 
characteristic of the 
LO ChPT rate. In order to compare the IAM result 
with the perturbative one (cf.~\fig{fig:hfuncs}), we also 
normalized $h_{\rm IAM}(m_\pi/T_c)=1$.  

\begin{figure}[t!]
\centering
\includegraphics[width=0.45\textwidth]{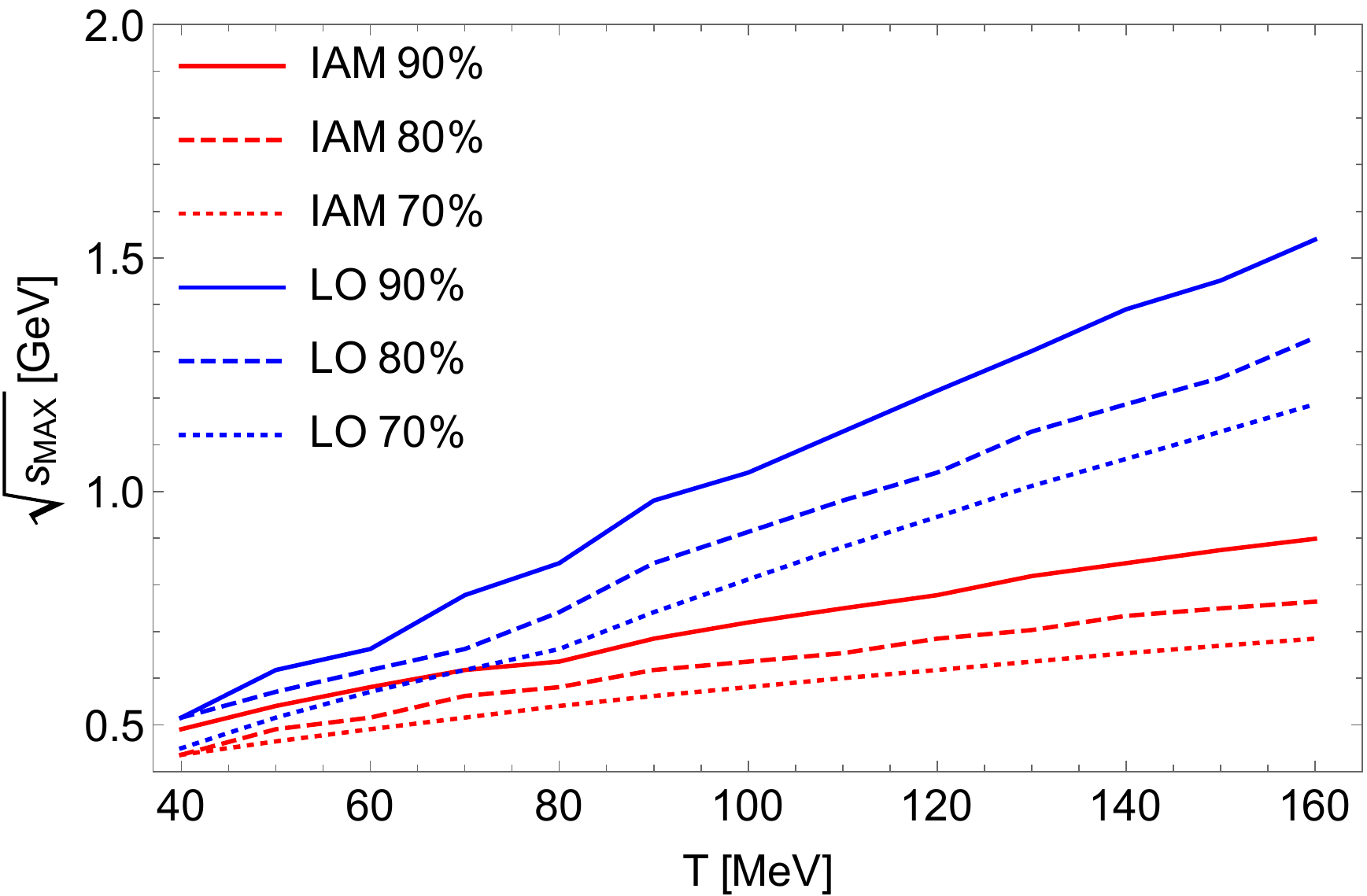}
\caption{Temperature dependence of
$\sqrt{s_{\rm \tiny MAX}}$
at which it is sufficient to cut off the integration of the thermal rate in order to get the 90\%, 80\%, 70\% of the total 
rate (without cutoff) for the LO and IAM cases. The plot shows the channel $\pi^+ \pi^-$.}
\label{fig:SqrtSMaxVST}       
\end{figure}

\begin{figure*}[t!]
\centering
\begin{tabular}{ccc}
\includegraphics[width=0.31\textwidth]{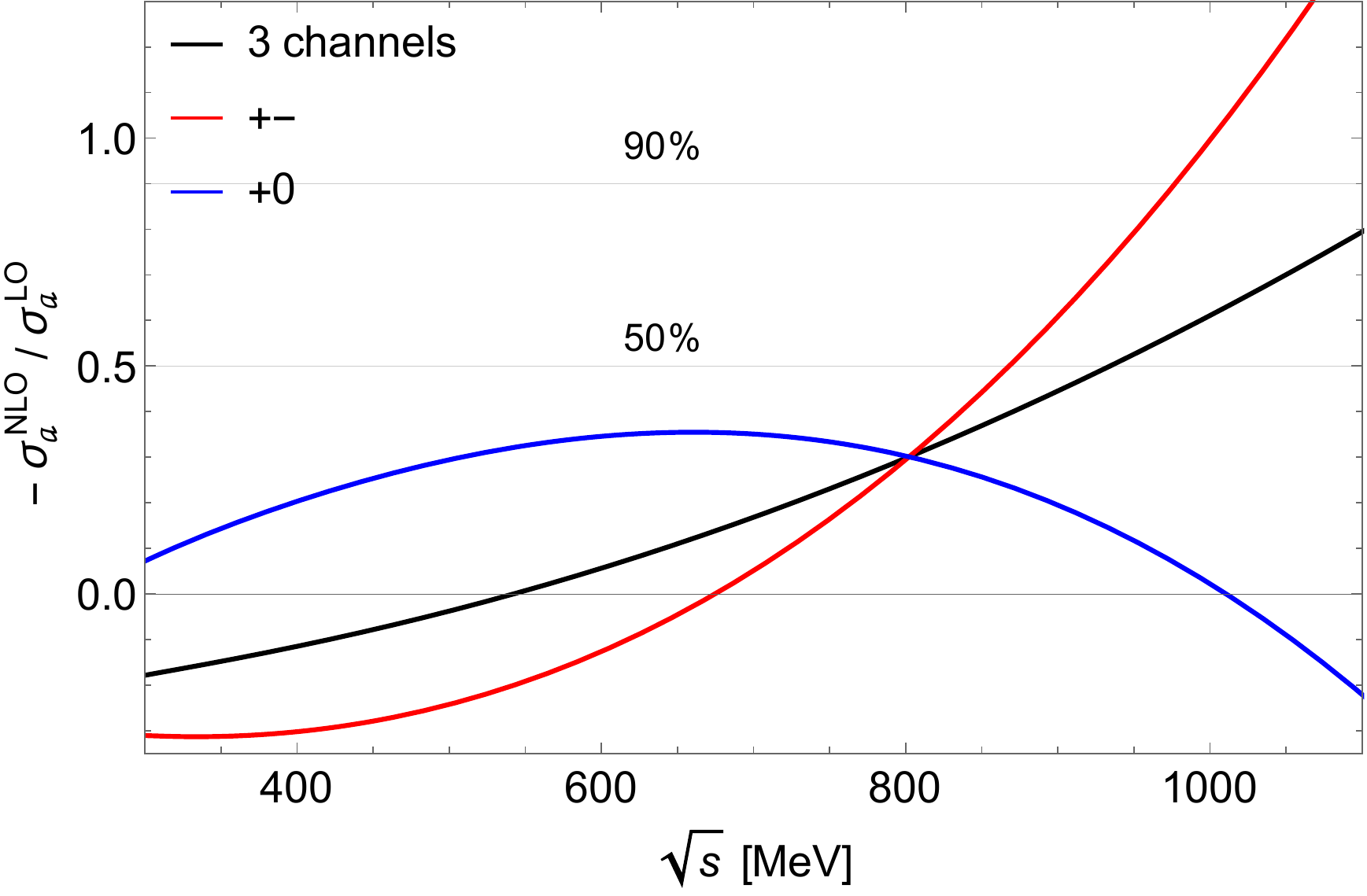} 
\ \
& 
\includegraphics[width=0.31\textwidth]{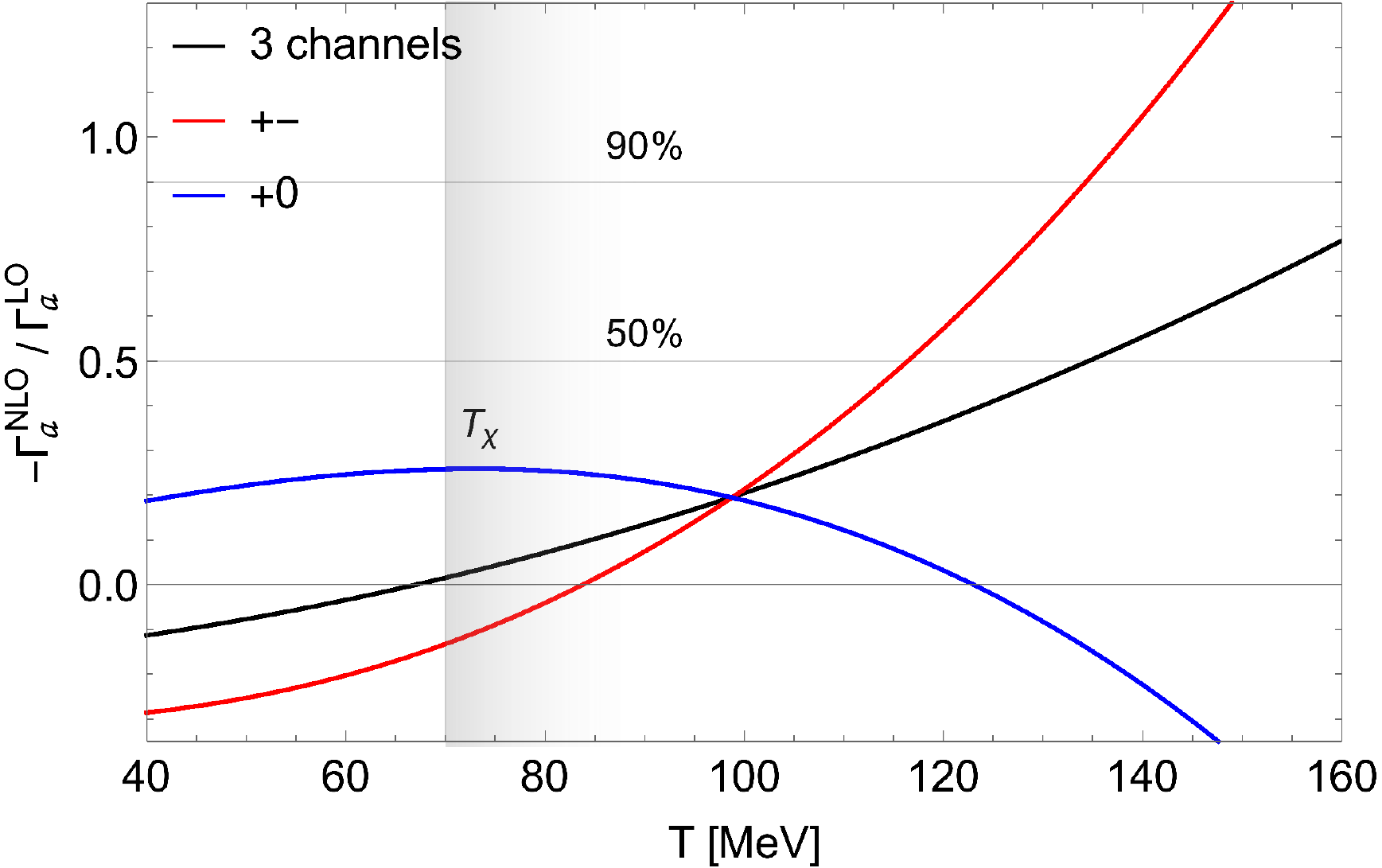}
\ \
& 
\includegraphics[width=0.31\textwidth]{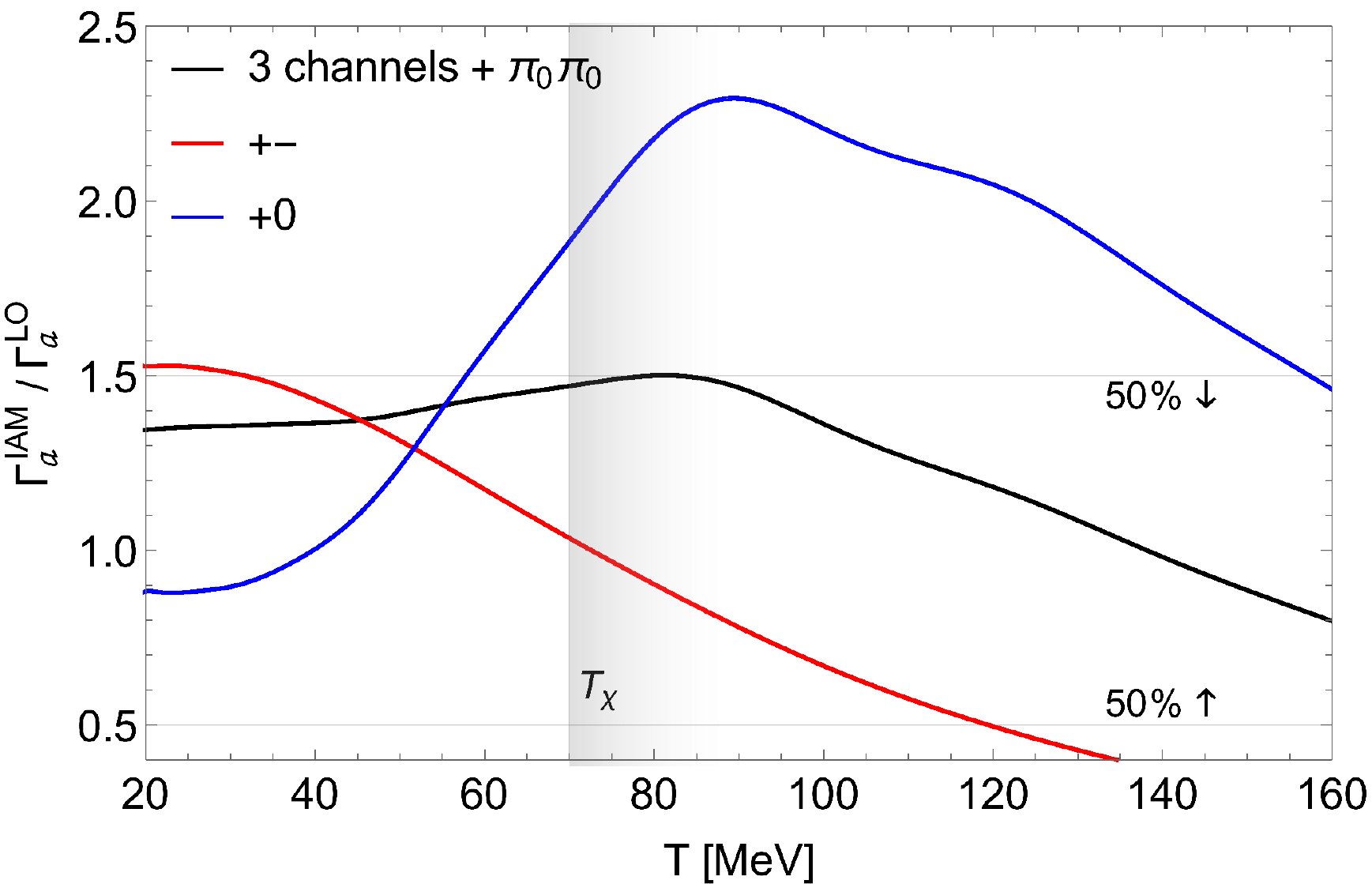}
\end{tabular}
\caption{Left (center) panel: 
ratio between the NLO correction 
and LO axion-pion cross-section 
(thermalization rate), considering the 
individual final-state channels 
$\pi^+\pi^-$ (red), $\pi^+\pi^0$ 
and $\pi^-\pi^0$ (blue, the latter two being equal) and their sum (black). Right panel shows the ratio between the IAM (cutoff at $1$ GeV)
and LO thermalization rate. } 
\label{fig:ratioNLOovLO}       
\end{figure*}

Integrals in Eq.~\eqref{Gamma1} cover a broad range of energies with contributions suppressed at high energies by the axion and pion Boltzmann factors. In order to assess the robustness of our predictions, especially at temperatures close to $T_c$, it is important to investigate the relative contributions to the thermal rate stemming from low-energies $\sqrt{s}\lesssim1$ GeV, which we deem is the upper energy limit of applicability for IAM (for further qualifications see Ref.~\cite{Salas-Bernardez:2020hua}). In \fig{fig:SqrtSMaxVST} we illustrate this by showing the temperature dependence of $\sqrt{s_{\rm \tiny MAX}}$ which is the cut-off (in $\sqrt{s}$) needed in Eq.~\eqref{Gamma1} for the low-energy contributon to describe the 70\%, 80\% or 90\% of the total thermal rate. By looking at the value of $\sqrt{s_{\rm \tiny MAX}}$ for $T\simeq T_c$ we find that $90\%$
of the contribution to the thermal rates in IAM stem from the low-energy region for all the temperatures of interest in our work.


In our analysis and in the parametrization shown in Eq.~\eqref{gammaIAM} we use the result of $\Gamma_a$ obtained by cutting off the contributions in Eq.~\eqref{Gamma1} at $\sqrt{s_{\rm \tiny MAX}}=1$ GeV. Moreover,
we use
as an estimate of our theoretical error
the difference between the thermal rate in IAM integrated with and without cutoff.

\section{On the breakdown of the chiral expansion}
\label{sec:breakdownChPT}

\begin{figure}[h!]
\centering
\includegraphics[width=0.45 \textwidth]{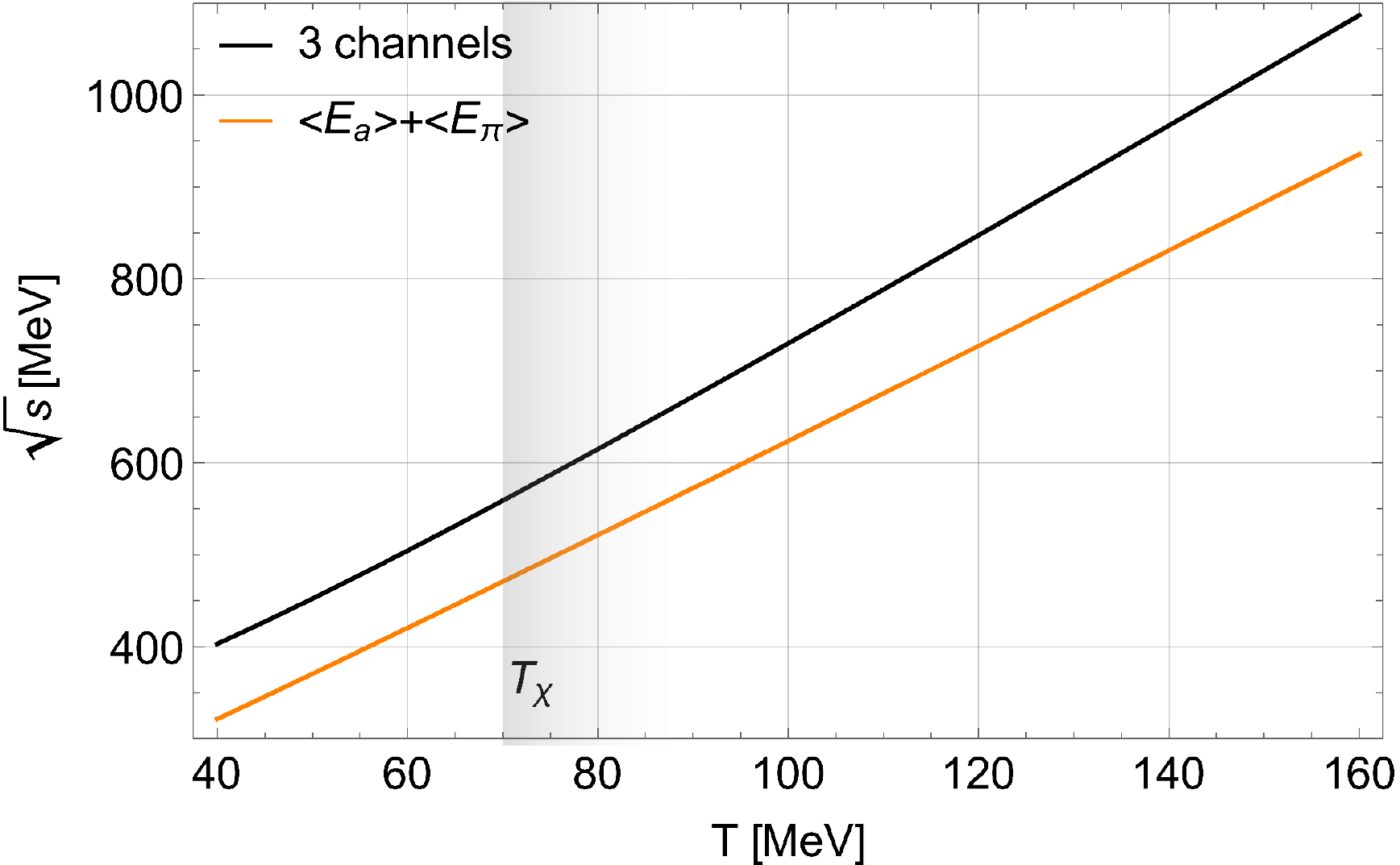}
\caption{$\sqrt{s}$-$T$ correspondence, using two different criteria: equating the $\%$ correction in the left and center panels of \fig{fig:ratioNLOovLO} (black line) and 
summing the axion-pion thermal energies 
in the initial state 
of the scattering (orange line).}
\label{fig:Tscriterium}       
\end{figure}

In Ref.~\cite{DiLuzio:2021vjd} the ratio between the NLO correction and the LO value of the axion-pion thermalization rate was taken as a criterion for the breakdown of ChPT, by requiring that $|\Gamma_a^{\rm NLO} / \Gamma_a^{\rm LO}| \lesssim 50 \%$. 
However, it is more instructive to inspect the breakdown of ChPT both at the level of cross sections and thermal rates, 
as well as for different final states separately.
This analysis 
is summarized in~\fig{fig:ratioNLOovLO}. 
Starting from the ratio of cross sections in the left panel we observe that for the $\pi^+\pi^0$ channel
it reaches a maximal value of $\sim 40 \%$ around $\sqrt{s}~\sim~0.6$~GeV, which agrees approximately with the energy at which NLO ChPT departs from the IAM prediction in \fig{fig:xsecs0ALL}. As discussed in Sec.~\ref{sec:unitarization}, this is due to large unitarity corrections and the emergence of the $\rho$ resonance, which is ultimately the cause of the breakdown of the chiral expansion in the $I=J=1$ channel at those energies. In the middle panel of \fig{fig:ratioNLOovLO} we show the temperature dependence of the ratio between the NLO and LO contributions to the thermal rates. In this case, the maximum is reached at $T_\chi\sim70$ MeV that, according to our discussion for the cross sections, we interpret as the temperature at which 
ChPT breaks down.
This correspondence between $\sqrt{s}$ and $T$ can be supported by different semiquantitative arguments. 
For instance, 
by equating the NLO/LO ratio of cross-sections and thermal rates given in \fig{fig:ratioNLOovLO}, one gets the 
correlation between $\sqrt{s}$ and $T$ shown in \fig{fig:Tscriterium}.
We have also checked that alternative criteria, like 
e.g.~taking 
$\sqrt{s} \sim \vev{E_\pi}_T + \vev{E_a}_T$ in terms 
of the 
thermal average $\vev{E}_T = \rho (T) / n (T)$, 
give similar results.  

Finally, on the right panel of \fig{fig:ratioNLOovLO} we show the ratio of the thermal rates between the results obtained with IAM and ChPT at LO. The differences in this case are more prominent and appear at lower temperatures. In fact, significant differences are visible even at $T=20$ MeV for the $\pi^+\pi^-$ channel. However, this is not surprising given that a similar effect at threshold is know from $\pi\pi$ scattering. Indeed, higher-orders corrections to the $I=J=0$ $\pi\pi$ scattering length at LO are around 25\% \cite{Leutwyler:2006qq}, that at the level of the cross sections implies a correction of around a 50\% near threshold.

\section{Cosmological implications}
\label{sec:cosmo}

We next discuss the 
cosmological implications 
of the newly computed 
axion-pion thermalization rate. 
While an exhaustive treatment of  
cosmological observables is beyond the 
scope of this paper (for recent 
analyses,  
see Refs.~\cite{Caloni:2022uya,DEramo:2022nvb}), 
we focus here on the 
axion contribution to the effective number of extra 
relativistic degrees of freedom \cite{Kolb:1990vq},
\beq 
\Delta N_{\rm eff} \simeq \frac{4}{7} 
\( \frac{43}{4 g_S(T_D)} \)^{4/3} \, , 
\eeq
with $g_S(T_D)$ the number of entropy degrees of freedom 
at the axion decoupling temperature, $T_D$.
The latter follows from the decoupling 
condition, 
$\Gamma_a(T_D) \simeq H(T_D)$,\footnote{A more 
refined determination of the 
axion thermal density, 
beyond the instantaneous decoupling approximation, 
would require the solution of the 
associated
Boltzmann equation 
(see e.g.~\cite{DEramo:2022nvb}).} 
in terms of the axion-pion thermalization rate in \eq{Gamma1} 
and the Hubble 
rate, $H(T) = \sqrt{4\pi^3 g_\star(T) / 45} \, T^2 / m_{\rm pl}$ 
(assuming a radiation dominated universe),
where $m_{\rm pl} = 1.22 \times 10^{19}$ GeV is 
the Planck mass and 
$g_\star(T)$ denotes the 
effective number of 
relativistic degrees of freedom. 
For the functions 
$g_S(T)$ and $g_\star(T)$ we employ 
the values provided by Ref.~\cite{Saikawa:2018rcs}.

\begin{figure}[t!]
\centering
\includegraphics[width=0.45 \textwidth]{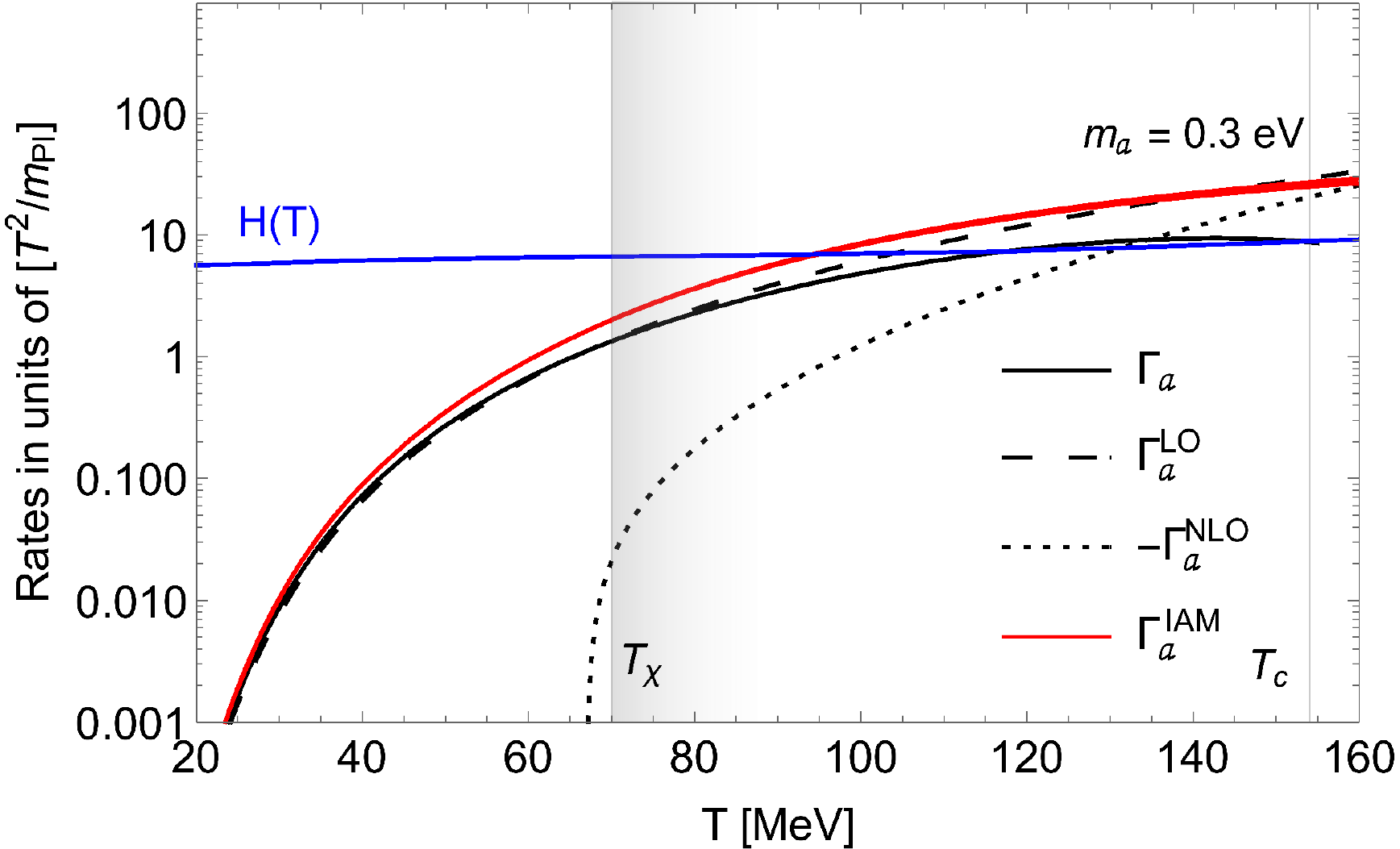}
\caption{Axion-pion thermalization rate vs.~Hubble rate 
(blue line)
for 
$m_a=0.3 \ \rm eV$. 
LO, NLO correction, 
and the total rate at NLO 
are 
denoted
respectively
by dashed, dotted and solid black lines, 
while the IAM rate 
is represented by a red band, where the upper line is the rate without cutoff, and the lower line the rate cut off at $\sqrt{s}=1$ GeV.}
\label{fig:rates}       
\end{figure}

In the following, we set to zero 
the 
model-dependent
axion couplings to quarks in \eq{eq:defCapi}, 
i.e.~$c^0_{u,\,d}=0$, 
in order to represent the bound from $\Delta N_{\rm eff}$ 
as a function of $m_a$ 
(the generalization to $c^0_{u,\,d} \neq 0$ being straightforward, see e.g.~\cite{DiLuzio:2023tqe}). 
The perturbative and unitarized 
rates 
are shown 
respectively in Fig.~\ref{fig:rates} 
for the reference axion mass value 
$m_a=0.3 \ \rm eV$. 
For the IAM rate we employ a theoretical 
error that is based on the 
criterion discussed 
at the end of \sect{sec:pheno}. 

\begin{figure}[t!]
\centering
\includegraphics[width=0.45 \textwidth]{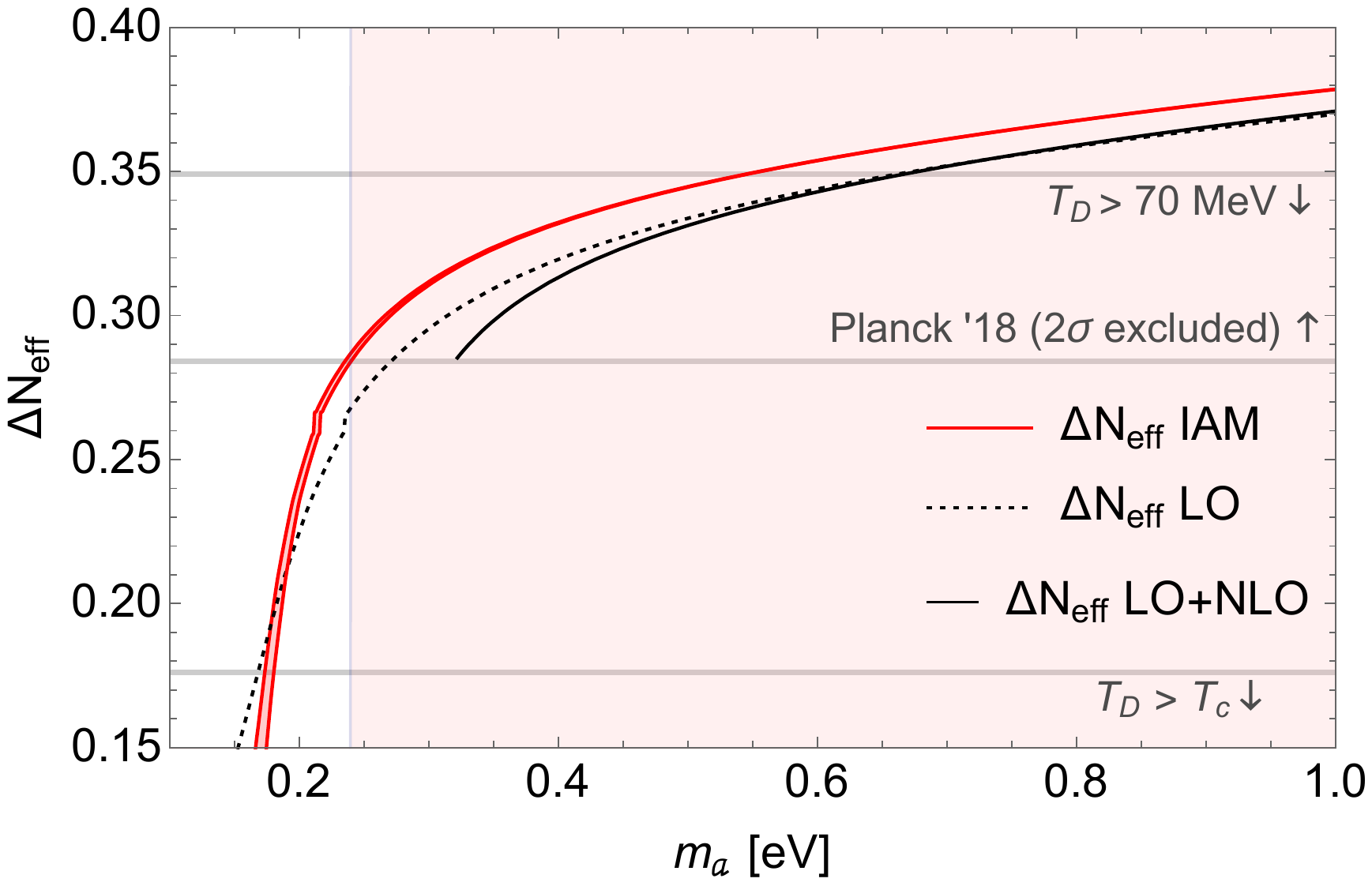}
\caption{$\Delta N_{\rm eff}$ as a function of $m_a$. 
The perturbative ChPT predictions are extrapolated 
for illustrative purposes
beyond 
the temperature, $T_\chi \sim 70$ MeV, 
where the chiral expansion fails. The LO+NLO curve is stopped at $m_a=0.31\ \rm eV$, corresponding to the minimum value of $m_a$ for which the total rate at NLO intersects the Hubble rate.}
\label{fig:deltaNeffm}       
\end{figure}

The bound of $\Delta N_{\rm eff}$ 
from Planck'18 data 
\cite{Akrami:2018vks,Aghanim:2018eyx}   
on the axion mass 
is finally displayed 
in \fig{fig:deltaNeffm}, employing different 
approximations for the 
ChPT calculation of the 
axion-pion thermalization rate. With the IAM computation, valid up to temperatures approaching $T_c$, we can extract the 
conservative bound $m_a \lesssim 0.24 \rm\ eV$.

To assess the impact of 
the high-energy discrepancy between the $\delta_{00}$ phase shift obtained from $\pi\pi$ data and the theoretical IAM prediction (see \fig{fig:phase_shifts}), we also computed the $\pi^+\pi^-$ and $\pi^0\pi^0$ rates by cutting off the energies above $\sqrt{s}\gtrsim0.8$ GeV. Under this condition, the total rate is reduced by $10\%$ at $T=150$ MeV, with an error band reaching $11\%$, in comparison to the $7\%$ represented by the red band in \fig{fig:rates}. The corresponding HDM bound would be $m_a \lesssim 0.25 \rm\ eV$.

We remark that in the region between $m_a = 0.1$ eV 
and 1 eV axions transit from behaving as dark radiation to 
hot dark matter, so a more refined cosmological analysis 
would be needed in this intermediate regime. 
On the other hand, for $m_a \lesssim 0.3$ eV where the 
bound is extracted, 
the use of $\Delta N_{\rm eff}$ 
is still adequate (see e.g.~Fig.~1 in \cite{Caloni:2022uya}). 
Note, finally, that the description in terms of 
axion dark radiation 
is well-justified in 
the presence of 
model-dependent axion couplings 
$c^0_{u,d} \gg 1$ (as in some 
axion models \cite{Darme:2020gyx}), 
since
in order to keep $C_{a\pi} /f_a$ constant, 
the relevant mass window gets shifted to lower values of 
$m_a$, or in symmetry-based models where the axion mass is 
exponentially suppressed \cite{Hook:2018jle,DiLuzio:2021pxd,DiLuzio:2021gos}.

\section{Conclusions}  
\label{sec:concl}

The purpose of this work was two-fold. 
On the one hand, to correct a mistake in 
Ref.~\cite{DiLuzio:2021vjd} about the NLO correction to 
$a \pi \to \pi \pi$ scattering 
and, on the other hand, to extend the 
validity of the chiral description of axion-pion scattering 
by means of a unitarization method known as IAM. 
While the axion-pion thermalization rate can be computed within 
ChPT up to temperatures of $T_\chi \sim 70$ MeV, 
the unitarization method allows one to extend this further 
up to temperatures approaching the 
QCD deconfinement, $T_c \simeq 155$ MeV. 
The IAM rate shows a sizeable deviation from the perturbative one 
for temperatures $T \gtrsim 40$ MeV, 
corresponding to the contribution of the 
$\sigma$ and $\rho$ resonances in the region 
$\sqrt{s} \gtrsim 400$ MeV for axion-pion scattering.

Further improvements 
of particle physics aspects 
for the calculation 
of the axion thermal relic could stem from extending the analysis to three flavours which, as discussed in Sect.~\ref{sec:unitarization}, can start producing large effects from energies $\sqrt{s}\simeq800$ MeV and higher due to the kaon threshold and the appearance of the $f_0(980)$. As discussed in Sect.~\ref{sec:pheno} and illustrated in~\fig{fig:SqrtSMaxVST}, these energies are only relevant for the higher temperatures, which could indeed become important to fully exploit future measurements of $\Delta N_{\rm eff}$ expected from the CMB-S4 experiments. In this context, one should also consider computing thermal corrections to the scattering amplitude (along the lines of the calculations done in Refs.~\cite{GomezNicola:2002tn,Dobado:2002xf}) and, eventually, develop techniques 
to describe axion thermal production 
in the intermediate region between 
$T_c \simeq 155$ MeV and 1 GeV. 

\section*{Note Added}

While completing this work, Ref.~\cite{Notari:2022zxo} 
appeared on the arXiv, 
where the validity of 
ChPT for axion-pion scattering 
is extended by
using $\pi\pi$ scattering data
via a rescaling of the corresponding cross sections.
In \app{sec:comparison} we provide a 
detailed comparison with the methodology of Ref.~\cite{Notari:2022zxo}, 
in which we show that we obtain 
a reasonable agreement, 
up to subleading $\mathcal{O}(8\%)$
corrections in the calculation of the thermal rate.

\begin{acknowledgments}

We thank Alessio Notari, Fabrizio Rompineve and 
Giovanni Villadoro for useful discussions regarding Ref.~\cite{Notari:2022zxo}.
The work of LDL is supported by the project ``CPV-Axion''
under the Supporting TAlent in ReSearch@University of Padova (STARS@UNIPD) and by the INFN Iniziative Specifica APINE.
The work of GP and LDL has received funding from the European Union's Horizon 2020 research and innovation programme under the 
Marie Sk\l{}odowska-Curie grant agreement N$^{\circ}$ 860881. Work by JMC is supported by PGC2018-102016-A-I00, and the 
``Ram\'on y Cajal'' program RYC-2016-20672. Work by JAO was partially supported by the MICINN AEI (Spain) Grant 
No.~PID2019–106080GB-C22/AEI/10.13039/501100011033,
and by the EU Horizon 2020 research and innovation programme, STRONG-2020 project, under Grant agreement No.~824093.
\end{acknowledgments}


\appendix

\section{NLO amplitudes} 
\label{app:NLOampl}

The full analytical expression of the 
renormalized 
NLO amplitude for the $a \piz\rightarrow\pip \pim$ process  
reads 

\begin{widetext}
\begin{align} 
\label{eq:Mapi0pippimNLO} 
\mathcal{M}^{\rm NLO}_{a \piz \rightarrow \pip \pim} &=
      \frac{ C_{a\pi}}{192 \pi ^2 f_{\pi }^3  f_a } 
      \Bigg\{ 15 m_{\pi }^2 (u+t)-11 u^2-8 u t-11 t^2 
      -6 \overline{\ell_1}  \left(m_{\pi }^2-s\right) \left(2 m_{\pi }^2-s\right) \nonumber \\
       & - 6 \overline{\ell_2} \left(-3 m_{\pi }^2 (u+t)+4 m_{\pi }^4+u^2+t^2\right) 
       +18  \overline{\ell_4} m_{\pi }^2 (m_{\pi }^2-s)
       \nonumber \\
       &+3 \[ 3  \sqrt{1 -\frac{4 m_{\pi }^2}{s}} s \left(m_{\pi }^2-s\right) \ln{\left(\frac{ \sigma (s)-1}{\sigma (s)+1}\right)} \right. \nonumber \\
       &+\sqrt{1- \frac{4 m_{\pi }^2}{t}} \left(m_{\pi }^2 (t-4 u)+3 m_{\pi }^4+t (u-t)\right)
      \ln{\left(\frac{ \sigma (t)-1}{\sigma (t)+1}\right)}\nonumber \\
       &\left. +  \sqrt{ 1-\frac{4 m_{\pi }^2}{u}} \left(m_{\pi }^2 (u-4 t)+3 m_{\pi }^4+u (t-u)\right) 
       \ln{\left(\frac{ \sigma (u)-1}{\sigma (u)+1}\right)}\] \Bigg\} \nonumber \\
       & -\frac{4  \ell_7 m_{\pi }^2 m_d \left(s-2 m_{\pi }^2 \right) m_u \left(m_d-m_u\right)}{ f_{\pi }^3 f_a \left(m_d+m_u\right){}^3} 
  \, ,
\end{align} 
where $\sigma(s)=(1-4m_\pi^2/s)^{1/2}$. 
Note that the term proportional to $\overline{\ell_4}$ 
in the second 
row arises from the NLO correction to $f_\pi$ 
in the LO amplitude (see e.g.~Ref.~\cite{Gasser:1983yg}).
The amplitudes for the crossed channels 
$a \pim\rightarrow\piz \pim$ and $a \pip\rightarrow\pip \piz$ 
are obtained by cross symmetry through the replacements 
$s \leftrightarrow t$ and $s \leftrightarrow u$, respectively.
Similarly, for the $a \piz \rightarrow \piz \piz$ 
amplitude that is needed for the IAM unitarization 
method we obtain
\begin{align}
   \mathcal{M}_{a \piz \rightarrow \piz \piz}&= \frac {3 C_{a\pi}} {96 \pi^2 f_ {\pi}^3 f_a}\Bigg\{-2(\overline {\ell_1}+2\overline {\ell_2}+3)\left (3 m_\pi^4-3 m_\pi^2 (t+u)+t^2+t u+u^2\right) \nonumber\\
     &- 3\Bigg(\sqrt {1-\frac{ 4 m_ {\pi}^2}{s}} \left (m_{\pi}^2 -s \right) {}^2\ln{\left(\frac{ \sigma (s)-1}{\sigma (s)+1}\right)} \nonumber \\
     &\ \ \ \ +\sqrt {1-\frac{4 m_ {\pi}^2}{t}}\left (m_ {\pi}^2 - t \right) {}^2\ln{\left(\frac{ \sigma (t)-1}{\sigma (t)+1}\right)} \nonumber \\
     &\ \ \ \ + \sqrt {1-\frac{4 m_ {\pi}^2}{u}}\left (m_ {\pi}^2 - u \right) {}^2\ln{\left(\frac{ \sigma (u)-1}{\sigma(u)+1}\right)} \Bigg) \Bigg\} \nonumber \\
    &+ \frac {36 \ell_ 7 m_ {\pi}^4 m_d m_u\left (m_d - m_u \right)} {f_ {\pi}^3 f_a\left (m_d + m_u \right) {}^3} \, .
\end{align}
\end{widetext}

\begin{figure*}[htb!]
\centering
\begin{tabular}{cc}
		\includegraphics[width=0.45\textwidth]{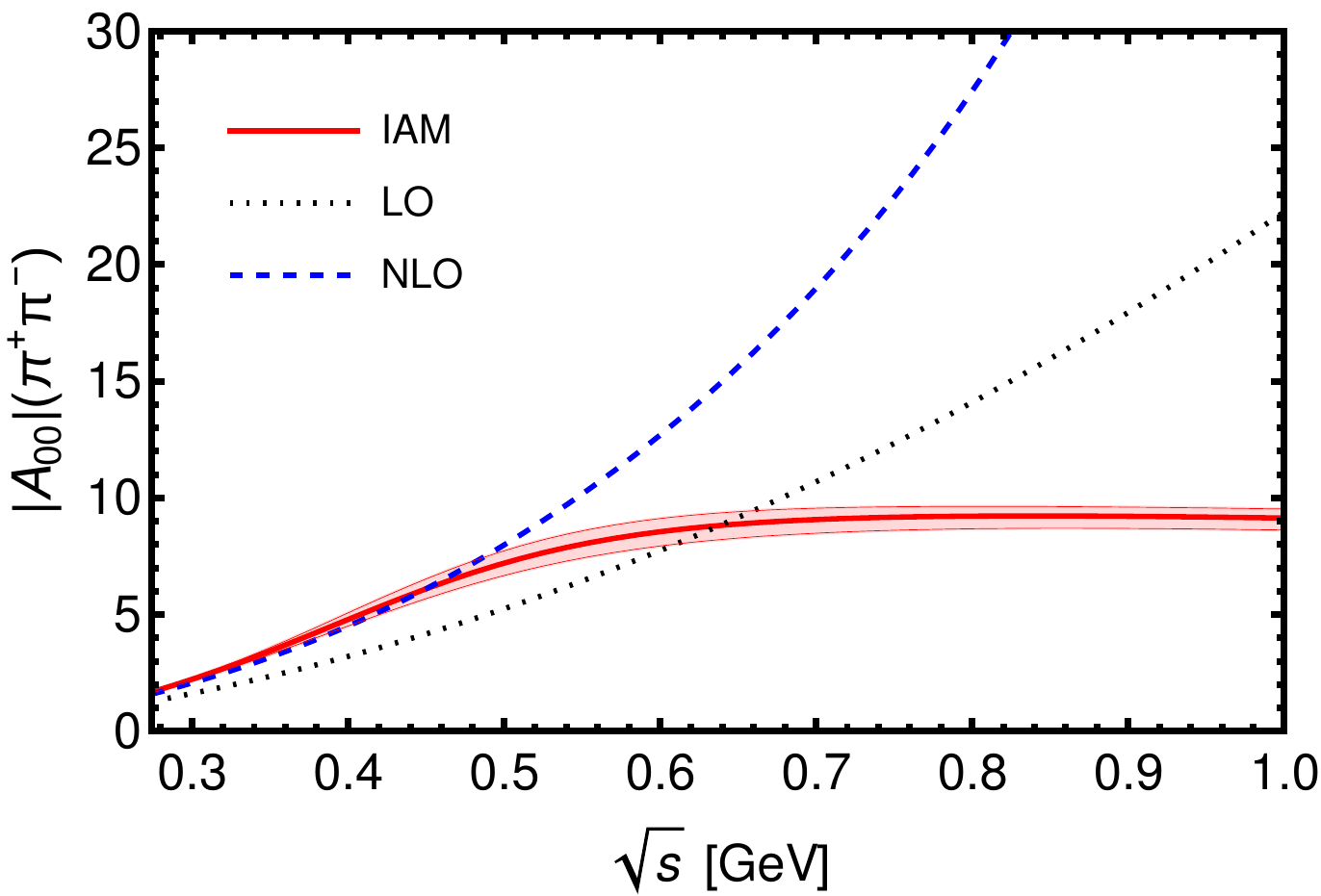}
&
		\includegraphics[width=0.45\textwidth]{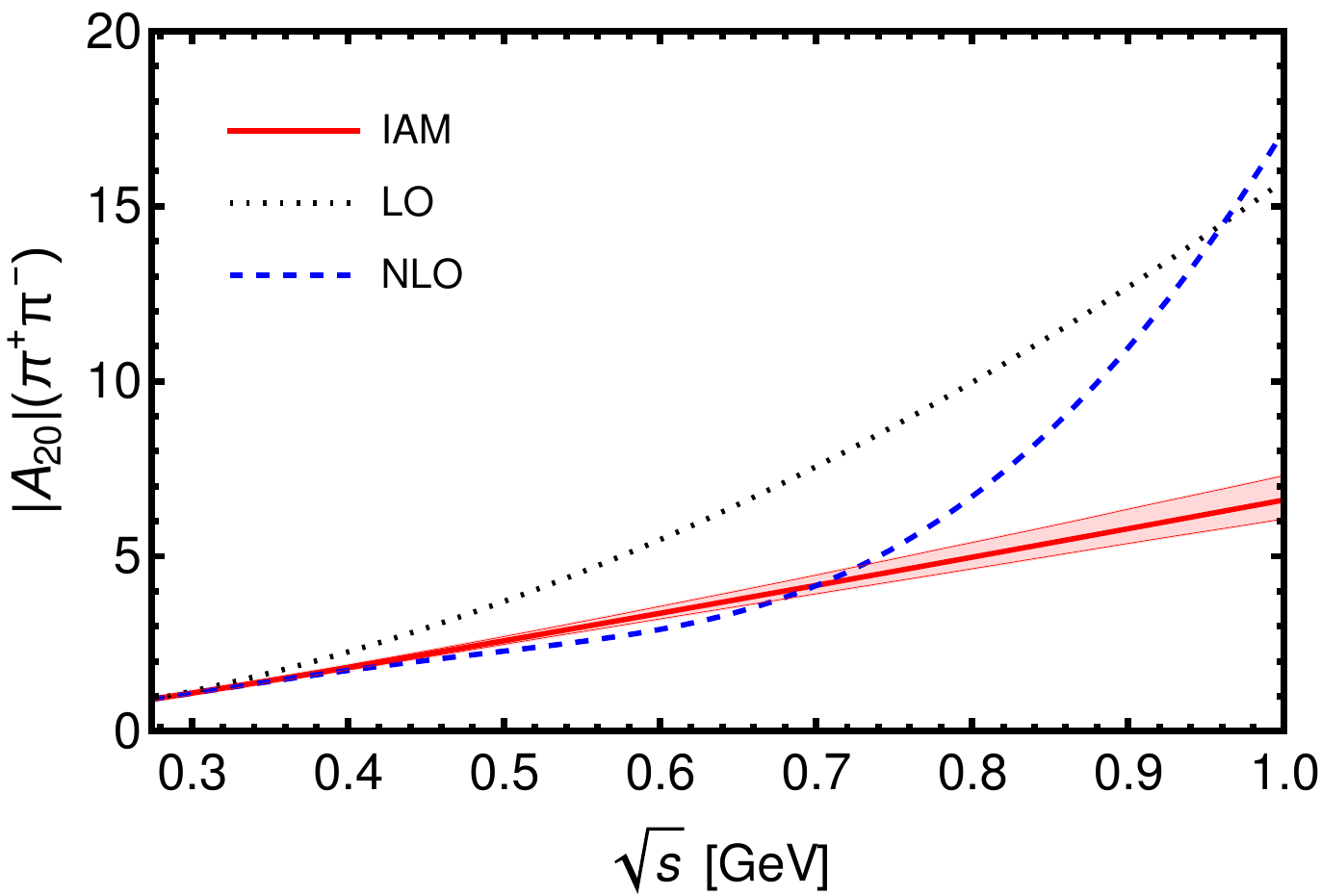}\\
		\includegraphics[width=0.45\textwidth]{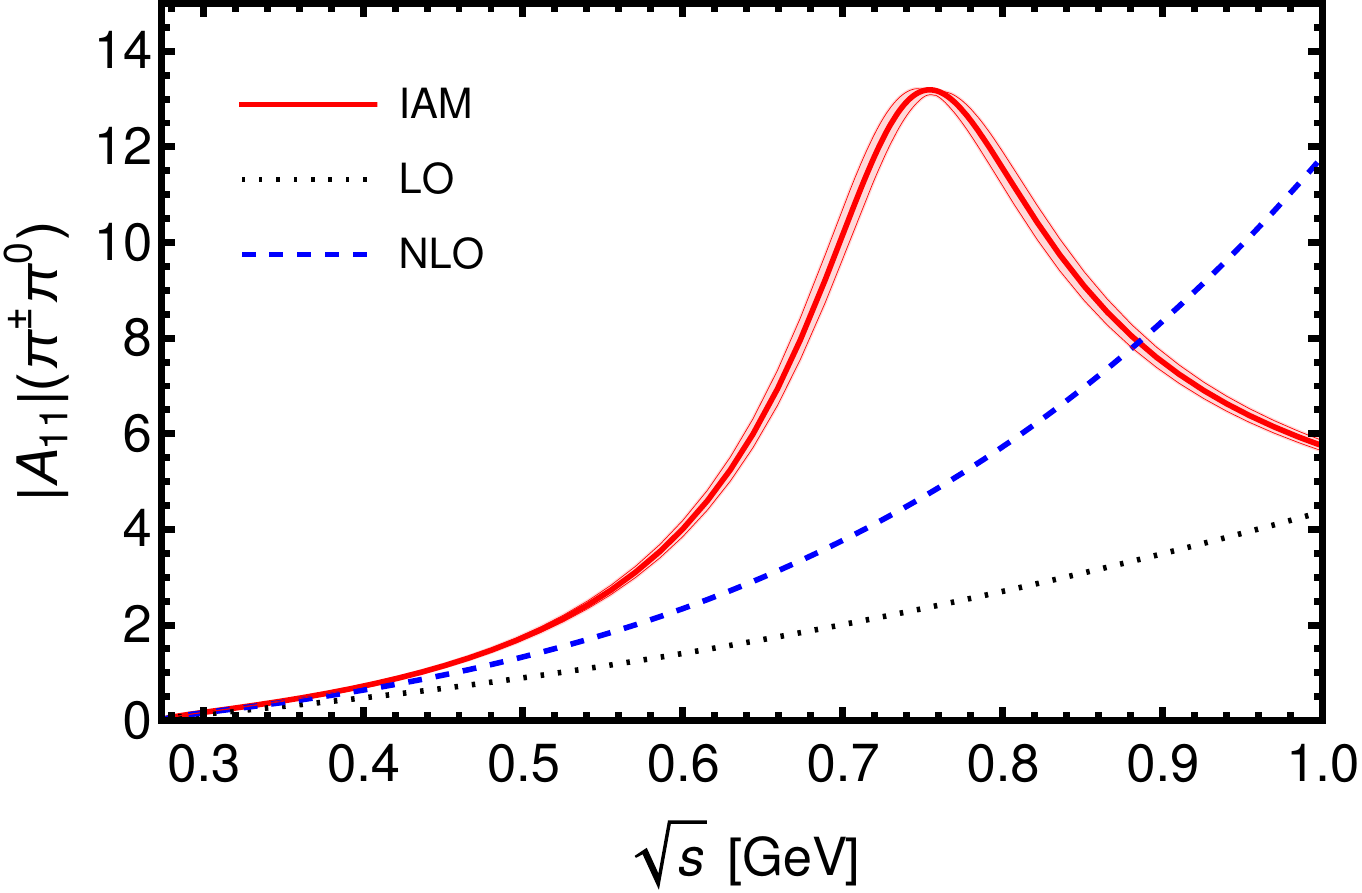}
&
		\includegraphics[width=0.45\textwidth]{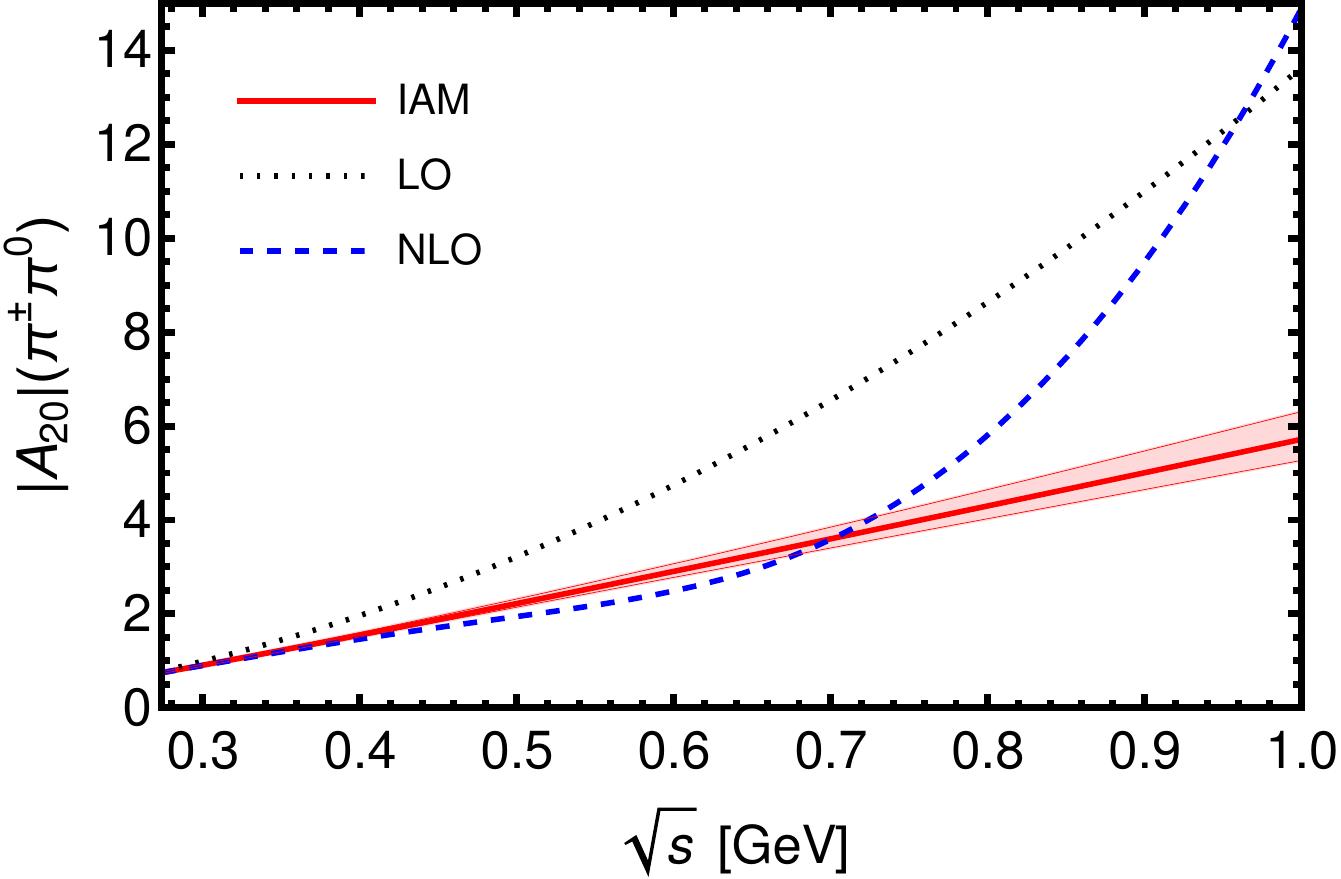}		
\end{tabular}
\caption{Absolute values of the PWAs in the different isospin and angular momentum channels considered in this work. The predictions in IAM, ChPT at LO and ChPT at NLO are shown in solid (red), dotted (black) and dashed (blue) lines, respectively. Error bands at 1$\sigma$ in the IAM stem from uncertainties in the LECs.}
\label{fig:AbsAIJ}
\end{figure*}

\begin{figure*}[htb!]
\centering
\begin{tabular}{cc}
\includegraphics[width=0.45\textwidth]{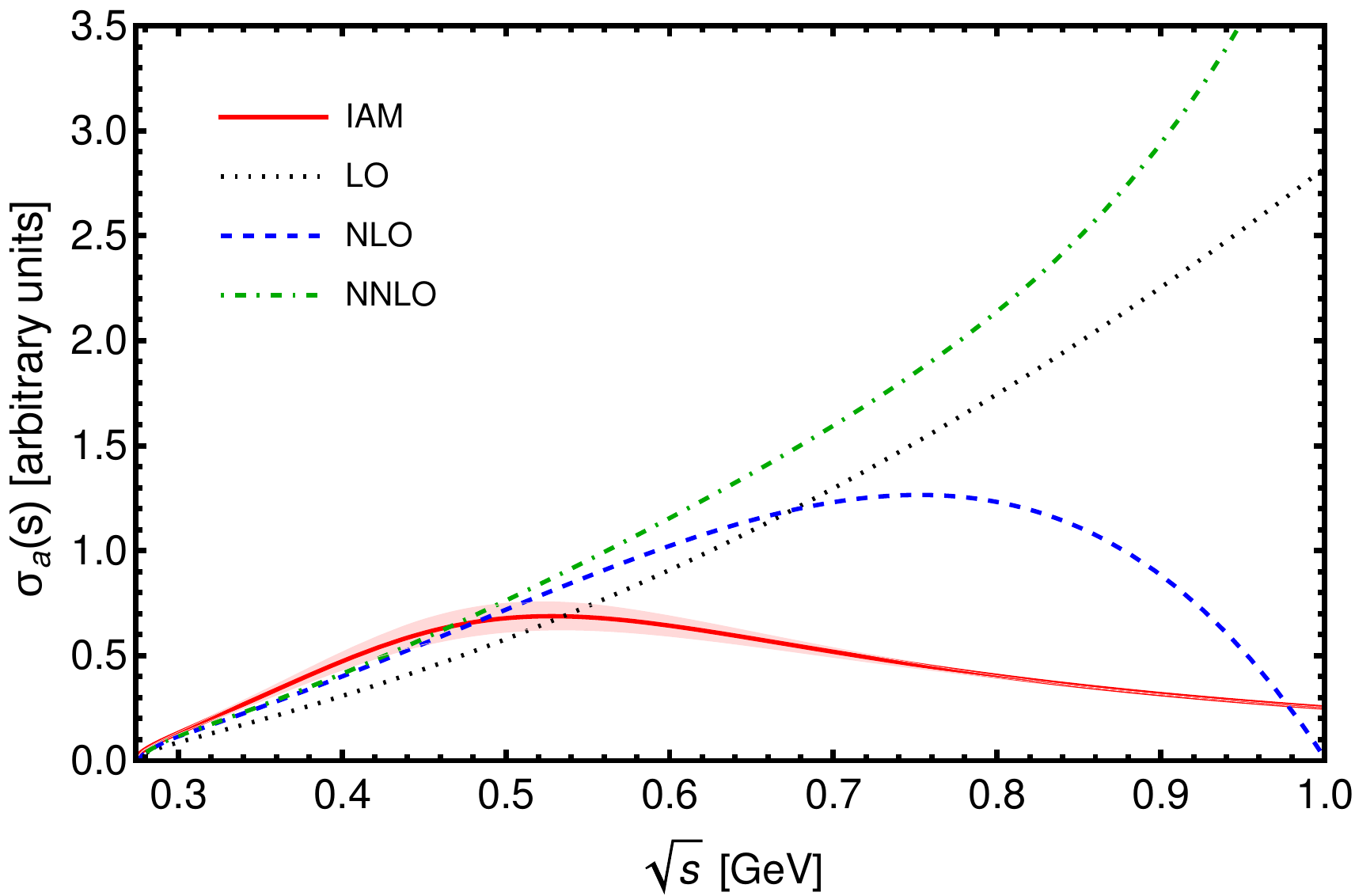}
&
\includegraphics[width=0.44\textwidth]{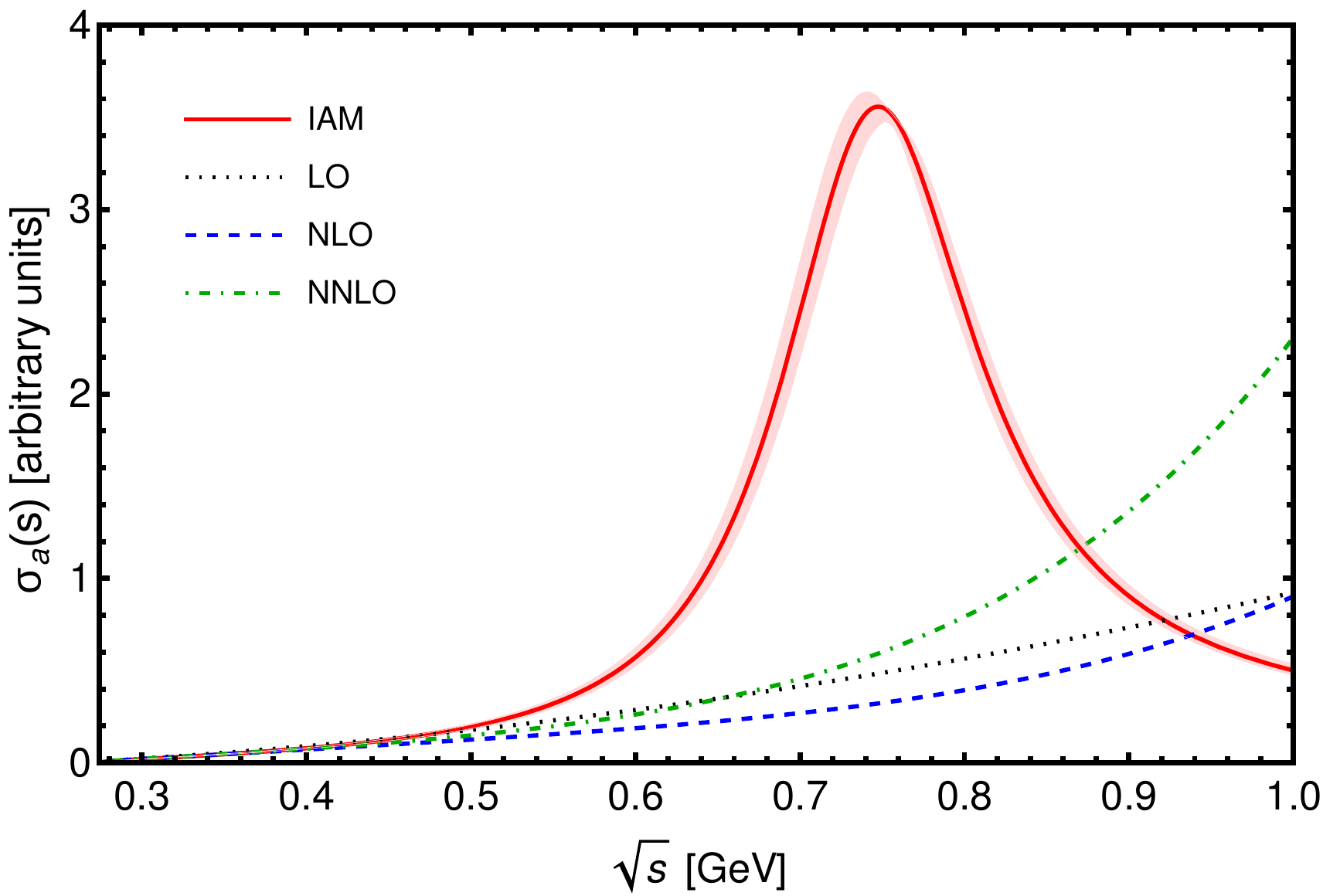}
\end{tabular}
\caption{Cross-sections of the $a\pi^0\to\pi^+\pi^-$ (left panel) and $a\pi^+\to\pi^0\pi^+$ (right panel) channels predicted by the IAM (solid, red) and in ChPT at LO (dotted, black), NLO (dashed, blue) and including NNLO pieces (dot-dashed, green). Uncertainties in the IAM predictions are 1$\sigma$ C.L. regions stemming from the errors in the LECs.} 
\label{fig:xsecs_channels}   
\end{figure*}

\section{Conventions and details of the IAM analysis} 
\label{sec:IAMdetails} 

The IAM analysis is performed at the level of PWAs, which requires the relations between $\pi\pi$ states in the isospin basis, labeled as $|I\,I_3\rangle$ for total isospin $I$ and third component $I_3$, and the charge basis. For the $\pi^+\pi^-$ final state,
\begin{align}
&|00\rangle=-\frac{1}{\sqrt{3}}\left(|\pi^+\pi^-\rangle+|\pi^-\pi^+\rangle+|\pi^0\pi^0\rangle\right),\nonumber\\
&|20\rangle=\frac{1}{\sqrt{6}}\left(2|\pi^0\pi^0\rangle-|\pi^+\pi^-\rangle-|\pi^-\pi^+\rangle\right).
\end{align}
For the $\pi^\pm \pi^0$ final state,
\begin{align}
&|2\pm1\rangle=\mp\frac{1}{\sqrt{2}}\left(|\pi^\pm\pi^0\rangle+|\pi^0\pi^\pm\rangle\right),\nonumber\\
&|1\pm1\rangle=\mp\frac{1}{\sqrt{2}}\left(|\pi^\pm\pi^0\rangle-|\pi^0\pi^\pm\rangle\right).
\end{align}
These relations have been used to project the chiral amplitudes (given in the charge basis) onto the isospin basis, leading to Eqs.~\eqref{eq:isospin_neutral} and \eqref{eq:isospin_charged}. 

In the following we present additional results comparing the different amplitudes included in our analysis. In Figs.~\ref{fig:AbsAIJ} we show the absolute values of the PWAs in ChPT at LO (black dotted), at NLO (blue dashed) and in the IAM (red solid lines). In turn, we show in  Figs.~\ref{fig:xsecs_channels} the contributions to the cross sections in separate channels (in the charge basis) contributing to the thermal rate. Besides the results in IAM (red solid), we show the ones in ChPT at LO (black dotted), contributing to the cross-section like LO$^2$, NLO (blue dashed), adding to the latter also the LO-NLO interference terms  and, finally, adding also the NNLO contributions to the rates (green dot-dashed lines).

\begin{figure*}[tb!]
\centering
\begin{tabular}{cc}
\includegraphics[width=0.45\textwidth]{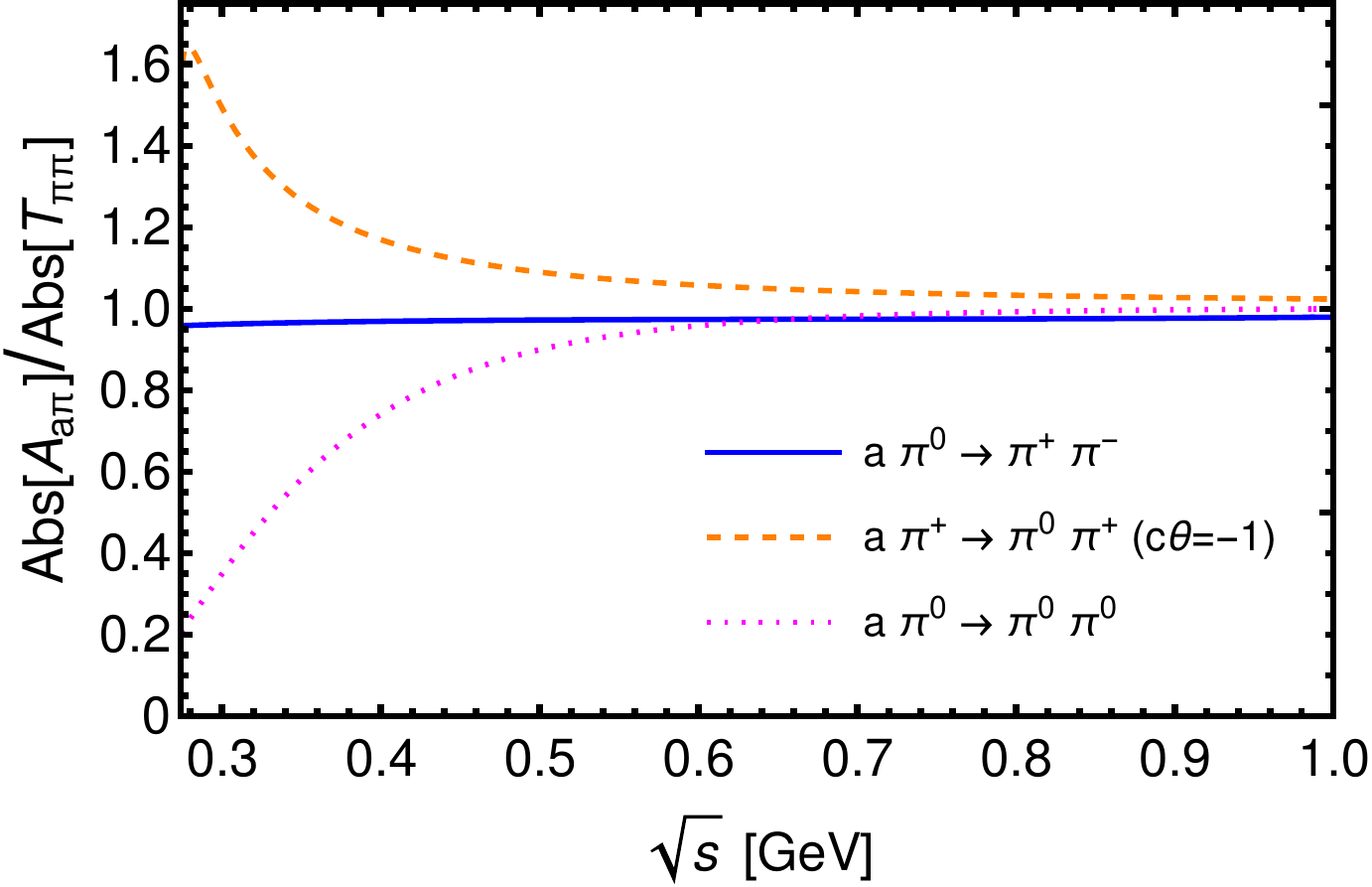}
&
\includegraphics[width=0.44\textwidth]{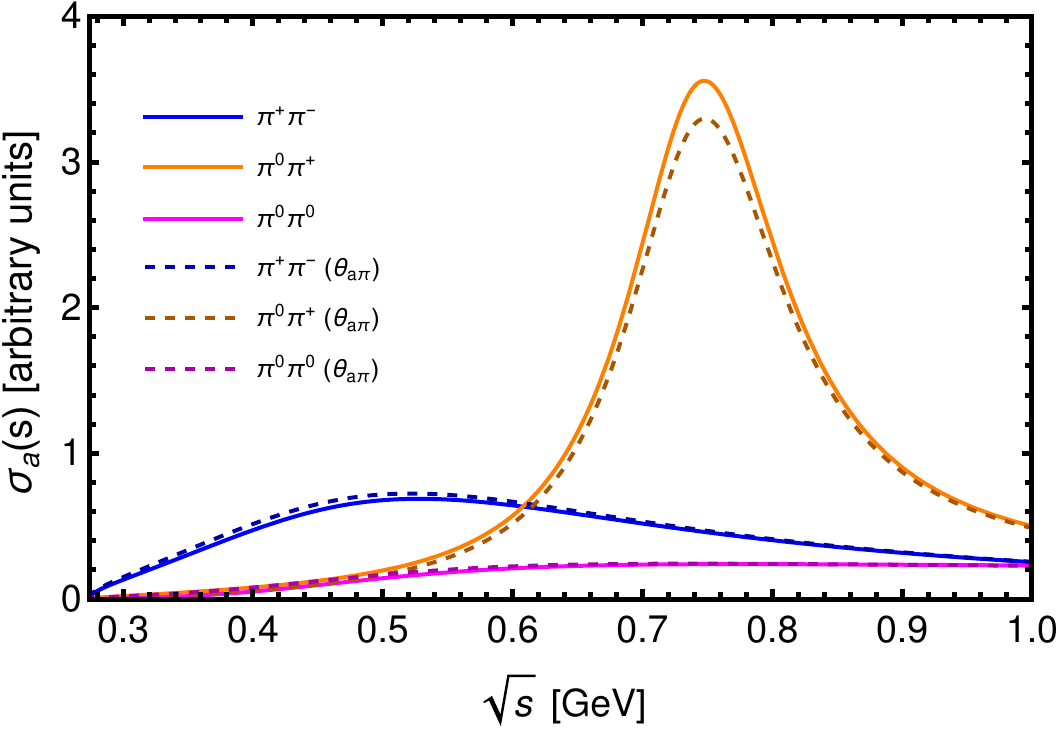}
\end{tabular}
\caption{Comparison of the amplitudes and cross sections for the full NLO ChPT calculation of $a\pi\to\pi\pi$ (amplitude denoted by $A_{a\pi}$) and for the NLO ChPT calculation of $\pi^0\pi\to\pi\pi$ rescaled by $\theta_{a\pi}$ (amplitude denoted by $T_{\pi\pi}$). In the left panel we show the ratio of the absolute values of the amplitudes while on the right panel we show a comparison of the cross sections unitarizing the corresponding perturbative amplitudes with IAM.} 
\label{fig:NRV}   
\end{figure*}

\section{ChPT expressions of phase shifts} 
\label{sec:pert_phase_shifts}

Let us describe a given $a\pi\to\pi\pi$ PWA (omitting indices $I$ and $J$) in ChPT up to NLO as
\begin{align}
\label{}
A=A_2+{\rm Re}(A_4)+i\rho T_2 A_2,
\end{align}
where we have labeled the amplitudes by their chiral order and $\rho\equiv\rho(s)=\sigma(s)/32\pi$. Then
\begin{align}
  A&=e^{i\delta}\sqrt{\left(A_2+{\rm Re}( A_4)\right)^2+\rho^2 T_2^2 A_2^2}\nonumber\\
  &=A_2+{\rm Re}(A_4)+i\delta_2 A_2+{\cal O}(p^6)~.
\end{align}
Comparing the two equations we obtain that 
\begin{align}
\label{221014.2}
\delta_2&=\rho T_2~.
\end{align}

A similar calculation can be done for $\pi\pi$ scattering PWAs that we denote as $T$. Given the corresponding element of the $S$-matrix, $S=e^{2i\delta}=1+2i\rho T$, with 
\begin{align}
  T&=\frac{1}{\rho}e^{i\delta}\sin\delta~.
\end{align}
By matching its perturbative  expansions, $T=T_2+T_4+{\cal O}(p^6)$, to $\delta=\delta_2+\delta_4+{\cal O}(p^6)$, one obtains 
\begin{align}
  \delta_2&=\rho T_2~,\\
  \delta_4&=\rho \Re T_4~.
  \end{align}
These are the expressions employed to obtain the ChPT phase shifts in Fig.~\ref{fig:phase_shifts}.
\\

\section{Comparison with Ref.~\cite{Notari:2022zxo}}
\label{sec:comparison}

A similar approach to treating the $a\pi\leftrightarrow \pi\pi$ rate below $T_c$ was followed in Ref.~\cite{Notari:2022zxo} that appeared concurrently with our work. This analysis uses a different chiral rotation of the quark fields to transfer the $a G\tilde G$ term into the quark mass matrix  in which the derivative axion coupling to the pion axial current vanishes and the complete axion-pion interactions are recovered by the rotation of the $a-\pi^0$ fields to the mass basis. 

In this framework it becomes clear that up to chiral-symmetry breaking terms $\propto m_\pi^2$, one can obtain the $a\pi\to \pi\pi$ scattering amplitude by rescaling the strong $\pi^0\pi\to \pi\pi$ amplitudes with the corresponding mixing angle $\theta_{a\pi}=3C_{a\pi}f_\pi/2f_a$.   Ref.~\cite{Notari:2022zxo} then uses this observation to obtain the axion-pion rates implementing amplitudes stemming from a set of Roy equations and dispersion-relations constraints calculated in Ref.~\cite{Garcia-Martin:2011iqs}. In comparison with a unitarization of the full NLO chiral amplitude such as the one developed in this paper, this procedure misses $\mathcal O(m_\pi^2/s)$ corrections that are expected to be important only at small energies (or temperatures). 

In Fig.~\ref{fig:NRV}, we illustrate this by comparing the results obtained for the different channels using the full NLO calculation of $a\pi\to\pi\pi$ in ChPT or using the NLO calculation of $\pi^0\pi\to\pi\pi$ scattering~\cite{Gasser:1983yg} multiplied by the mixing angle $\theta_{a\pi}$. From the left panel, showing the ratio of the amplitudes in the two methods, we observe that the $\mathcal O(m_\pi^2/s)$ corrections to the $a\pi^+\to\pi^0\pi^+$ and $a\pi^0\to\pi^0\pi^0$ are quite significant, up to $50\%-75\%$ for $\sqrt{s}\lesssim 0.5$ GeV, while they are small 
(of order $5\%$ in the same energy range)
for the $a\pi^0\to\pi^-\pi^+$ channel.\footnote{For instance, note that in the basis of Ref.~\cite{Notari:2022zxo}, the rotation by $\theta_{a\pi}$ generates a $a\pi^0\to\pi^0\pi^0$ term from the LO $m_\pi^2/f_\pi^2(\pi^0)^4$ term in the Lagrangian. This term would be canceled in the full calculation by an $a(\pi^0)^3$ piece directly stemming from the quark mass term. Related to this, the error estimate ${\cal O}(m_\pi^2/s)$ from Ref.~\cite{Notari:2022zxo} for $a\pi^0\to \pi^0\pi^0$ fails short for this case because it really scales as ${\cal O}(m_\pi^2(4\pi f_\pi)^2/s^2)$, which in the EFT region of convergence is not small.} 
However,
for $\sqrt{s}\lesssim 0.5$ GeV, the $\pi^0\pi^+$ and $\pi^0\pi^0$ channels are subdominant with respect to $\pi^+\pi^-$, thus rendering the differences in the total rate to be small.
 
\begin{figure}[tb!]
\centering
\includegraphics[width=0.45\textwidth]{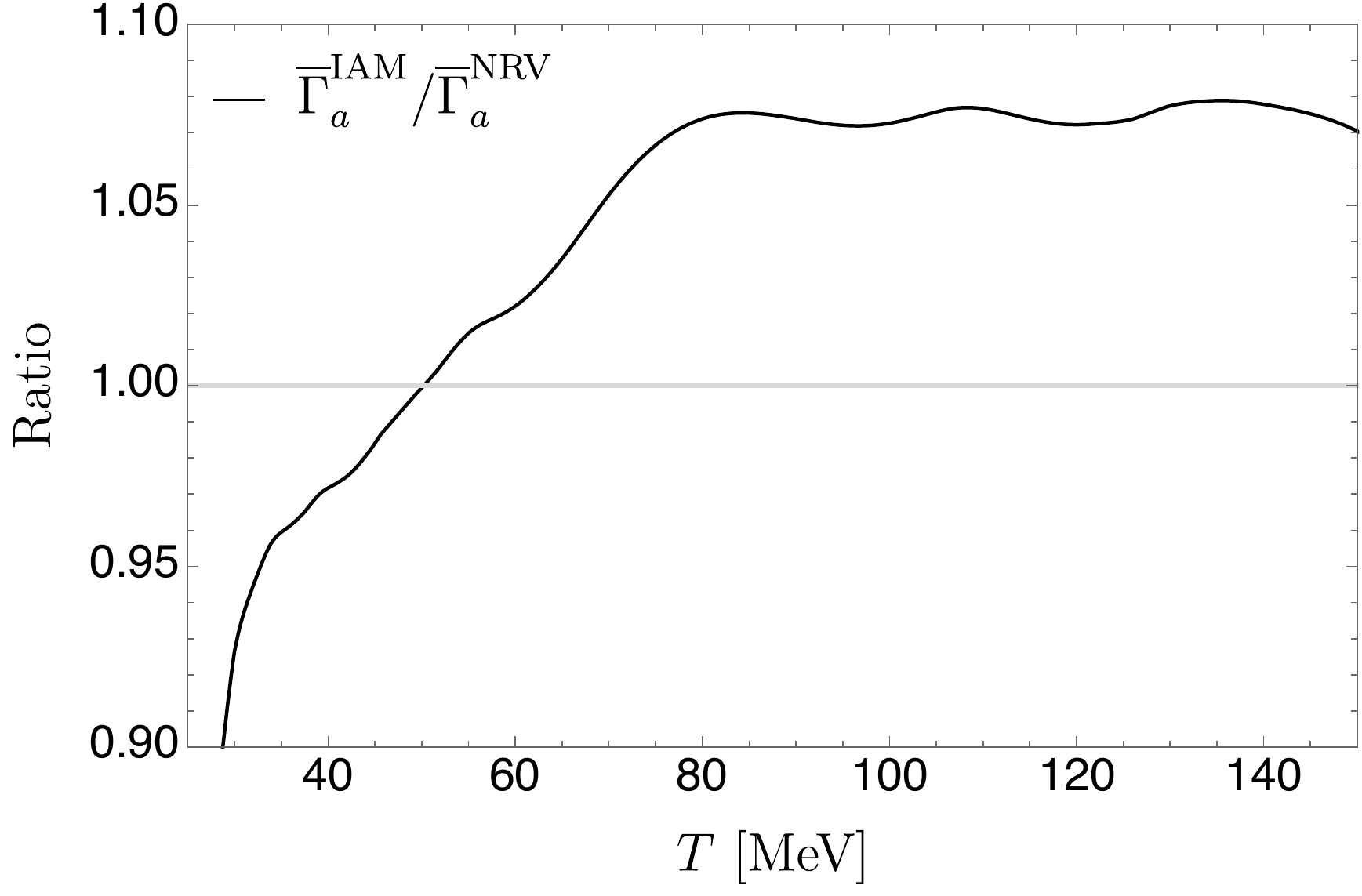}
\caption{Ratio between the IAM rate computed in this work and the $\bar{\Gamma}$ defined in \cite{Notari:2022zxo}. To uniform with the $\bar{\Gamma}$ definition in \cite{Notari:2022zxo}, we show here the IAM rate integrated with the modified Boltzmann factors
$ e^{-E_a/T} f_2 (1 + f_3)(1 + f_4)$.}
\label{fig:ratioNRV}   
\end{figure}

This is observed in the right-hand panel of Fig.~\ref{fig:NRV} where we show the total cross sections obtained for the different channels in the IAM using as perturbative input the full NLO ChPT calculation or the one derived from NLO ChPT of $\pi^0\pi\to\pi\pi$ rescaled by $\theta_{a\pi}$.
 In Fig.~\ref{fig:ratioNRV} we compare the full thermal rates in the two approaches, where we show that they agree within 8$\%$ (with higher discrepancy at higher $T$) in the temperature range between $40$ and $150$ MeV. This translates into a maximum $10\%$ difference in the instantaneous decoupling 
temperature.

One could use different non-perturbative methods that at low energies recover the chiral expansion up to some order in ChPT and end up with unitarized partial-wave amplitudes with the correct analytical properties \cite{Oller:2020guq}. A full analysis of the differences in the prediction of the rate with the IAM method is beyond the scope of our work. However, let us briefly discuss the differences stemming from using  
another popular approach called the $N/D$ method \cite{Oller:1998zr} in  meson-meson, meson-baryon and baryon-baryon scattering. A figure of merit in the comparison between IAM and $N/D$ in these cases is the spread in the central values of  the pole positions of the $\sigma$ and $\rho(770)$ resonances at different orders and in different number flavors of ChPT. For the $\sigma$ we have a spread in the real and imaginary parts of the pole position in $\sqrt{s}$ of only a 1.2\% and 2.4\%, respectively. We have taken the pole positions reported by applying, on the one hand, the IAM implemented from the NLO SU(2) \cite{Dobado:1996ps}, NNLO SU(2) \cite{Hannah:1999ev} and NLO SU(3) ChPT \cite{Oller:1998hw,Dobado:1996ps},  and, on the other hand, the $N/D$  method applied from the NLO SU(2) \cite{Albaladejo:2012te}, NNLO U(3) \cite{Guo:2012yt}, and tree-level ChPT \cite{Albaladejo:2008qa}. 
Similarly, for the $\rho(770)$ pole position in the $\sqrt{s}$ plane we find less than 1\% and 2.7\% of spread for the real and imaginary parts of the pole positions, respectively. Here, we have taken the pole positions from Refs.~\cite{Oller:1998hw,Dobado:1996ps,Guo:2012yt}.
These variations are representative of the differences one typically encounters between different methods to unitarize ChPT and we take them as indicative of the corresponding uncertainties in our approach. Note that these estimates are smaller than the uncertainties due to the variation of the cutoff discussed in Sec.~\ref{sec:cosmo}.

\bibliographystyle{apsrev4-1.bst}
\bibliography{bibliography}

\begin{thebibliography}{89}%
\makeatletter
\providecommand \@ifxundefined [1]{%
 \@ifx{#1\undefined}
}%
\providecommand \@ifnum [1]{%
 \ifnum #1\expandafter \@firstoftwo
 \else \expandafter \@secondoftwo
 \fi
}%
\providecommand \@ifx [1]{%
 \ifx #1\expandafter \@firstoftwo
 \else \expandafter \@secondoftwo
 \fi
}%
\providecommand \natexlab [1]{#1}%
\providecommand \enquote  [1]{``#1''}%
\providecommand \bibnamefont  [1]{#1}%
\providecommand \bibfnamefont [1]{#1}%
\providecommand \citenamefont [1]{#1}%
\providecommand \href@noop [0]{\@secondoftwo}%
\providecommand \href [0]{\begingroup \@sanitize@url \@href}%
\providecommand \@href[1]{\@@startlink{#1}\@@href}%
\providecommand \@@href[1]{\endgroup#1\@@endlink}%
\providecommand \@sanitize@url [0]{\catcode `\\12\catcode `\$12\catcode
  `\&12\catcode `\#12\catcode `\^12\catcode `\_12\catcode `\%12\relax}%
\providecommand \@@startlink[1]{}%
\providecommand \@@endlink[0]{}%
\providecommand \url  [0]{\begingroup\@sanitize@url \@url }%
\providecommand \@url [1]{\endgroup\@href {#1}{\urlprefix }}%
\providecommand \urlprefix  [0]{URL }%
\providecommand \Eprint [0]{\href }%
\providecommand \doibase [0]{http://dx.doi.org/}%
\providecommand \selectlanguage [0]{\@gobble}%
\providecommand \bibinfo  [0]{\@secondoftwo}%
\providecommand \bibfield  [0]{\@secondoftwo}%
\providecommand \translation [1]{[#1]}%
\providecommand \BibitemOpen [0]{}%
\providecommand \bibitemStop [0]{}%
\providecommand \bibitemNoStop [0]{.\EOS\space}%
\providecommand \EOS [0]{\spacefactor3000\relax}%
\providecommand \BibitemShut  [1]{\csname bibitem#1\endcsname}%
\let\auto@bib@innerbib\@empty
\bibitem [{\citenamefont {Peccei}\ and\ \citenamefont
  {Quinn}(1977{\natexlab{a}})}]{Peccei:1977hh}%
  \BibitemOpen
  \bibfield  {author} {\bibinfo {author} {\bibfnamefont {R.~D.}\ \bibnamefont
  {Peccei}}\ and\ \bibinfo {author} {\bibfnamefont {H.~R.}\ \bibnamefont
  {Quinn}},\ }\href {\doibase 10.1103/PhysRevLett.38.1440} {\bibfield
  {journal} {\bibinfo  {journal} {Phys. Rev. Lett.}\ }\textbf {\bibinfo
  {volume} {38}},\ \bibinfo {pages} {1440} (\bibinfo {year}
  {1977}{\natexlab{a}})}\BibitemShut {NoStop}%
\bibitem [{\citenamefont {Peccei}\ and\ \citenamefont
  {Quinn}(1977{\natexlab{b}})}]{Peccei:1977ur}%
  \BibitemOpen
  \bibfield  {author} {\bibinfo {author} {\bibfnamefont {R.~D.}\ \bibnamefont
  {Peccei}}\ and\ \bibinfo {author} {\bibfnamefont {H.~R.}\ \bibnamefont
  {Quinn}},\ }\href {\doibase 10.1103/PhysRevD.16.1791} {\bibfield  {journal}
  {\bibinfo  {journal} {Phys. Rev.}\ }\textbf {\bibinfo {volume} {D16}},\
  \bibinfo {pages} {1791} (\bibinfo {year} {1977}{\natexlab{b}})}\BibitemShut
  {NoStop}%
\bibitem [{\citenamefont {Wilczek}(1978)}]{Wilczek:1977pj}%
  \BibitemOpen
  \bibfield  {author} {\bibinfo {author} {\bibfnamefont {F.}~\bibnamefont
  {Wilczek}},\ }\href {\doibase 10.1103/PhysRevLett.40.279} {\bibfield
  {journal} {\bibinfo  {journal} {Phys. Rev. Lett.}\ }\textbf {\bibinfo
  {volume} {40}},\ \bibinfo {pages} {279} (\bibinfo {year} {1978})}\BibitemShut
  {NoStop}%
\bibitem [{\citenamefont {Weinberg}(1978)}]{Weinberg:1977ma}%
  \BibitemOpen
  \bibfield  {author} {\bibinfo {author} {\bibfnamefont {S.}~\bibnamefont
  {Weinberg}},\ }\href {\doibase 10.1103/PhysRevLett.40.223} {\bibfield
  {journal} {\bibinfo  {journal} {Phys. Rev. Lett.}\ }\textbf {\bibinfo
  {volume} {40}},\ \bibinfo {pages} {223} (\bibinfo {year} {1978})}\BibitemShut
  {NoStop}%
\bibitem [{\citenamefont {Preskill}\ \emph {et~al.}(1983)\citenamefont
  {Preskill}, \citenamefont {Wise},\ and\ \citenamefont
  {Wilczek}}]{Preskill:1982cy}%
  \BibitemOpen
  \bibfield  {author} {\bibinfo {author} {\bibfnamefont {J.}~\bibnamefont
  {Preskill}}, \bibinfo {author} {\bibfnamefont {M.~B.}\ \bibnamefont {Wise}},
  \ and\ \bibinfo {author} {\bibfnamefont {F.}~\bibnamefont {Wilczek}},\ }\href
  {\doibase 10.1016/0370-2693(83)90637-8} {\bibfield  {journal} {\bibinfo
  {journal} {Phys. Lett.}\ }\textbf {\bibinfo {volume} {120B}},\ \bibinfo
  {pages} {127} (\bibinfo {year} {1983})}\BibitemShut {NoStop}%
\bibitem [{\citenamefont {Abbott}\ and\ \citenamefont
  {Sikivie}(1983)}]{Abbott:1982af}%
  \BibitemOpen
  \bibfield  {author} {\bibinfo {author} {\bibfnamefont {L.~F.}\ \bibnamefont
  {Abbott}}\ and\ \bibinfo {author} {\bibfnamefont {P.}~\bibnamefont
  {Sikivie}},\ }\href {\doibase 10.1016/0370-2693(83)90638-X} {\bibfield
  {journal} {\bibinfo  {journal} {Phys. Lett.}\ }\textbf {\bibinfo {volume}
  {120B}},\ \bibinfo {pages} {133} (\bibinfo {year} {1983})}\BibitemShut
  {NoStop}%
\bibitem [{\citenamefont {Dine}\ and\ \citenamefont
  {Fischler}(1983)}]{Dine:1982ah}%
  \BibitemOpen
  \bibfield  {author} {\bibinfo {author} {\bibfnamefont {M.}~\bibnamefont
  {Dine}}\ and\ \bibinfo {author} {\bibfnamefont {W.}~\bibnamefont
  {Fischler}},\ }\href {\doibase 10.1016/0370-2693(83)90639-1} {\bibfield
  {journal} {\bibinfo  {journal} {Phys. Lett.}\ }\textbf {\bibinfo {volume}
  {120B}},\ \bibinfo {pages} {137} (\bibinfo {year} {1983})}\BibitemShut
  {NoStop}%
\bibitem [{\citenamefont {Davis}(1986)}]{Davis:1986xc}%
  \BibitemOpen
  \bibfield  {author} {\bibinfo {author} {\bibfnamefont {R.~L.}\ \bibnamefont
  {Davis}},\ }\href {\doibase 10.1016/0370-2693(86)90300-X} {\bibfield
  {journal} {\bibinfo  {journal} {Phys. Lett. B}\ }\textbf {\bibinfo {volume}
  {180}},\ \bibinfo {pages} {225} (\bibinfo {year} {1986})}\BibitemShut
  {NoStop}%
\bibitem [{\citenamefont {Turner}(1987)}]{Turner:1986tb}%
  \BibitemOpen
  \bibfield  {author} {\bibinfo {author} {\bibfnamefont {M.~S.}\ \bibnamefont
  {Turner}},\ }\href {\doibase 10.1103/PhysRevLett.59.2489} {\bibfield
  {journal} {\bibinfo  {journal} {Phys. Rev. Lett.}\ }\textbf {\bibinfo
  {volume} {59}},\ \bibinfo {pages} {2489} (\bibinfo {year} {1987})},\ \bibinfo
  {note} {[Erratum: Phys.Rev.Lett. 60, 1101 (1988)]}\BibitemShut {NoStop}%
\bibitem [{\citenamefont {Aghanim}\ \emph
  {et~al.}(2020{\natexlab{a}})\citenamefont {Aghanim} \emph
  {et~al.}}]{Akrami:2018vks}%
  \BibitemOpen
  \bibfield  {author} {\bibinfo {author} {\bibfnamefont {N.}~\bibnamefont
  {Aghanim}} \emph {et~al.} (\bibinfo {collaboration} {Planck}),\ }\href
  {\doibase 10.1051/0004-6361/201833880} {\bibfield  {journal} {\bibinfo
  {journal} {Astron. Astrophys.}\ }\textbf {\bibinfo {volume} {641}},\ \bibinfo
  {pages} {A1} (\bibinfo {year} {2020}{\natexlab{a}})},\ \Eprint
  {http://arxiv.org/abs/1807.06205} {arXiv:1807.06205 [astro-ph.CO]}
  \BibitemShut {NoStop}%
\bibitem [{\citenamefont {Aghanim}\ \emph
  {et~al.}(2020{\natexlab{b}})\citenamefont {Aghanim} \emph
  {et~al.}}]{Aghanim:2018eyx}%
  \BibitemOpen
  \bibfield  {author} {\bibinfo {author} {\bibfnamefont {N.}~\bibnamefont
  {Aghanim}} \emph {et~al.} (\bibinfo {collaboration} {Planck}),\ }\href
  {\doibase 10.1051/0004-6361/201833910} {\bibfield  {journal} {\bibinfo
  {journal} {Astron. Astrophys.}\ }\textbf {\bibinfo {volume} {641}},\ \bibinfo
  {pages} {A6} (\bibinfo {year} {2020}{\natexlab{b}})},\ \Eprint
  {http://arxiv.org/abs/1807.06209} {arXiv:1807.06209 [astro-ph.CO]}
  \BibitemShut {NoStop}%
\bibitem [{\citenamefont {Abazajian}\ \emph {et~al.}(2016)\citenamefont
  {Abazajian} \emph {et~al.}}]{Abazajian:2016yjj}%
  \BibitemOpen
  \bibfield  {author} {\bibinfo {author} {\bibfnamefont {K.~N.}\ \bibnamefont
  {Abazajian}} \emph {et~al.} (\bibinfo {collaboration} {CMB-S4}),\ }\href@noop
  {} {\  (\bibinfo {year} {2016})},\ \Eprint {http://arxiv.org/abs/1610.02743}
  {arXiv:1610.02743 [astro-ph.CO]} \BibitemShut {NoStop}%
\bibitem [{\citenamefont {Masso}\ \emph {et~al.}(2002)\citenamefont {Masso},
  \citenamefont {Rota},\ and\ \citenamefont {Zsembinszki}}]{Masso:2002np}%
  \BibitemOpen
  \bibfield  {author} {\bibinfo {author} {\bibfnamefont {E.}~\bibnamefont
  {Masso}}, \bibinfo {author} {\bibfnamefont {F.}~\bibnamefont {Rota}}, \ and\
  \bibinfo {author} {\bibfnamefont {G.}~\bibnamefont {Zsembinszki}},\ }\href
  {\doibase 10.1103/PhysRevD.66.023004} {\bibfield  {journal} {\bibinfo
  {journal} {Phys. Rev. D}\ }\textbf {\bibinfo {volume} {66}},\ \bibinfo
  {pages} {023004} (\bibinfo {year} {2002})},\ \Eprint
  {http://arxiv.org/abs/hep-ph/0203221} {arXiv:hep-ph/0203221} \BibitemShut
  {NoStop}%
\bibitem [{\citenamefont {Graf}\ and\ \citenamefont
  {Steffen}(2011)}]{Graf:2010tv}%
  \BibitemOpen
  \bibfield  {author} {\bibinfo {author} {\bibfnamefont {P.}~\bibnamefont
  {Graf}}\ and\ \bibinfo {author} {\bibfnamefont {F.~D.}\ \bibnamefont
  {Steffen}},\ }\href {\doibase 10.1103/PhysRevD.83.075011} {\bibfield
  {journal} {\bibinfo  {journal} {Phys. Rev. D}\ }\textbf {\bibinfo {volume}
  {83}},\ \bibinfo {pages} {075011} (\bibinfo {year} {2011})},\ \Eprint
  {http://arxiv.org/abs/1008.4528} {arXiv:1008.4528 [hep-ph]} \BibitemShut
  {NoStop}%
\bibitem [{\citenamefont {Berezhiani}\ \emph {et~al.}(1992)\citenamefont
  {Berezhiani}, \citenamefont {Sakharov},\ and\ \citenamefont
  {Khlopov}}]{Berezhiani:1992rk}%
  \BibitemOpen
  \bibfield  {author} {\bibinfo {author} {\bibfnamefont {Z.}~\bibnamefont
  {Berezhiani}}, \bibinfo {author} {\bibfnamefont {A.}~\bibnamefont
  {Sakharov}}, \ and\ \bibinfo {author} {\bibfnamefont {M.}~\bibnamefont
  {Khlopov}},\ }\href@noop {} {\bibfield  {journal} {\bibinfo  {journal} {Sov.
  J. Nucl. Phys.}\ }\textbf {\bibinfo {volume} {55}},\ \bibinfo {pages} {1063}
  (\bibinfo {year} {1992})}\BibitemShut {NoStop}%
\bibitem [{\citenamefont {Chang}\ and\ \citenamefont
  {Choi}(1993)}]{Chang:1993gm}%
  \BibitemOpen
  \bibfield  {author} {\bibinfo {author} {\bibfnamefont {S.}~\bibnamefont
  {Chang}}\ and\ \bibinfo {author} {\bibfnamefont {K.}~\bibnamefont {Choi}},\
  }\href {\doibase 10.1016/0370-2693(93)90656-3} {\bibfield  {journal}
  {\bibinfo  {journal} {Phys. Lett.}\ }\textbf {\bibinfo {volume} {B316}},\
  \bibinfo {pages} {51} (\bibinfo {year} {1993})},\ \Eprint
  {http://arxiv.org/abs/hep-ph/9306216} {arXiv:hep-ph/9306216 [hep-ph]}
  \BibitemShut {NoStop}%
\bibitem [{\citenamefont {Hannestad}\ \emph {et~al.}(2005)\citenamefont
  {Hannestad}, \citenamefont {Mirizzi},\ and\ \citenamefont
  {Raffelt}}]{Hannestad:2005df}%
  \BibitemOpen
  \bibfield  {author} {\bibinfo {author} {\bibfnamefont {S.}~\bibnamefont
  {Hannestad}}, \bibinfo {author} {\bibfnamefont {A.}~\bibnamefont {Mirizzi}},
  \ and\ \bibinfo {author} {\bibfnamefont {G.}~\bibnamefont {Raffelt}},\ }\href
  {\doibase 10.1088/1475-7516/2005/07/002} {\bibfield  {journal} {\bibinfo
  {journal} {JCAP}\ }\textbf {\bibinfo {volume} {07}},\ \bibinfo {pages} {002}
  (\bibinfo {year} {2005})},\ \Eprint {http://arxiv.org/abs/hep-ph/0504059}
  {arXiv:hep-ph/0504059} \BibitemShut {NoStop}%
\bibitem [{\citenamefont {D'Eramo}\ \emph
  {et~al.}(2022{\natexlab{a}})\citenamefont {D'Eramo}, \citenamefont
  {Hajkarim},\ and\ \citenamefont {Yun}}]{DEramo:2021psx}%
  \BibitemOpen
  \bibfield  {author} {\bibinfo {author} {\bibfnamefont {F.}~\bibnamefont
  {D'Eramo}}, \bibinfo {author} {\bibfnamefont {F.}~\bibnamefont {Hajkarim}}, \
  and\ \bibinfo {author} {\bibfnamefont {S.}~\bibnamefont {Yun}},\ }\href
  {\doibase 10.1103/PhysRevLett.128.152001} {\bibfield  {journal} {\bibinfo
  {journal} {Phys. Rev. Lett.}\ }\textbf {\bibinfo {volume} {128}},\ \bibinfo
  {pages} {152001} (\bibinfo {year} {2022}{\natexlab{a}})},\ \Eprint
  {http://arxiv.org/abs/2108.04259} {arXiv:2108.04259 [hep-ph]} \BibitemShut
  {NoStop}%
\bibitem [{\citenamefont {D'Eramo}\ \emph {et~al.}(2021)\citenamefont
  {D'Eramo}, \citenamefont {Hajkarim},\ and\ \citenamefont
  {Yun}}]{DEramo:2021lgb}%
  \BibitemOpen
  \bibfield  {author} {\bibinfo {author} {\bibfnamefont {F.}~\bibnamefont
  {D'Eramo}}, \bibinfo {author} {\bibfnamefont {F.}~\bibnamefont {Hajkarim}}, \
  and\ \bibinfo {author} {\bibfnamefont {S.}~\bibnamefont {Yun}},\ }\href
  {\doibase 10.1007/JHEP10(2021)224} {\bibfield  {journal} {\bibinfo  {journal}
  {JHEP}\ }\textbf {\bibinfo {volume} {10}},\ \bibinfo {pages} {224} (\bibinfo
  {year} {2021})},\ \Eprint {http://arxiv.org/abs/2108.05371} {arXiv:2108.05371
  [hep-ph]} \BibitemShut {NoStop}%
\bibitem [{\citenamefont {Aoki}\ \emph {et~al.}(2006)\citenamefont {Aoki},
  \citenamefont {Fodor}, \citenamefont {Katz},\ and\ \citenamefont
  {Szabo}}]{Aoki:2006br}%
  \BibitemOpen
  \bibfield  {author} {\bibinfo {author} {\bibfnamefont {Y.}~\bibnamefont
  {Aoki}}, \bibinfo {author} {\bibfnamefont {Z.}~\bibnamefont {Fodor}},
  \bibinfo {author} {\bibfnamefont {S.~D.}\ \bibnamefont {Katz}}, \ and\
  \bibinfo {author} {\bibfnamefont {K.~K.}\ \bibnamefont {Szabo}},\ }\href
  {\doibase 10.1016/j.physletb.2006.10.021} {\bibfield  {journal} {\bibinfo
  {journal} {Phys. Lett. B}\ }\textbf {\bibinfo {volume} {643}},\ \bibinfo
  {pages} {46} (\bibinfo {year} {2006})},\ \Eprint
  {http://arxiv.org/abs/hep-lat/0609068} {arXiv:hep-lat/0609068} \BibitemShut
  {NoStop}%
\bibitem [{\citenamefont {Borsanyi}\ \emph {et~al.}(2010)\citenamefont
  {Borsanyi}, \citenamefont {Fodor}, \citenamefont {Hoelbling}, \citenamefont
  {Katz}, \citenamefont {Krieg}, \citenamefont {Ratti},\ and\ \citenamefont
  {Szabo}}]{Borsanyi:2010bp}%
  \BibitemOpen
  \bibfield  {author} {\bibinfo {author} {\bibfnamefont {S.}~\bibnamefont
  {Borsanyi}}, \bibinfo {author} {\bibfnamefont {Z.}~\bibnamefont {Fodor}},
  \bibinfo {author} {\bibfnamefont {C.}~\bibnamefont {Hoelbling}}, \bibinfo
  {author} {\bibfnamefont {S.~D.}\ \bibnamefont {Katz}}, \bibinfo {author}
  {\bibfnamefont {S.}~\bibnamefont {Krieg}}, \bibinfo {author} {\bibfnamefont
  {C.}~\bibnamefont {Ratti}}, \ and\ \bibinfo {author} {\bibfnamefont {K.~K.}\
  \bibnamefont {Szabo}} (\bibinfo {collaboration} {Wuppertal-Budapest}),\
  }\href {\doibase 10.1007/JHEP09(2010)073} {\bibfield  {journal} {\bibinfo
  {journal} {JHEP}\ }\textbf {\bibinfo {volume} {09}},\ \bibinfo {pages} {073}
  (\bibinfo {year} {2010})},\ \Eprint {http://arxiv.org/abs/1005.3508}
  {arXiv:1005.3508 [hep-lat]} \BibitemShut {NoStop}%
\bibitem [{\citenamefont {Bazavov}\ \emph {et~al.}(2012)\citenamefont {Bazavov}
  \emph {et~al.}}]{Bazavov:2011nk}%
  \BibitemOpen
  \bibfield  {author} {\bibinfo {author} {\bibfnamefont {A.}~\bibnamefont
  {Bazavov}} \emph {et~al.},\ }\href {\doibase 10.1103/PhysRevD.85.054503}
  {\bibfield  {journal} {\bibinfo  {journal} {Phys. Rev. D}\ }\textbf {\bibinfo
  {volume} {85}},\ \bibinfo {pages} {054503} (\bibinfo {year} {2012})},\
  \Eprint {http://arxiv.org/abs/1111.1710} {arXiv:1111.1710 [hep-lat]}
  \BibitemShut {NoStop}%
\bibitem [{\citenamefont {Caloni}\ \emph {et~al.}(2022)\citenamefont {Caloni},
  \citenamefont {Gerbino}, \citenamefont {Lattanzi},\ and\ \citenamefont
  {Visinelli}}]{Caloni:2022uya}%
  \BibitemOpen
  \bibfield  {author} {\bibinfo {author} {\bibfnamefont {L.}~\bibnamefont
  {Caloni}}, \bibinfo {author} {\bibfnamefont {M.}~\bibnamefont {Gerbino}},
  \bibinfo {author} {\bibfnamefont {M.}~\bibnamefont {Lattanzi}}, \ and\
  \bibinfo {author} {\bibfnamefont {L.}~\bibnamefont {Visinelli}},\ }\href
  {\doibase 10.1088/1475-7516/2022/09/021} {\bibfield  {journal} {\bibinfo
  {journal} {JCAP}\ }\textbf {\bibinfo {volume} {09}},\ \bibinfo {pages} {021}
  (\bibinfo {year} {2022})},\ \Eprint {http://arxiv.org/abs/2205.01637}
  {arXiv:2205.01637 [astro-ph.CO]} \BibitemShut {NoStop}%
\bibitem [{\citenamefont {D'Eramo}\ \emph
  {et~al.}(2022{\natexlab{b}})\citenamefont {D'Eramo}, \citenamefont
  {Di~Valentino}, \citenamefont {Giar\`e}, \citenamefont {Hajkarim},
  \citenamefont {Melchiorri}, \citenamefont {Mena}, \citenamefont {Renzi},\
  and\ \citenamefont {Yun}}]{DEramo:2022nvb}%
  \BibitemOpen
  \bibfield  {author} {\bibinfo {author} {\bibfnamefont {F.}~\bibnamefont
  {D'Eramo}}, \bibinfo {author} {\bibfnamefont {E.}~\bibnamefont
  {Di~Valentino}}, \bibinfo {author} {\bibfnamefont {W.}~\bibnamefont
  {Giar\`e}}, \bibinfo {author} {\bibfnamefont {F.}~\bibnamefont {Hajkarim}},
  \bibinfo {author} {\bibfnamefont {A.}~\bibnamefont {Melchiorri}}, \bibinfo
  {author} {\bibfnamefont {O.}~\bibnamefont {Mena}}, \bibinfo {author}
  {\bibfnamefont {F.}~\bibnamefont {Renzi}}, \ and\ \bibinfo {author}
  {\bibfnamefont {S.}~\bibnamefont {Yun}},\ }\href {\doibase
  10.1088/1475-7516/2022/09/022} {\bibfield  {journal} {\bibinfo  {journal}
  {JCAP}\ }\textbf {\bibinfo {volume} {09}},\ \bibinfo {pages} {022} (\bibinfo
  {year} {2022}{\natexlab{b}})},\ \Eprint {http://arxiv.org/abs/2205.07849}
  {arXiv:2205.07849 [astro-ph.CO]} \BibitemShut {NoStop}%
\bibitem [{\citenamefont {Di~Luzio}\ \emph
  {et~al.}(2021{\natexlab{a}})\citenamefont {Di~Luzio}, \citenamefont
  {Martinelli},\ and\ \citenamefont {Piazza}}]{DiLuzio:2021vjd}%
  \BibitemOpen
  \bibfield  {author} {\bibinfo {author} {\bibfnamefont {L.}~\bibnamefont
  {Di~Luzio}}, \bibinfo {author} {\bibfnamefont {G.}~\bibnamefont
  {Martinelli}}, \ and\ \bibinfo {author} {\bibfnamefont {G.}~\bibnamefont
  {Piazza}},\ }\href {\doibase 10.1103/PhysRevLett.126.241801} {\bibfield
  {journal} {\bibinfo  {journal} {Phys. Rev. Lett.}\ }\textbf {\bibinfo
  {volume} {126}},\ \bibinfo {pages} {241801} (\bibinfo {year}
  {2021}{\natexlab{a}})},\ \Eprint {http://arxiv.org/abs/2101.10330}
  {arXiv:2101.10330 [hep-ph]} \BibitemShut {NoStop}%
\bibitem [{\citenamefont {Lehmann}(1972)}]{Lehmann:1972kv}%
  \BibitemOpen
  \bibfield  {author} {\bibinfo {author} {\bibfnamefont {H.}~\bibnamefont
  {Lehmann}},\ }\href {\doibase 10.1016/0370-2693(72)90691-0} {\bibfield
  {journal} {\bibinfo  {journal} {Phys. Lett. B}\ }\textbf {\bibinfo {volume}
  {41}},\ \bibinfo {pages} {529} (\bibinfo {year} {1972})}\BibitemShut
  {NoStop}%
\bibitem [{\citenamefont {Truong}(1988)}]{Truong:1988zp}%
  \BibitemOpen
  \bibfield  {author} {\bibinfo {author} {\bibfnamefont {T.~N.}\ \bibnamefont
  {Truong}},\ }\href {\doibase 10.1103/PhysRevLett.61.2526} {\bibfield
  {journal} {\bibinfo  {journal} {Phys. Rev. Lett.}\ }\textbf {\bibinfo
  {volume} {61}},\ \bibinfo {pages} {2526} (\bibinfo {year}
  {1988})}\BibitemShut {NoStop}%
\bibitem [{\citenamefont {Oller}(2020{\natexlab{a}})}]{Oller:2020guq}%
  \BibitemOpen
  \bibfield  {author} {\bibinfo {author} {\bibfnamefont {J.~A.}\ \bibnamefont
  {Oller}},\ }\href {\doibase 10.3390/sym12071114} {\bibfield  {journal}
  {\bibinfo  {journal} {Symmetry}\ }\textbf {\bibinfo {volume} {12}},\ \bibinfo
  {pages} {1114} (\bibinfo {year} {2020}{\natexlab{a}})},\ \Eprint
  {http://arxiv.org/abs/2005.14417} {arXiv:2005.14417 [hep-ph]} \BibitemShut
  {NoStop}%
\bibitem [{\citenamefont {Georgi}\ \emph {et~al.}(1986)\citenamefont {Georgi},
  \citenamefont {Kaplan},\ and\ \citenamefont {Randall}}]{Georgi:1986df}%
  \BibitemOpen
  \bibfield  {author} {\bibinfo {author} {\bibfnamefont {H.}~\bibnamefont
  {Georgi}}, \bibinfo {author} {\bibfnamefont {D.~B.}\ \bibnamefont {Kaplan}},
  \ and\ \bibinfo {author} {\bibfnamefont {L.}~\bibnamefont {Randall}},\ }\href
  {\doibase 10.1016/0370-2693(86)90688-X} {\bibfield  {journal} {\bibinfo
  {journal} {Phys. Lett. B}\ }\textbf {\bibinfo {volume} {169}},\ \bibinfo
  {pages} {73} (\bibinfo {year} {1986})}\BibitemShut {NoStop}%
\bibitem [{\citenamefont {Di~Luzio}\ \emph {et~al.}(2020)\citenamefont
  {Di~Luzio}, \citenamefont {Giannotti}, \citenamefont {Nardi},\ and\
  \citenamefont {Visinelli}}]{DiLuzio:2020wdo}%
  \BibitemOpen
  \bibfield  {author} {\bibinfo {author} {\bibfnamefont {L.}~\bibnamefont
  {Di~Luzio}}, \bibinfo {author} {\bibfnamefont {M.}~\bibnamefont {Giannotti}},
  \bibinfo {author} {\bibfnamefont {E.}~\bibnamefont {Nardi}}, \ and\ \bibinfo
  {author} {\bibfnamefont {L.}~\bibnamefont {Visinelli}},\ }\href {\doibase
  10.1016/j.physrep.2020.06.002} {\bibfield  {journal} {\bibinfo  {journal}
  {Phys. Rept.}\ }\textbf {\bibinfo {volume} {870}},\ \bibinfo {pages} {1}
  (\bibinfo {year} {2020})},\ \Eprint {http://arxiv.org/abs/2003.01100}
  {arXiv:2003.01100 [hep-ph]} \BibitemShut {NoStop}%
\bibitem [{\citenamefont {Kim}(1979)}]{Kim:1979if}%
  \BibitemOpen
  \bibfield  {author} {\bibinfo {author} {\bibfnamefont {J.~E.}\ \bibnamefont
  {Kim}},\ }\href {\doibase 10.1103/PhysRevLett.43.103} {\bibfield  {journal}
  {\bibinfo  {journal} {Phys. Rev. Lett.}\ }\textbf {\bibinfo {volume} {43}},\
  \bibinfo {pages} {103} (\bibinfo {year} {1979})}\BibitemShut {NoStop}%
\bibitem [{\citenamefont {Shifman}\ \emph {et~al.}(1980)\citenamefont
  {Shifman}, \citenamefont {Vainshtein},\ and\ \citenamefont
  {Zakharov}}]{Shifman:1979if}%
  \BibitemOpen
  \bibfield  {author} {\bibinfo {author} {\bibfnamefont {M.~A.}\ \bibnamefont
  {Shifman}}, \bibinfo {author} {\bibfnamefont {A.~I.}\ \bibnamefont
  {Vainshtein}}, \ and\ \bibinfo {author} {\bibfnamefont {V.~I.}\ \bibnamefont
  {Zakharov}},\ }\href {\doibase 10.1016/0550-3213(80)90209-6} {\bibfield
  {journal} {\bibinfo  {journal} {Nucl. Phys.}\ }\textbf {\bibinfo {volume}
  {B166}},\ \bibinfo {pages} {493} (\bibinfo {year} {1980})}\BibitemShut
  {NoStop}%
\bibitem [{\citenamefont {Zhitnitsky}(1980)}]{Zhitnitsky:1980tq}%
  \BibitemOpen
  \bibfield  {author} {\bibinfo {author} {\bibfnamefont {A.~R.}\ \bibnamefont
  {Zhitnitsky}},\ }\href@noop {} {\bibfield  {journal} {\bibinfo  {journal}
  {Sov. J. Nucl. Phys.}\ }\textbf {\bibinfo {volume} {31}},\ \bibinfo {pages}
  {260} (\bibinfo {year} {1980})},\ \bibinfo {note} {[Yad.
  Fiz.31,497(1980)]}\BibitemShut {NoStop}%
\bibitem [{\citenamefont {Dine}\ \emph {et~al.}(1981)\citenamefont {Dine},
  \citenamefont {Fischler},\ and\ \citenamefont {Srednicki}}]{Dine:1981rt}%
  \BibitemOpen
  \bibfield  {author} {\bibinfo {author} {\bibfnamefont {M.}~\bibnamefont
  {Dine}}, \bibinfo {author} {\bibfnamefont {W.}~\bibnamefont {Fischler}}, \
  and\ \bibinfo {author} {\bibfnamefont {M.}~\bibnamefont {Srednicki}},\ }\href
  {\doibase 10.1016/0370-2693(81)90590-6} {\bibfield  {journal} {\bibinfo
  {journal} {Phys. Lett.}\ }\textbf {\bibinfo {volume} {B104}},\ \bibinfo
  {pages} {199} (\bibinfo {year} {1981})}\BibitemShut {NoStop}%
\bibitem [{\citenamefont {Di~Luzio}\ and\ \citenamefont
  {Piazza}(2022)}]{DiLuzio:2022tbb}%
  \BibitemOpen
  \bibfield  {author} {\bibinfo {author} {\bibfnamefont {L.}~\bibnamefont
  {Di~Luzio}}\ and\ \bibinfo {author} {\bibfnamefont {G.}~\bibnamefont
  {Piazza}},\ }\href@noop {} {\  (\bibinfo {year} {2022})},\ \Eprint
  {http://arxiv.org/abs/2206.04061} {arXiv:2206.04061 [hep-ph]} \BibitemShut
  {NoStop}%
\bibitem [{\citenamefont {Gasser}\ and\ \citenamefont
  {Leutwyler}(1984)}]{Gasser:1983yg}%
  \BibitemOpen
  \bibfield  {author} {\bibinfo {author} {\bibfnamefont {J.}~\bibnamefont
  {Gasser}}\ and\ \bibinfo {author} {\bibfnamefont {H.}~\bibnamefont
  {Leutwyler}},\ }\href {\doibase 10.1016/0003-4916(84)90242-2} {\bibfield
  {journal} {\bibinfo  {journal} {Annals Phys.}\ }\textbf {\bibinfo {volume}
  {158}},\ \bibinfo {pages} {142} (\bibinfo {year} {1984})}\BibitemShut
  {NoStop}%
\bibitem [{\citenamefont {Lehmann}\ \emph {et~al.}(1955)\citenamefont
  {Lehmann}, \citenamefont {Symanzik},\ and\ \citenamefont
  {Zimmermann}}]{Lehmann:1954rq}%
  \BibitemOpen
  \bibfield  {author} {\bibinfo {author} {\bibfnamefont {H.}~\bibnamefont
  {Lehmann}}, \bibinfo {author} {\bibfnamefont {K.}~\bibnamefont {Symanzik}}, \
  and\ \bibinfo {author} {\bibfnamefont {W.}~\bibnamefont {Zimmermann}},\
  }\href {\doibase 10.1007/BF02731765} {\bibfield  {journal} {\bibinfo
  {journal} {Nuovo Cim.}\ }\textbf {\bibinfo {volume} {1}},\ \bibinfo {pages}
  {205} (\bibinfo {year} {1955})}\BibitemShut {NoStop}%
\bibitem [{\citenamefont {Colangelo}\ \emph {et~al.}(2001)\citenamefont
  {Colangelo}, \citenamefont {Gasser},\ and\ \citenamefont
  {Leutwyler}}]{Colangelo:2001df}%
  \BibitemOpen
  \bibfield  {author} {\bibinfo {author} {\bibfnamefont {G.}~\bibnamefont
  {Colangelo}}, \bibinfo {author} {\bibfnamefont {J.}~\bibnamefont {Gasser}}, \
  and\ \bibinfo {author} {\bibfnamefont {H.}~\bibnamefont {Leutwyler}},\ }\href
  {\doibase 10.1016/S0550-3213(01)00147-X} {\bibfield  {journal} {\bibinfo
  {journal} {Nucl. Phys. B}\ }\textbf {\bibinfo {volume} {603}},\ \bibinfo
  {pages} {125} (\bibinfo {year} {2001})},\ \Eprint
  {http://arxiv.org/abs/hep-ph/0103088} {arXiv:hep-ph/0103088} \BibitemShut
  {NoStop}%
\bibitem [{\citenamefont {Aoki}\ \emph {et~al.}(2022)\citenamefont {Aoki} \emph
  {et~al.}}]{FlavourLatticeAveragingGroupFLAG:2021npn}%
  \BibitemOpen
  \bibfield  {author} {\bibinfo {author} {\bibfnamefont {Y.}~\bibnamefont
  {Aoki}} \emph {et~al.} (\bibinfo {collaboration} {Flavour Lattice Averaging
  Group (FLAG)}),\ }\href {\doibase 10.1140/epjc/s10052-022-10536-1} {\bibfield
   {journal} {\bibinfo  {journal} {Eur. Phys. J. C}\ }\textbf {\bibinfo
  {volume} {82}},\ \bibinfo {pages} {869} (\bibinfo {year} {2022})},\ \Eprint
  {http://arxiv.org/abs/2111.09849} {arXiv:2111.09849 [hep-lat]} \BibitemShut
  {NoStop}%
\bibitem [{\citenamefont {Grilli~di Cortona}\ \emph {et~al.}(2016)\citenamefont
  {Grilli~di Cortona}, \citenamefont {Hardy}, \citenamefont {Pardo~Vega},\ and\
  \citenamefont {Villadoro}}]{diCortona:2015ldu}%
  \BibitemOpen
  \bibfield  {author} {\bibinfo {author} {\bibfnamefont {G.}~\bibnamefont
  {Grilli~di Cortona}}, \bibinfo {author} {\bibfnamefont {E.}~\bibnamefont
  {Hardy}}, \bibinfo {author} {\bibfnamefont {J.}~\bibnamefont {Pardo~Vega}}, \
  and\ \bibinfo {author} {\bibfnamefont {G.}~\bibnamefont {Villadoro}},\ }\href
  {\doibase 10.1007/JHEP01(2016)034} {\bibfield  {journal} {\bibinfo  {journal}
  {JHEP}\ }\textbf {\bibinfo {volume} {01}},\ \bibinfo {pages} {034} (\bibinfo
  {year} {2016})},\ \Eprint {http://arxiv.org/abs/1511.02867} {arXiv:1511.02867
  [hep-ph]} \BibitemShut {NoStop}%
\bibitem [{\citenamefont {Zyla}\ \emph {et~al.}(2020)\citenamefont {Zyla} \emph
  {et~al.}}]{Zyla:2020zbs}%
  \BibitemOpen
  \bibfield  {author} {\bibinfo {author} {\bibfnamefont {P.}~\bibnamefont
  {Zyla}} \emph {et~al.} (\bibinfo {collaboration} {Particle Data Group}),\
  }\href {\doibase 10.1093/ptep/ptaa104} {\bibfield  {journal} {\bibinfo
  {journal} {PTEP}\ }\textbf {\bibinfo {volume} {2020}},\ \bibinfo {pages}
  {083C01} (\bibinfo {year} {2020})}\BibitemShut {NoStop}%
\bibitem [{\citenamefont {Lindenbaum}\ and\ \citenamefont
  {Longacre}(1992)}]{Lindenbaum:1991tq}%
  \BibitemOpen
  \bibfield  {author} {\bibinfo {author} {\bibfnamefont {S.~J.}\ \bibnamefont
  {Lindenbaum}}\ and\ \bibinfo {author} {\bibfnamefont {R.~S.}\ \bibnamefont
  {Longacre}},\ }\href {\doibase 10.1016/0370-2693(92)92022-9} {\bibfield
  {journal} {\bibinfo  {journal} {Phys. Lett. B}\ }\textbf {\bibinfo {volume}
  {274}},\ \bibinfo {pages} {492} (\bibinfo {year} {1992})}\BibitemShut
  {NoStop}%
\bibitem [{\citenamefont {Estabrooks}\ and\ \citenamefont
  {Martin}(1974)}]{Estabrooks:1974vu}%
  \BibitemOpen
  \bibfield  {author} {\bibinfo {author} {\bibfnamefont {P.}~\bibnamefont
  {Estabrooks}}\ and\ \bibinfo {author} {\bibfnamefont {A.~D.}\ \bibnamefont
  {Martin}},\ }\href {\doibase 10.1016/0550-3213(74)90488-X} {\bibfield
  {journal} {\bibinfo  {journal} {Nucl. Phys. B}\ }\textbf {\bibinfo {volume}
  {79}},\ \bibinfo {pages} {301} (\bibinfo {year} {1974})}\BibitemShut
  {NoStop}%
\bibitem [{\citenamefont {Losty}\ \emph {et~al.}(1974)\citenamefont {Losty},
  \citenamefont {Chaloupka}, \citenamefont {Ferrando}, \citenamefont
  {Montanet}, \citenamefont {Paul}, \citenamefont {Yaffe}, \citenamefont
  {Zieminski}, \citenamefont {Alitti}, \citenamefont {Gandois},\ and\
  \citenamefont {Louie}}]{Losty:1973et}%
  \BibitemOpen
  \bibfield  {author} {\bibinfo {author} {\bibfnamefont {M.~J.}\ \bibnamefont
  {Losty}}, \bibinfo {author} {\bibfnamefont {V.}~\bibnamefont {Chaloupka}},
  \bibinfo {author} {\bibfnamefont {A.}~\bibnamefont {Ferrando}}, \bibinfo
  {author} {\bibfnamefont {L.}~\bibnamefont {Montanet}}, \bibinfo {author}
  {\bibfnamefont {E.}~\bibnamefont {Paul}}, \bibinfo {author} {\bibfnamefont
  {D.}~\bibnamefont {Yaffe}}, \bibinfo {author} {\bibfnamefont
  {A.}~\bibnamefont {Zieminski}}, \bibinfo {author} {\bibfnamefont
  {J.}~\bibnamefont {Alitti}}, \bibinfo {author} {\bibfnamefont
  {B.}~\bibnamefont {Gandois}}, \ and\ \bibinfo {author} {\bibfnamefont
  {J.}~\bibnamefont {Louie}},\ }\href {\doibase 10.1016/0550-3213(74)90131-X}
  {\bibfield  {journal} {\bibinfo  {journal} {Nucl. Phys. B}\ }\textbf
  {\bibinfo {volume} {69}},\ \bibinfo {pages} {185} (\bibinfo {year}
  {1974})}\BibitemShut {NoStop}%
\bibitem [{\citenamefont {Hoogland}\ \emph {et~al.}(1977)\citenamefont
  {Hoogland} \emph {et~al.}}]{Hoogland:1977kt}%
  \BibitemOpen
  \bibfield  {author} {\bibinfo {author} {\bibfnamefont {W.}~\bibnamefont
  {Hoogland}} \emph {et~al.},\ }\href {\doibase 10.1016/0550-3213(77)90154-7}
  {\bibfield  {journal} {\bibinfo  {journal} {Nucl. Phys. B}\ }\textbf
  {\bibinfo {volume} {126}},\ \bibinfo {pages} {109} (\bibinfo {year}
  {1977})}\BibitemShut {NoStop}%
\bibitem [{\citenamefont {Batley}\ \emph {et~al.}(2008)\citenamefont {Batley}
  \emph {et~al.}}]{NA482:2007xvj}%
  \BibitemOpen
  \bibfield  {author} {\bibinfo {author} {\bibfnamefont {J.~R.}\ \bibnamefont
  {Batley}} \emph {et~al.} (\bibinfo {collaboration} {NA48/2}),\ }\href
  {\doibase 10.1140/epjc/s10052-008-0547-0} {\bibfield  {journal} {\bibinfo
  {journal} {Eur. Phys. J. C}\ }\textbf {\bibinfo {volume} {54}},\ \bibinfo
  {pages} {411} (\bibinfo {year} {2008})}\BibitemShut {NoStop}%
\bibitem [{\citenamefont {Froggatt}\ and\ \citenamefont
  {Petersen}(1977)}]{Froggatt:1977hu}%
  \BibitemOpen
  \bibfield  {author} {\bibinfo {author} {\bibfnamefont {C.~D.}\ \bibnamefont
  {Froggatt}}\ and\ \bibinfo {author} {\bibfnamefont {J.~L.}\ \bibnamefont
  {Petersen}},\ }\href {\doibase 10.1016/0550-3213(77)90021-9} {\bibfield
  {journal} {\bibinfo  {journal} {Nucl. Phys. B}\ }\textbf {\bibinfo {volume}
  {129}},\ \bibinfo {pages} {89} (\bibinfo {year} {1977})}\BibitemShut
  {NoStop}%
\bibitem [{\citenamefont {Ochs}(1974)}]{Ochs:thesis1974}%
  \BibitemOpen
  \bibfield  {author} {\bibinfo {author} {\bibfnamefont {W.}~\bibnamefont
  {Ochs}},\ }\href@noop {} {\bibfield  {journal} {\bibinfo  {journal} {Ph.D.
  thesis, University of Munich}\ } (\bibinfo {year} {1974})}\BibitemShut
  {NoStop}%
\bibitem [{\citenamefont {Hyams}\ \emph {et~al.}(1973)\citenamefont {Hyams}
  \emph {et~al.}}]{Hyams:1973zf}%
  \BibitemOpen
  \bibfield  {author} {\bibinfo {author} {\bibfnamefont {B.}~\bibnamefont
  {Hyams}} \emph {et~al.},\ }\href {\doibase 10.1016/0550-3213(73)90618-4}
  {\bibfield  {journal} {\bibinfo  {journal} {Nucl. Phys. B}\ }\textbf
  {\bibinfo {volume} {64}},\ \bibinfo {pages} {134} (\bibinfo {year}
  {1973})}\BibitemShut {NoStop}%
\bibitem [{\citenamefont {Protopopescu}\ and\ \citenamefont {{\it et
  al.}}(1973)}]{Protopopescu:1973sh}%
  \BibitemOpen
  \bibfield  {author} {\bibinfo {author} {\bibfnamefont {S.~D.}\ \bibnamefont
  {Protopopescu}}\ and\ \bibinfo {author} {\bibnamefont {{\it et al.}}},\
  }\href {\doibase 10.1103/PhysRevD.7.1279} {\bibfield  {journal} {\bibinfo
  {journal} {Phys. Rev. D}\ }\textbf {\bibinfo {volume} {7}},\ \bibinfo {pages}
  {1279} (\bibinfo {year} {1973})}\BibitemShut {NoStop}%
\bibitem [{\citenamefont {Estabrooks}\ and\ \citenamefont {{\it et
  al.}}(1973)}]{Estabrooks:1973}%
  \BibitemOpen
  \bibfield  {author} {\bibinfo {author} {\bibfnamefont {P.}~\bibnamefont
  {Estabrooks}}\ and\ \bibinfo {author} {\bibnamefont {{\it et al.}}},\
  }\href@noop {} {\bibfield  {journal} {\bibinfo  {journal} {AIP Conf. Proc.}\
  }\textbf {\bibinfo {volume} {13}},\ \bibinfo {pages} {37} (\bibinfo {year}
  {1973})}\BibitemShut {NoStop}%
\bibitem [{\citenamefont {Grayer}\ and\ \citenamefont {{\it et
  al.}}(1972)}]{Grayer:1972}%
  \BibitemOpen
  \bibfield  {author} {\bibinfo {author} {\bibfnamefont {G.}~\bibnamefont
  {Grayer}}\ and\ \bibinfo {author} {\bibnamefont {{\it et al.}}},\ }\href@noop
  {} {\bibfield  {journal} {\bibinfo  {journal} {Proceedings of the 3rd
  Philadephia Conference on Experimental Meson Spectroscopy (American Institute
  of Physics, Philadephia, New York)}\ } (\bibinfo {year} {1972})}\BibitemShut
  {NoStop}%
\bibitem [{\citenamefont {Kaminski}\ \emph {et~al.}(1997)\citenamefont
  {Kaminski}, \citenamefont {Lesniak},\ and\ \citenamefont
  {Rybicki}}]{Kaminski:1996da}%
  \BibitemOpen
  \bibfield  {author} {\bibinfo {author} {\bibfnamefont {R.}~\bibnamefont
  {Kaminski}}, \bibinfo {author} {\bibfnamefont {L.}~\bibnamefont {Lesniak}}, \
  and\ \bibinfo {author} {\bibfnamefont {K.}~\bibnamefont {Rybicki}},\ }\href
  {\doibase 10.1007/s002880050372} {\bibfield  {journal} {\bibinfo  {journal}
  {Z. Phys. C}\ }\textbf {\bibinfo {volume} {74}},\ \bibinfo {pages} {79}
  (\bibinfo {year} {1997})},\ \Eprint {http://arxiv.org/abs/hep-ph/9606362}
  {arXiv:hep-ph/9606362} \BibitemShut {NoStop}%
\bibitem [{\citenamefont {Oller}\ and\ \citenamefont
  {Oset}(1999)}]{Oller:1998zr}%
  \BibitemOpen
  \bibfield  {author} {\bibinfo {author} {\bibfnamefont {J.~A.}\ \bibnamefont
  {Oller}}\ and\ \bibinfo {author} {\bibfnamefont {E.}~\bibnamefont {Oset}},\
  }\href {\doibase 10.1103/PhysRevD.60.074023} {\bibfield  {journal} {\bibinfo
  {journal} {Phys. Rev. D}\ }\textbf {\bibinfo {volume} {60}},\ \bibinfo
  {pages} {074023} (\bibinfo {year} {1999})},\ \Eprint
  {http://arxiv.org/abs/hep-ph/9809337} {arXiv:hep-ph/9809337} \BibitemShut
  {NoStop}%
\bibitem [{\citenamefont {Oller}(2020{\natexlab{b}})}]{Oller:2019opk}%
  \BibitemOpen
  \bibfield  {author} {\bibinfo {author} {\bibfnamefont {J.~A.}\ \bibnamefont
  {Oller}},\ }\href {\doibase 10.1016/j.ppnp.2019.103728} {\bibfield  {journal}
  {\bibinfo  {journal} {Prog. Part. Nucl. Phys.}\ }\textbf {\bibinfo {volume}
  {110}},\ \bibinfo {pages} {103728} (\bibinfo {year} {2020}{\natexlab{b}})},\
  \Eprint {http://arxiv.org/abs/1909.00370} {arXiv:1909.00370 [hep-ph]}
  \BibitemShut {NoStop}%
\bibitem [{\citenamefont {Watson}(1952)}]{Watson:1952ji}%
  \BibitemOpen
  \bibfield  {author} {\bibinfo {author} {\bibfnamefont {K.~M.}\ \bibnamefont
  {Watson}},\ }\href {\doibase 10.1103/PhysRev.88.1163} {\bibfield  {journal}
  {\bibinfo  {journal} {Phys. Rev.}\ }\textbf {\bibinfo {volume} {88}},\
  \bibinfo {pages} {1163} (\bibinfo {year} {1952})}\BibitemShut {NoStop}%
\bibitem [{\citenamefont {Pelaez}(2016)}]{Pelaez:2015qba}%
  \BibitemOpen
  \bibfield  {author} {\bibinfo {author} {\bibfnamefont {J.~R.}\ \bibnamefont
  {Pelaez}},\ }\href {\doibase 10.1016/j.physrep.2016.09.001} {\bibfield
  {journal} {\bibinfo  {journal} {Phys. Rept.}\ }\textbf {\bibinfo {volume}
  {658}},\ \bibinfo {pages} {1} (\bibinfo {year} {2016})},\ \Eprint
  {http://arxiv.org/abs/1510.00653} {arXiv:1510.00653 [hep-ph]} \BibitemShut
  {NoStop}%
\bibitem [{\citenamefont {Oller}(2019)}]{Oller:2019rej}%
  \BibitemOpen
  \bibfield  {author} {\bibinfo {author} {\bibfnamefont {J.~A.}\ \bibnamefont
  {Oller}},\ }\href {\doibase 10.1007/978-3-030-13582-9} {\emph {\bibinfo
  {title} {{A Brief Introduction to Dispersion Relations}}}},\ SpringerBriefs
  in Physics\ (\bibinfo  {publisher} {Springer},\ \bibinfo {year}
  {2019})\BibitemShut {NoStop}%
\bibitem [{\citenamefont {Dobado}\ and\ \citenamefont
  {Pelaez}(1997)}]{Dobado:1996ps}%
  \BibitemOpen
  \bibfield  {author} {\bibinfo {author} {\bibfnamefont {A.}~\bibnamefont
  {Dobado}}\ and\ \bibinfo {author} {\bibfnamefont {J.~R.}\ \bibnamefont
  {Pelaez}},\ }\href {\doibase 10.1103/PhysRevD.56.3057} {\bibfield  {journal}
  {\bibinfo  {journal} {Phys. Rev. D}\ }\textbf {\bibinfo {volume} {56}},\
  \bibinfo {pages} {3057} (\bibinfo {year} {1997})},\ \Eprint
  {http://arxiv.org/abs/hep-ph/9604416} {arXiv:hep-ph/9604416} \BibitemShut
  {NoStop}%
\bibitem [{\citenamefont {Oller}\ and\ \citenamefont
  {Oset}(1997)}]{Oller:1997ti}%
  \BibitemOpen
  \bibfield  {author} {\bibinfo {author} {\bibfnamefont {J.~A.}\ \bibnamefont
  {Oller}}\ and\ \bibinfo {author} {\bibfnamefont {E.}~\bibnamefont {Oset}},\
  }\href {\doibase 10.1016/S0375-9474(97)00160-7} {\bibfield  {journal}
  {\bibinfo  {journal} {Nucl. Phys. A}\ }\textbf {\bibinfo {volume} {620}},\
  \bibinfo {pages} {438} (\bibinfo {year} {1997})},\ \bibinfo {note} {[Erratum:
  Nucl.Phys.A 652, 407--409 (1999)]},\ \Eprint
  {http://arxiv.org/abs/hep-ph/9702314} {arXiv:hep-ph/9702314} \BibitemShut
  {NoStop}%
\bibitem [{\citenamefont {Oller}\ \emph {et~al.}(1999)\citenamefont {Oller},
  \citenamefont {Oset},\ and\ \citenamefont {Pelaez}}]{Oller:1998hw}%
  \BibitemOpen
  \bibfield  {author} {\bibinfo {author} {\bibfnamefont {J.~A.}\ \bibnamefont
  {Oller}}, \bibinfo {author} {\bibfnamefont {E.}~\bibnamefont {Oset}}, \ and\
  \bibinfo {author} {\bibfnamefont {J.~R.}\ \bibnamefont {Pelaez}},\ }\href
  {\doibase 10.1103/PhysRevD.59.074001} {\bibfield  {journal} {\bibinfo
  {journal} {Phys. Rev. D}\ }\textbf {\bibinfo {volume} {59}},\ \bibinfo
  {pages} {074001} (\bibinfo {year} {1999})},\ \bibinfo {note} {[Erratum:
  Phys.Rev.D 60, 099906 (1999), Erratum: Phys.Rev.D 75, 099903 (2007)]},\
  \Eprint {http://arxiv.org/abs/hep-ph/9804209} {arXiv:hep-ph/9804209}
  \BibitemShut {NoStop}%
\bibitem [{\citenamefont {Nieves}\ and\ \citenamefont
  {Ruiz~Arriola}(1999)}]{Nieves:1998hp}%
  \BibitemOpen
  \bibfield  {author} {\bibinfo {author} {\bibfnamefont {J.}~\bibnamefont
  {Nieves}}\ and\ \bibinfo {author} {\bibfnamefont {E.}~\bibnamefont
  {Ruiz~Arriola}},\ }\href {\doibase 10.1016/S0370-2693(99)00461-X} {\bibfield
  {journal} {\bibinfo  {journal} {Phys. Lett. B}\ }\textbf {\bibinfo {volume}
  {455}},\ \bibinfo {pages} {30} (\bibinfo {year} {1999})},\ \Eprint
  {http://arxiv.org/abs/nucl-th/9807035} {arXiv:nucl-th/9807035} \BibitemShut
  {NoStop}%
\bibitem [{\citenamefont {Dobado}\ \emph {et~al.}(1990)\citenamefont {Dobado},
  \citenamefont {Herrero},\ and\ \citenamefont {Truong}}]{Dobado:1989qm}%
  \BibitemOpen
  \bibfield  {author} {\bibinfo {author} {\bibfnamefont {A.}~\bibnamefont
  {Dobado}}, \bibinfo {author} {\bibfnamefont {M.~J.}\ \bibnamefont {Herrero}},
  \ and\ \bibinfo {author} {\bibfnamefont {T.~N.}\ \bibnamefont {Truong}},\
  }\href {\doibase 10.1016/0370-2693(90)90109-J} {\bibfield  {journal}
  {\bibinfo  {journal} {Phys. Lett. B}\ }\textbf {\bibinfo {volume} {235}},\
  \bibinfo {pages} {134} (\bibinfo {year} {1990})}\BibitemShut {NoStop}%
\bibitem [{\citenamefont {Truong}(1991)}]{Truong:1991gv}%
  \BibitemOpen
  \bibfield  {author} {\bibinfo {author} {\bibfnamefont {T.~N.}\ \bibnamefont
  {Truong}},\ }\href {\doibase 10.1103/PhysRevLett.67.2260} {\bibfield
  {journal} {\bibinfo  {journal} {Phys. Rev. Lett.}\ }\textbf {\bibinfo
  {volume} {67}},\ \bibinfo {pages} {2260} (\bibinfo {year}
  {1991})}\BibitemShut {NoStop}%
\bibitem [{\citenamefont {Dobado}\ and\ \citenamefont
  {Pelaez}(1993)}]{Dobado:1992ha}%
  \BibitemOpen
  \bibfield  {author} {\bibinfo {author} {\bibfnamefont {A.}~\bibnamefont
  {Dobado}}\ and\ \bibinfo {author} {\bibfnamefont {J.~R.}\ \bibnamefont
  {Pelaez}},\ }\href {\doibase 10.1103/PhysRevD.47.4883} {\bibfield  {journal}
  {\bibinfo  {journal} {Phys. Rev. D}\ }\textbf {\bibinfo {volume} {47}},\
  \bibinfo {pages} {4883} (\bibinfo {year} {1993})},\ \Eprint
  {http://arxiv.org/abs/hep-ph/9301276} {arXiv:hep-ph/9301276} \BibitemShut
  {NoStop}%
\bibitem [{\citenamefont {Salas-Bern\'ardez}\ \emph {et~al.}(2021)\citenamefont
  {Salas-Bern\'ardez}, \citenamefont {Llanes-Estrada}, \citenamefont
  {Escudero-Pedrosa},\ and\ \citenamefont {Oller}}]{Salas-Bernardez:2020hua}%
  \BibitemOpen
  \bibfield  {author} {\bibinfo {author} {\bibfnamefont {A.}~\bibnamefont
  {Salas-Bern\'ardez}}, \bibinfo {author} {\bibfnamefont {F.~J.}\ \bibnamefont
  {Llanes-Estrada}}, \bibinfo {author} {\bibfnamefont {J.}~\bibnamefont
  {Escudero-Pedrosa}}, \ and\ \bibinfo {author} {\bibfnamefont {J.~A.}\
  \bibnamefont {Oller}},\ }\href {\doibase 10.21468/SciPostPhys.11.2.020}
  {\bibfield  {journal} {\bibinfo  {journal} {SciPost Phys.}\ }\textbf
  {\bibinfo {volume} {11}},\ \bibinfo {pages} {020} (\bibinfo {year} {2021})},\
  \Eprint {http://arxiv.org/abs/2010.13709} {arXiv:2010.13709 [hep-ph]}
  \BibitemShut {NoStop}%
\bibitem [{\citenamefont {Workman}(2022)}]{Workman:2022ynf}%
  \BibitemOpen
  \bibfield  {author} {\bibinfo {author} {\bibfnamefont {R.~L.}\ \bibnamefont
  {Workman}} (\bibinfo {collaboration} {Particle Data Group}),\ }\href
  {\doibase 10.1093/ptep/ptac097} {\bibfield  {journal} {\bibinfo  {journal}
  {PTEP}\ }\textbf {\bibinfo {volume} {2022}},\ \bibinfo {pages} {083C01}
  (\bibinfo {year} {2022})}\BibitemShut {NoStop}%
\bibitem [{\citenamefont {Janssen}\ \emph {et~al.}(1995)\citenamefont
  {Janssen}, \citenamefont {Pearce}, \citenamefont {Holinde},\ and\
  \citenamefont {Speth}}]{Janssen:1994wn}%
  \BibitemOpen
  \bibfield  {author} {\bibinfo {author} {\bibfnamefont {G.}~\bibnamefont
  {Janssen}}, \bibinfo {author} {\bibfnamefont {B.~C.}\ \bibnamefont {Pearce}},
  \bibinfo {author} {\bibfnamefont {K.}~\bibnamefont {Holinde}}, \ and\
  \bibinfo {author} {\bibfnamefont {J.}~\bibnamefont {Speth}},\ }\href
  {\doibase 10.1103/PhysRevD.52.2690} {\bibfield  {journal} {\bibinfo
  {journal} {Phys. Rev. D}\ }\textbf {\bibinfo {volume} {52}},\ \bibinfo
  {pages} {2690} (\bibinfo {year} {1995})},\ \Eprint
  {http://arxiv.org/abs/nucl-th/9411021} {arXiv:nucl-th/9411021} \BibitemShut
  {NoStop}%
\bibitem [{\citenamefont {Albaladejo}\ and\ \citenamefont
  {Oller}(2012)}]{Albaladejo:2012te}%
  \BibitemOpen
  \bibfield  {author} {\bibinfo {author} {\bibfnamefont {M.}~\bibnamefont
  {Albaladejo}}\ and\ \bibinfo {author} {\bibfnamefont {J.~A.}\ \bibnamefont
  {Oller}},\ }\href {\doibase 10.1103/PhysRevD.86.034003} {\bibfield  {journal}
  {\bibinfo  {journal} {Phys. Rev. D}\ }\textbf {\bibinfo {volume} {86}},\
  \bibinfo {pages} {034003} (\bibinfo {year} {2012})},\ \Eprint
  {http://arxiv.org/abs/1205.6606} {arXiv:1205.6606 [hep-ph]} \BibitemShut
  {NoStop}%
\bibitem [{\citenamefont {Guerrero}\ and\ \citenamefont
  {Oller}(1999)}]{Guerrero:1998ei}%
  \BibitemOpen
  \bibfield  {author} {\bibinfo {author} {\bibfnamefont {F.}~\bibnamefont
  {Guerrero}}\ and\ \bibinfo {author} {\bibfnamefont {J.~A.}\ \bibnamefont
  {Oller}},\ }\href {\doibase 10.1016/S0550-3213(98)00663-4} {\bibfield
  {journal} {\bibinfo  {journal} {Nucl. Phys. B}\ }\textbf {\bibinfo {volume}
  {537}},\ \bibinfo {pages} {459} (\bibinfo {year} {1999})},\ \bibinfo {note}
  {[Erratum: Nucl.Phys.B 602, 641--643 (2001)]},\ \Eprint
  {http://arxiv.org/abs/hep-ph/9805334} {arXiv:hep-ph/9805334} \BibitemShut
  {NoStop}%
\bibitem [{\citenamefont {Gasser}\ and\ \citenamefont
  {Leutwyler}(1987{\natexlab{a}})}]{Gasser:1986vb}%
  \BibitemOpen
  \bibfield  {author} {\bibinfo {author} {\bibfnamefont {J.}~\bibnamefont
  {Gasser}}\ and\ \bibinfo {author} {\bibfnamefont {H.}~\bibnamefont
  {Leutwyler}},\ }\href {\doibase 10.1016/0370-2693(87)90492-8} {\bibfield
  {journal} {\bibinfo  {journal} {Phys. Lett. B}\ }\textbf {\bibinfo {volume}
  {184}},\ \bibinfo {pages} {83} (\bibinfo {year}
  {1987}{\natexlab{a}})}\BibitemShut {NoStop}%
\bibitem [{\citenamefont {Gasser}\ and\ \citenamefont
  {Leutwyler}(1987{\natexlab{b}})}]{Gasser:1987ah}%
  \BibitemOpen
  \bibfield  {author} {\bibinfo {author} {\bibfnamefont {J.}~\bibnamefont
  {Gasser}}\ and\ \bibinfo {author} {\bibfnamefont {H.}~\bibnamefont
  {Leutwyler}},\ }\href {\doibase 10.1016/0370-2693(87)91652-2} {\bibfield
  {journal} {\bibinfo  {journal} {Phys. Lett. B}\ }\textbf {\bibinfo {volume}
  {188}},\ \bibinfo {pages} {477} (\bibinfo {year}
  {1987}{\natexlab{b}})}\BibitemShut {NoStop}%
\bibitem [{\citenamefont {Gerber}\ and\ \citenamefont
  {Leutwyler}(1989)}]{Gerber:1988tt}%
  \BibitemOpen
  \bibfield  {author} {\bibinfo {author} {\bibfnamefont {P.}~\bibnamefont
  {Gerber}}\ and\ \bibinfo {author} {\bibfnamefont {H.}~\bibnamefont
  {Leutwyler}},\ }\href {\doibase 10.1016/0550-3213(89)90349-0} {\bibfield
  {journal} {\bibinfo  {journal} {Nucl. Phys. B}\ }\textbf {\bibinfo {volume}
  {321}},\ \bibinfo {pages} {387} (\bibinfo {year} {1989})}\BibitemShut
  {NoStop}%
\bibitem [{\citenamefont {Hannestad}\ and\ \citenamefont
  {Madsen}(1995)}]{Hannestad:1995rs}%
  \BibitemOpen
  \bibfield  {author} {\bibinfo {author} {\bibfnamefont {S.}~\bibnamefont
  {Hannestad}}\ and\ \bibinfo {author} {\bibfnamefont {J.}~\bibnamefont
  {Madsen}},\ }\href {\doibase 10.1103/PhysRevD.52.1764} {\bibfield  {journal}
  {\bibinfo  {journal} {Phys. Rev. D}\ }\textbf {\bibinfo {volume} {52}},\
  \bibinfo {pages} {1764} (\bibinfo {year} {1995})},\ \Eprint
  {http://arxiv.org/abs/astro-ph/9506015} {arXiv:astro-ph/9506015} \BibitemShut
  {NoStop}%
\bibitem [{\citenamefont {Leutwyler}(2006)}]{Leutwyler:2006qq}%
  \BibitemOpen
  \bibfield  {author} {\bibinfo {author} {\bibfnamefont {H.}~\bibnamefont
  {Leutwyler}},\ }in\ \href {\doibase 10.1142/9789812790804_0002} {\emph
  {\bibinfo {booktitle} {{5th International Workshop on Chiral Dynamics, theory
  and Experiment}}}}\ (\bibinfo {year} {2006})\ \Eprint
  {http://arxiv.org/abs/hep-ph/0612112} {arXiv:hep-ph/0612112} \BibitemShut
  {NoStop}%
\bibitem [{\citenamefont {Kolb}\ and\ \citenamefont
  {Turner}(1990)}]{Kolb:1990vq}%
  \BibitemOpen
  \bibfield  {author} {\bibinfo {author} {\bibfnamefont {E.~W.}\ \bibnamefont
  {Kolb}}\ and\ \bibinfo {author} {\bibfnamefont {M.~S.}\ \bibnamefont
  {Turner}},\ }\href@noop {} {\emph {\bibinfo {title} {{The Early
  Universe}}}},\ Vol.~\bibinfo {volume} {69}\ (\bibinfo {year}
  {1990})\BibitemShut {NoStop}%
\bibitem [{\citenamefont {Saikawa}\ and\ \citenamefont
  {Shirai}(2018)}]{Saikawa:2018rcs}%
  \BibitemOpen
  \bibfield  {author} {\bibinfo {author} {\bibfnamefont {K.}~\bibnamefont
  {Saikawa}}\ and\ \bibinfo {author} {\bibfnamefont {S.}~\bibnamefont
  {Shirai}},\ }\href {\doibase 10.1088/1475-7516/2018/05/035} {\bibfield
  {journal} {\bibinfo  {journal} {JCAP}\ }\textbf {\bibinfo {volume} {05}},\
  \bibinfo {pages} {035} (\bibinfo {year} {2018})},\ \Eprint
  {http://arxiv.org/abs/1803.01038} {arXiv:1803.01038 [hep-ph]} \BibitemShut
  {NoStop}%
\bibitem [{\citenamefont {Di~Luzio}\ \emph {et~al.}(2023)\citenamefont
  {Di~Luzio}, \citenamefont {Giannotti}, \citenamefont {Mescia}, \citenamefont
  {Nardi}, \citenamefont {Okawa},\ and\ \citenamefont
  {Piazza}}]{DiLuzio:2023tqe}%
  \BibitemOpen
  \bibfield  {author} {\bibinfo {author} {\bibfnamefont {L.}~\bibnamefont
  {Di~Luzio}}, \bibinfo {author} {\bibfnamefont {M.}~\bibnamefont {Giannotti}},
  \bibinfo {author} {\bibfnamefont {F.}~\bibnamefont {Mescia}}, \bibinfo
  {author} {\bibfnamefont {E.}~\bibnamefont {Nardi}}, \bibinfo {author}
  {\bibfnamefont {S.}~\bibnamefont {Okawa}}, \ and\ \bibinfo {author}
  {\bibfnamefont {G.}~\bibnamefont {Piazza}},\ }\href@noop {} {\  (\bibinfo
  {year} {2023})},\ \Eprint {http://arxiv.org/abs/2305.11958} {arXiv:2305.11958
  [hep-ph]} \BibitemShut {NoStop}%
\bibitem [{\citenamefont {Darm\'e}\ \emph {et~al.}(2020)\citenamefont
  {Darm\'e}, \citenamefont {Di~Luzio}, \citenamefont {Giannotti},\ and\
  \citenamefont {Nardi}}]{Darme:2020gyx}%
  \BibitemOpen
  \bibfield  {author} {\bibinfo {author} {\bibfnamefont {L.}~\bibnamefont
  {Darm\'e}}, \bibinfo {author} {\bibfnamefont {L.}~\bibnamefont {Di~Luzio}},
  \bibinfo {author} {\bibfnamefont {M.}~\bibnamefont {Giannotti}}, \ and\
  \bibinfo {author} {\bibfnamefont {E.}~\bibnamefont {Nardi}},\ }\href@noop {}
  {\  (\bibinfo {year} {2020})},\ \Eprint {http://arxiv.org/abs/2010.15846}
  {arXiv:2010.15846 [hep-ph]} \BibitemShut {NoStop}%
\bibitem [{\citenamefont {Hook}(2018)}]{Hook:2018jle}%
  \BibitemOpen
  \bibfield  {author} {\bibinfo {author} {\bibfnamefont {A.}~\bibnamefont
  {Hook}},\ }\href {\doibase 10.1103/PhysRevLett.120.261802} {\bibfield
  {journal} {\bibinfo  {journal} {Phys. Rev. Lett.}\ }\textbf {\bibinfo
  {volume} {120}},\ \bibinfo {pages} {261802} (\bibinfo {year} {2018})},\
  \Eprint {http://arxiv.org/abs/1802.10093} {arXiv:1802.10093 [hep-ph]}
  \BibitemShut {NoStop}%
\bibitem [{\citenamefont {Di~Luzio}\ \emph
  {et~al.}(2021{\natexlab{b}})\citenamefont {Di~Luzio}, \citenamefont {Gavela},
  \citenamefont {Quilez},\ and\ \citenamefont {Ringwald}}]{DiLuzio:2021pxd}%
  \BibitemOpen
  \bibfield  {author} {\bibinfo {author} {\bibfnamefont {L.}~\bibnamefont
  {Di~Luzio}}, \bibinfo {author} {\bibfnamefont {B.}~\bibnamefont {Gavela}},
  \bibinfo {author} {\bibfnamefont {P.}~\bibnamefont {Quilez}}, \ and\ \bibinfo
  {author} {\bibfnamefont {A.}~\bibnamefont {Ringwald}},\ }\href {\doibase
  10.1007/JHEP05(2021)184} {\bibfield  {journal} {\bibinfo  {journal} {JHEP}\
  }\textbf {\bibinfo {volume} {05}},\ \bibinfo {pages} {184} (\bibinfo {year}
  {2021}{\natexlab{b}})},\ \Eprint {http://arxiv.org/abs/2102.00012}
  {arXiv:2102.00012 [hep-ph]} \BibitemShut {NoStop}%
\bibitem [{\citenamefont {Di~Luzio}\ \emph
  {et~al.}(2021{\natexlab{c}})\citenamefont {Di~Luzio}, \citenamefont {Gavela},
  \citenamefont {Quilez},\ and\ \citenamefont {Ringwald}}]{DiLuzio:2021gos}%
  \BibitemOpen
  \bibfield  {author} {\bibinfo {author} {\bibfnamefont {L.}~\bibnamefont
  {Di~Luzio}}, \bibinfo {author} {\bibfnamefont {B.}~\bibnamefont {Gavela}},
  \bibinfo {author} {\bibfnamefont {P.}~\bibnamefont {Quilez}}, \ and\ \bibinfo
  {author} {\bibfnamefont {A.}~\bibnamefont {Ringwald}},\ }\href {\doibase
  10.1088/1475-7516/2021/10/001} {\bibfield  {journal} {\bibinfo  {journal}
  {JCAP}\ }\textbf {\bibinfo {volume} {10}},\ \bibinfo {pages} {001} (\bibinfo
  {year} {2021}{\natexlab{c}})},\ \Eprint {http://arxiv.org/abs/2102.01082}
  {arXiv:2102.01082 [hep-ph]} \BibitemShut {NoStop}%
\bibitem [{\citenamefont {Gomez~Nicola}\ \emph {et~al.}(2002)\citenamefont
  {Gomez~Nicola}, \citenamefont {Llanes-Estrada},\ and\ \citenamefont
  {Pelaez}}]{GomezNicola:2002tn}%
  \BibitemOpen
  \bibfield  {author} {\bibinfo {author} {\bibfnamefont {A.}~\bibnamefont
  {Gomez~Nicola}}, \bibinfo {author} {\bibfnamefont {F.~J.}\ \bibnamefont
  {Llanes-Estrada}}, \ and\ \bibinfo {author} {\bibfnamefont {J.~R.}\
  \bibnamefont {Pelaez}},\ }\href {\doibase 10.1016/S0370-2693(02)02959-3}
  {\bibfield  {journal} {\bibinfo  {journal} {Phys. Lett. B}\ }\textbf
  {\bibinfo {volume} {550}},\ \bibinfo {pages} {55} (\bibinfo {year} {2002})},\
  \Eprint {http://arxiv.org/abs/hep-ph/0203134} {arXiv:hep-ph/0203134}
  \BibitemShut {NoStop}%
\bibitem [{\citenamefont {Dobado}\ \emph {et~al.}(2002)\citenamefont {Dobado},
  \citenamefont {Gomez~Nicola}, \citenamefont {Llanes-Estrada},\ and\
  \citenamefont {Pelaez}}]{Dobado:2002xf}%
  \BibitemOpen
  \bibfield  {author} {\bibinfo {author} {\bibfnamefont {A.}~\bibnamefont
  {Dobado}}, \bibinfo {author} {\bibfnamefont {A.}~\bibnamefont
  {Gomez~Nicola}}, \bibinfo {author} {\bibfnamefont {F.~J.}\ \bibnamefont
  {Llanes-Estrada}}, \ and\ \bibinfo {author} {\bibfnamefont {J.~R.}\
  \bibnamefont {Pelaez}},\ }\href {\doibase 10.1103/PhysRevC.66.055201}
  {\bibfield  {journal} {\bibinfo  {journal} {Phys. Rev. C}\ }\textbf {\bibinfo
  {volume} {66}},\ \bibinfo {pages} {055201} (\bibinfo {year} {2002})},\
  \Eprint {http://arxiv.org/abs/hep-ph/0206238} {arXiv:hep-ph/0206238}
  \BibitemShut {NoStop}%
\bibitem [{\citenamefont {Notari}\ \emph {et~al.}(2022)\citenamefont {Notari},
  \citenamefont {Rompineve},\ and\ \citenamefont {Villadoro}}]{Notari:2022zxo}%
  \BibitemOpen
  \bibfield  {author} {\bibinfo {author} {\bibfnamefont {A.}~\bibnamefont
  {Notari}}, \bibinfo {author} {\bibfnamefont {F.}~\bibnamefont {Rompineve}}, \
  and\ \bibinfo {author} {\bibfnamefont {G.}~\bibnamefont {Villadoro}},\
  }\href@noop {} {\  (\bibinfo {year} {2022})},\ \Eprint
  {http://arxiv.org/abs/2211.03799} {arXiv:2211.03799 [hep-ph]} \BibitemShut
  {NoStop}%
\bibitem [{\citenamefont {Garcia-Martin}\ \emph {et~al.}(2011)\citenamefont
  {Garcia-Martin}, \citenamefont {Kaminski}, \citenamefont {Pelaez},
  \citenamefont {Ruiz~de Elvira},\ and\ \citenamefont
  {Yndurain}}]{Garcia-Martin:2011iqs}%
  \BibitemOpen
  \bibfield  {author} {\bibinfo {author} {\bibfnamefont {R.}~\bibnamefont
  {Garcia-Martin}}, \bibinfo {author} {\bibfnamefont {R.}~\bibnamefont
  {Kaminski}}, \bibinfo {author} {\bibfnamefont {J.~R.}\ \bibnamefont
  {Pelaez}}, \bibinfo {author} {\bibfnamefont {J.}~\bibnamefont {Ruiz~de
  Elvira}}, \ and\ \bibinfo {author} {\bibfnamefont {F.~J.}\ \bibnamefont
  {Yndurain}},\ }\href {\doibase 10.1103/PhysRevD.83.074004} {\bibfield
  {journal} {\bibinfo  {journal} {Phys. Rev. D}\ }\textbf {\bibinfo {volume}
  {83}},\ \bibinfo {pages} {074004} (\bibinfo {year} {2011})},\ \Eprint
  {http://arxiv.org/abs/1102.2183} {arXiv:1102.2183 [hep-ph]} \BibitemShut
  {NoStop}%
\bibitem [{\citenamefont {Hannah}(1999)}]{Hannah:1999ev}%
  \BibitemOpen
  \bibfield  {author} {\bibinfo {author} {\bibfnamefont {T.}~\bibnamefont
  {Hannah}},\ }\href {\doibase 10.1103/PhysRevD.60.017502} {\bibfield
  {journal} {\bibinfo  {journal} {Phys. Rev. D}\ }\textbf {\bibinfo {volume}
  {60}},\ \bibinfo {pages} {017502} (\bibinfo {year} {1999})},\ \Eprint
  {http://arxiv.org/abs/hep-ph/9905236} {arXiv:hep-ph/9905236} \BibitemShut
  {NoStop}%
\bibitem [{\citenamefont {Guo}\ \emph {et~al.}(2012)\citenamefont {Guo},
  \citenamefont {Oller},\ and\ \citenamefont {Ruiz~de Elvira}}]{Guo:2012yt}%
  \BibitemOpen
  \bibfield  {author} {\bibinfo {author} {\bibfnamefont {Z.-H.}\ \bibnamefont
  {Guo}}, \bibinfo {author} {\bibfnamefont {J.~A.}\ \bibnamefont {Oller}}, \
  and\ \bibinfo {author} {\bibfnamefont {J.}~\bibnamefont {Ruiz~de Elvira}},\
  }\href {\doibase 10.1103/PhysRevD.86.054006} {\bibfield  {journal} {\bibinfo
  {journal} {Phys. Rev. D}\ }\textbf {\bibinfo {volume} {86}},\ \bibinfo
  {pages} {054006} (\bibinfo {year} {2012})},\ \Eprint
  {http://arxiv.org/abs/1206.4163} {arXiv:1206.4163 [hep-ph]} \BibitemShut
  {NoStop}%
\bibitem [{\citenamefont {Albaladejo}\ and\ \citenamefont
  {Oller}(2008)}]{Albaladejo:2008qa}%
  \BibitemOpen
  \bibfield  {author} {\bibinfo {author} {\bibfnamefont {M.}~\bibnamefont
  {Albaladejo}}\ and\ \bibinfo {author} {\bibfnamefont {J.~A.}\ \bibnamefont
  {Oller}},\ }\href {\doibase 10.1103/PhysRevLett.101.252002} {\bibfield
  {journal} {\bibinfo  {journal} {Phys. Rev. Lett.}\ }\textbf {\bibinfo
  {volume} {101}},\ \bibinfo {pages} {252002} (\bibinfo {year} {2008})},\
  \Eprint {http://arxiv.org/abs/0801.4929} {arXiv:0801.4929 [hep-ph]}
  \BibitemShut {NoStop}%
\end{thebibliography}%

\end{document}